\documentclass[article]{emulateapj}
\usepackage{natbib}
\newcommand{\hi}{\mbox{\rm \ion{H}{1}}}

\newcommand{\htwo}{\mbox{\rm H$_2$}}
\newcommand{\jtwo}{\mbox{$2\rightarrow1$}}
\newcommand{\jone}{\mbox{$1\rightarrow0$}}

\newcommand{\xcounits}{\mbox{cm$^{-2}$ (K km s$^{-1}$)$^{-1}$}}

\shorttitle{HERACLES: The HERA CO--Line Extragalactic Survey}
\shortauthors{Leroy et al.}
\slugcomment{Published in \emph{The Astronomical Journal} (2009AJ, 137, 4670L}

\begin{document}
\title{HERACLES: The HERA CO--Line Extragalactic Survey}

\author{Adam K. Leroy\altaffilmark{1}, Fabian Walter\altaffilmark{1}, Frank
  Bigiel\altaffilmark{1}, Antonio Usero\altaffilmark{2,3}, Axel
  Weiss\altaffilmark{4}, Elias Brinks\altaffilmark{2}, W.J.G. de
  Blok\altaffilmark{5,6}, Robert C.  Kennicutt\altaffilmark{7}, Karl-Friedrich
  Schuster\altaffilmark{8}, Carsten Kramer\altaffilmark{9}, H.W.
  Wiesemeyer\altaffilmark{8,9}, H\'el\`ene Roussel\altaffilmark{1,10}}

\altaffiltext{1}{Max-Planck-Institut f{\"u}r Astronomie, K{\"o}nigstuhl 17,
  D-69117, Heidelberg, Germany}
\altaffiltext{2}{Centre for Astrophysics Research, University of
  Hertfordshire, Hatfield AL10 9AB, U.K.}
\altaffiltext{3}{Observatorio Astron\'omico Nacional, C/ Alfonso XII, 3, 28014, Madrid, 
Spain}
\altaffiltext{4}{MPIfR, Auf dem H\"{u}gel 69, 53121, Bonn, Germany}
\altaffiltext{5}{Research School of Astronomy \& Astrophysics, Mount Stromlo
  Observatory, Cotter Road, Weston ACT 2611, Australia}
\altaffiltext{6}{Department of Astronomy, University of Cape Town, Private Bag
  X3, Rondebosch 7701, South Africa}
\altaffiltext{7}{University of Cambridge, Institute of Astronomy, Madingley
  Road, Cambridge CB3 0HA, UK}
\altaffiltext{8}{IRAM, 300 rue de la Piscine, 38406 St. Martin d'H\`eres}
\altaffiltext{9}{IRAM, Avenida Divina Pastora 7, E-18012 Granada, Spain}
\altaffiltext{10}{Institut d'Astrophysique de Paris, CNRS et Universite P. \&
  M. Curie, 98 bis Blvd Arago, 75014 Paris, France}

\begin{abstract}
  We present the HERA CO--Line Extragalactic Survey (HERACLES), an atlas of CO
  emission from $18$ nearby galaxies that are also part of The \hi\ Nearby
  Galaxy Survey (THINGS) and the {\em Spitzer} Infrared Nearby Galaxies Survey
  (SINGS). We used the HERA multi-pixel receiver on the IRAM~30-m telescope to
  map the CO $J=\jtwo$ line over the full optical disk (defined by the
  isophotal radius $r_{25}$) of each target, at $13\arcsec$ angular resolution
  and $2.6$~km~s$^{-1}$ velocity resolution. Here we describe the observations
  and reduction of the data and show channel maps, azimuthally averaged
  profiles, integrated intensity maps, and peak intensity maps. The implied
  \htwo\ masses range from $7 \times 10^6$ to $6 \times 10^9$~M$_{\odot}$,
  with four low metallicity dwarf irregular galaxies yielding only upper
  limits. In the cases where CO is detected, the integrated \htwo
  -to-\hi\ ratios range from $0.02$ -- $1.13$ and \htwo -to-stellar mass
  ratios from $0.01$ to $0.25$.  Exponential scale lengths of the CO emission
  for our targets are in the range $0.8$ -- $3.2$ kpc, or $0.2 \pm
  0.05~r_{25}$. The intensity-weighted mean velocity of CO matches that of
  \hi\ very well, with a $1\sigma$ scatter of only $6$~km~s$^{-1}$. The CO
  $J=\jtwo / J=\jone$ line ratio varies over a range similar to that found in
  the Milky Way and other nearby galaxies, $\sim 0.6$--$1.0$, with higher
  values found in the centers of galaxies. The typical line ratio, $\sim 0.8$,
  could be produced by optically thick gas with an excitation temperature of
  $\sim 10$~K.
\end{abstract}


\section{Introduction}
\label{INTRO}

Molecular hydrogen, \htwo , is the phase of the interstellar medium (ISM) most
closely related to star formation. Clouds of molecular gas are thought to be
birthplaces of virtually all stars and \htwo\ often dominates the mass budget
of the interstellar medium (ISM) in the inner, most vigorously star-forming
parts of spiral galaxies ($\lesssim 0.5~r_{25}$). Unfortunately, \htwo\ lacks
a dipole moment and typical temperatures in giant molecular clouds (GMCs) are
too low to excite quadrupole or vibrational transitions. Therefore, indirect
approaches are required to estimate the distribution of molecular hydrogen.
Although not without associated uncertainties, CO line emission remains the
most straightforward and reliable tracer of \htwo\ in galaxies. It is
relatively bright and its ability to trace the bulk distribution of \htwo\ has
been confirmed via comparisons with gamma rays
\citep[e.g.,][]{LEBRUN83,STRONG88} and  dust emission
\citep[e.g.,][]{DESERT88,DAME01}. 

Since the first detections of CO emission from Milky Way molecular clouds
\citep{WILSON70} and other galaxies \citep{RICKARD75,SOLOMON75}, a number of
surveys have used CO to characterize the molecular content of galaxies.
\citet{YOUNG91} summarize the first two decades of such observations, during
which CO was observed in $\sim 100$ galaxies. These data yielded a basic
understanding of the radial distribution of CO in disk galaxies and the
relationship between \hi\ (atomic hydrogen), H$_2$, and star formation as a
function of morphology. This number expanded to $\sim 300$ with the
publication of The Five Colleges Radio Astronomy Observatory (FCRAO)
Extragalactic CO Survey \citep[henceforth, the ``FCRAO survey,''][]{YOUNG95},
which remains the definitive survey of CO in the local volume. 

The FCRAO survey and most of its predecessors focused on the CO $J=\jone$
transition, i.e. the transition between the first rotationally excited state
and the ground state. \citet{BRAINE93} observed both CO $J=\jtwo$ and
$J=\jone$ emission from the centers of $81$ galaxies (finding a typical line
ratio of $0.89 \pm 0.06$).  These data support the idea that CO $J=\jtwo$
emission is a viable tracer of \htwo\ in other galaxies; \citet{ISRAEL84} and
\citet{SAKAMOTO95} showed that the distribution of CO $J=\jtwo$ follows that
of CO $J=\jone$ in the Milky Way.

In the past decade, technical improvements --- especially the construction of
several millimeter-wave interferometers --- have allowed higher resolution
imaging of CO in galaxies. These have obtained maps of Local Group galaxies
with spatial resolution matched to the sizes of individual GMCs
\citep[e.g.,][]{FUKUI99,ENGARGIOLA03,LEROY06,ROSOLOWSKY07} and observations of
large samples of more distant galaxies with resolutions of only a few hundred
parsecs \citep[e.g.,][]{WALTER01,SAKAMOTO99}. Of particular note, the Berkeley
Illinois Maryland Association Survey of Nearby Galaxies \citep[BIMA
  SONG;][]{HELFER03} mapped CO emission at $\sim 6\arcsec$ resolution in $44$
nearby spirals. The relatively wide extent of the BIMA SONG maps (typical full
width $\sim 190\arcsec$) make this survey the first to systematically map the
disks of normal spiral galaxies with resolution matched to the scales over
which GMCs, the basic units of the molecular ISM, are believed to form.

Over the last few years another technical improvement, heterodyne receiver
arrays on single-dish telescopes observing at $1$--$3$~mm, has made it
possible to efficiently map CO emission from large areas of the sky. This
allows complete inventories of molecular gas in samples of nearby galaxies
that were prohibitive with single-pixel receivers (the FCRAO survey, by
contrast, only sampled the major axes of its targets).  \citet{SCHUSTER07}
demonstrated this capability by using the Heterodyne Receiver Array
\citep[HERA,][]{SCHUSTER04} on the IRAM 30-m to map CO $J=\jtwo$
emission from the whole disk of M~51.  \citet{KUNO07} presented CO
$J=\jone$ maps of 40 nearby spirals, largely carried out using the
BEARS focal plane array on the Nobeyama 45-m. In both cases, the large
diameter of the telescopes (and the fact that HERA observes at $230$~GHz)
ensure a resolution of $11$--$15\arcsec$, adequate to resolve spiral
structure, bars, rings, and large star-forming complexes.

Here we present the HERA CO--Line Extragalactic Survey (HERACLES). HERACLES is
a new survey of CO emission in nearby galaxies carried out with the HERA
receiver array on the IRAM 30-m telescope\footnote{IRAM is supported by
  CNRS/INSU (France), the MPG (Germany) and the IGN (Spain)}. The main
motivation for this new survey is to quantify the relationship between
atomic gas, molecular gas, and star formation in a significant sample of
galaxies. To meet this goal, HERACLES differs from previous CO surveys in two
main ways: the area surveyed and the sample.

In contrast to the FCRAO survey or BIMA SONG, we map the whole optical disk of
each galaxy with full spatial sampling and good sensitivity.  Despite the
large area surveyed, we still achieve a relatively good resolution of
$13\arcsec$, $\sim 500$~pc at the median distance of our sample. Because we
use a single dish telescope rather than an interferometer, we are sensitive to
extended structure that may be missed by the latter (which do not recover
extended structure on the sky by design).

The HERACLES sample overlaps that of The \hi\ Nearby Galaxy Survey
\citep[][]{WALTER08} and the {\em Spitzer} Infrared Nearby Galaxies Survey
\citep[][]{KENNICUTT03}, ensuring an immediately accessible multiwavelength
database spanning from radio to UV. These data offer windows on the diffuse
ISM, dust, embedded star formation, photodissociation regions, and both young
and old stars.

This paper describes the HERACLES observing (\S \ref{OBSERVE}) and data
reduction (\S \ref{REDUCE}) strategies. We show the distribution of CO
emission (\S \ref{RESULTS}) and compare HERACLES results to previous
observations of the \hi\ 21-cm (\S \ref{HICOMP}) and the CO $J=\jone$ (\S
\ref{COMPARE}) lines in the same galaxies. Two papers present the first round
of in-depth scientific analysis. In \citet{BIGIEL08}, we combine HERACLES with
{\sc Hi}, IR, and UV data to measure the relationship between the surface
densities of \hi , \htwo , and star formation rate (the ``Schmidt Law''). In
\citet{LEROY08} we use the same data to measure the star formation per unit
gas as a function of environment, comparing these measurements to proposed
theories.

\section{Observations}
\label{OBSERVE}

\begin{deluxetable}{l c c c c}
  \tabletypesize{\small} \tablewidth{0pt} \tablecolumns{5}
  \tablecaption{\label{SAMPLETAB} HERACLES Sample} \tablehead{
    \colhead{Galaxy} & \colhead{Dist.} & \colhead{Incl.} & \colhead{P.A.}\tablenotemark{a} & \colhead{$r_{25}$}\tablenotemark{b} \\
    & (Mpc) & ($\arcdeg$) & ($\arcdeg$) & ($\arcmin$)} \startdata
  NGC~628 & 7.3 & 7 & 20 & 4.9 \\
  NGC~925 & 9.2 & 66 & 287 & 5.3 \\
  Holmberg~II & 3.4 & 41 & 177 & 3.8 \\
  NGC~2841 & 14.1 & 74 & 154 & 5.3 \\
  NGC~2903 & 8.9 & 65 & 204 & 5.9 \\
  Holmberg~I & 3.8 & 12 & 50 & 1.6 \\
  NGC~2976 & 3.6 & 65 & 335 & 3.6 \\
  NGC~3184 & 11.1 & 16 & 179 & 3.7 \\
  NGC~3198 & 13.8 & 72 & 215 & 3.2 \\
  IC~2574 & 4.0 & 53 & 56 & 6.4 \\
  NGC~3351 & 10.1 & 41 & 192 & 3.6 \\
  NGC~3521 & 10.7 & 73 & 340 & 4.2 \\
  NGC~4214 & 2.9 & 44 & 65 & 3.4 \\
  NGC~4736 & 4.7 & 41 & 296 & 3.9 \\
  DDO~154 & 4.3 & 66 & 230 & 1.0 \\
  NGC~5055 & 10.1 & 59 & 102 & 5.9 \\
  NGC~6946 & 5.9 & 33 & 243 & 5.7 \\
  NGC~7331 & 14.7 & 76 & 168 & 4.6 \\
\enddata
\tablenotetext{a}{Position angle of major axis, measured north through east.}
\tablenotetext{b}{Radius of the $B$-band 25~mag/arcsec$^2$ isophote.}
\end{deluxetable}

\begin{deluxetable*}{l c c l c}
  \tabletypesize{\small} \tablewidth{0pt} \tablecolumns{5}
  \tablecaption{\label{OBSTAB} Observing Dates and Orientation} \tablehead{
    \colhead{Galaxy} & \colhead{Scan P.A.\tablenotemark{a}} & \colhead{Scan
      Legs\tablenotemark{b}} & \colhead{Dates Observed} &
    \colhead{Spectra\tablenotemark{c}} \\
  & ($\arcdeg$) & & & ($10^4$) }
 \startdata
  NGC~628 & $0$ & $9 \times 9$ & 10, 12, 13, 14 Jan 2007 & 105 \\
  NGC~925 & $287$ & $9 \times 4$ & 23\tablenotemark{d}, 30 Oct 2007;
  12\tablenotemark{e}, 13 Jan 2008; 25, 26 Mar 2008 & 55 \\
  Holmberg~II & $0$ & $5 \times 5$ & 21, 22 Feb 2008 & 28 \\
NGC~2841 & $153$ & $6 \times 2$ & 26, 27, 30 Oct 2007 & 31 \\
NGC~2903 & $204$ & $10 \times 4$ & 27\tablenotemark{d} Nov 2007; 13,
16, 17, 21 Feb 2008\tablenotemark{e} & 87 \\
Holmberg~I & $0$ & $3 \times 3$ & 30 Oct 2007; 1, 3 Nov 2007 & 18 \\
NGC~2976 & $335$ & $6 \times 3$ & 27, 28, 31 Oct 2007; 3 Nov 2007 & 44 \\
NGC~3184 & $0$ & $6 \times 6$ & 31 Oct 2007; 1, 4, 5 Nov 2007 & 66 \\
NGC~3198 & $215$ & $6 \times 2$ & 26, 28 Oct 2007; 2 Nov 2007 & 31 \\
IC~2574 & $55$ & $11 \times 6$ & 17, 21, 22, 25, 27 Feb 2008; 25, 26
Mar 2008 & 125 \\
NGC~3351 & $0$ & $6 \times 6$ & 31 Oct 2007; 1, 2 Nov 2007; 17 Jan 2008
& 70 \\
NGC~3521 & $340$ & $7 \times 3$ & 2, 3, 4 Nov 2007 & 50 \\
NGC~4214 & $65$ & $4 \times 4$ & 5 Nov 2007; 10, 11, 12 Jan 2008 & 38 \\
NGC~4736 & $0$ & $8 \times 4$ & 19, 20 Jan 2007; 21,
24\tablenotemark{d}, 25, 26 Feb 2007 & 75 \\
DDO~154 & $230$ & $4 \times 2$ & 13, 16 Feb 2008 & 16 \\
NGC~5055 & $0$ & $13 \times 5$ & 13\tablenotemark{e}, 14, 15, 16 Jan
2007 & 95 \\
NGC~6946 & $0$ & $11 \times 11$ & 13, 14, 15, 17, 18, 19 Jan 2007 & 207
\\
NGC~7331 & $168$ & $8 \times 3$ & 23\tablenotemark{d},
24\tablenotemark{d}, 26, 27 Oct 2007; 10, 11, 12 Jan 2008 & 63 \\
\enddata
\tablenotetext{a}{Orientation of major axis scans measured from north through
  east. See Figure \ref{STRATEGY}.}  
\tablenotetext{b}{Number of fully-sampled (i.e., back and forth)
  scans along short $\times$ long edge. Width of one scan is $\sim
  68\arcsec$.}  
\tablenotetext{c}{Number of spectra in the final reduction. Each represents an
  0.5 second integration by one polarization on one receiver. One hour
  integrating on source yields $\approx 13 \times 10^4$ spectra.}
\tablenotetext{d}{For this day, we adjusted the rejection threshold to
  remove more than the 10\% of spectra with the highest RMS about the baseline
  fit (see \S \ref{REJECT}).}
\tablenotetext{e}{For this day, only one polarization was usable.}
\end{deluxetable*}

\begin{figure}
  \plotone{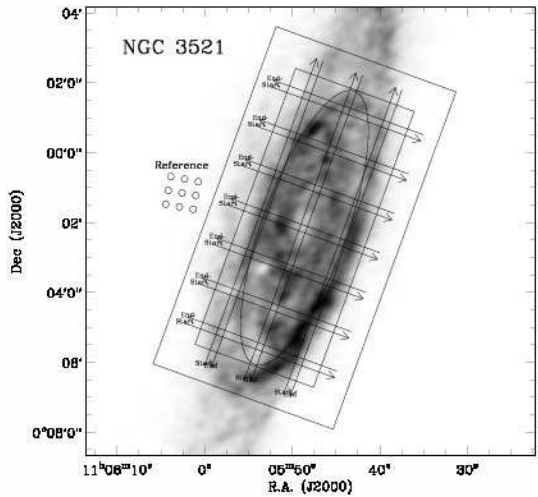}
  \caption{\label{STRATEGY} Our observing strategy illustrated for the
    inclined spiral galaxy NGC~3521.  The gray scale image shows the THINGS
    \hi\ column density map \citep{WALTER08} and the oval indicates $r_{25}$,
    our target area. Arrows indicate the length and direction of individual
    scan legs, which form back--and--forth pairs that together yield a
    fully-sampled map. To cover the galaxy we use $7$ back--and--forth scans
    along the short edge and $3$ along the long edge (indicated by the entry
    $7 \times 3$ in column 3 of Table \ref{OBSTAB}).  The beam print of HERA
    appears at the reference position, which was chosen to be free of gas but
    near the galaxy. Concentric rectangles show the areas of full (inner) and
    partial (outer) sensitivity.}
\end{figure}

The HERACLES sample consists of galaxies that are targets of THINGS, far
enough north to be easily observed by the 30-m, and not prohibitively large
($\lesssim 12 \times 12\arcmin$).  These $18$ spiral and irregular galaxies
are listed in Table \ref{SAMPLETAB}.  We draw the inclination, position angle,
and distances in this table from \citet[][]{WALTER08} and the isophotal
radius, $r_{25}$, from the LEDA database.

From January 2006 through March 2008, we used the HERA on the IRAM~30--m
telescope to map CO $J=2\rightarrow 1$ emission from these galaxies. HERA is a
multi--pixel receiver that simultaneously observes 9 positions on the sky at
two orthogonal linear polarizations.

We tuned HERA to observe the CO $J=\jtwo$ transition near $230$~GHz
and attached the HERA receivers to the Wideband Line Multiple Autocorrelator
(WILMA) backend. WILMA consists of $18$ units, each with 2~MHz channel width
and 930~MHz bandwidth, yielding a velocity resolution of $2.6$~km~s$^{-1}$ and
velocity coverage of $1200$~km~s$^{-1}$ at the wavelength of the CO
$J=\jtwo$ line.

Table \ref{OBSTAB} (column 4) gives the dates when individual galaxies were
observed. We generally observed only under good winter conditions, meaning
zenith opacities $\tau < 0.3$ at 225~GHz.  During most runs, the opacity was
better than this, often $\lesssim 0.1$.  System temperatures varied with
pixel, receiver, and semester but typical values were $275$--$350$~K for the 9
receiver pixels of the first polarization (HERA1) and $350$-$450$~K for the 9
receiver pixels of the second polarization (HERA2). On average, data from one
pixel (not always the same one) could not be used due to unreliable baselines.
During a few days, indicated in Table \ref{OBSTAB}, only one polarization was
usable.

We used on-the-fly mapping mode and scanned across the galaxy at
$8\arcsec$~sec$^{-1}$, writing out a spectrum every $0.5$ seconds, i.e.,
integrating over $4\arcsec$ of scanning. Approximately every $2$ minutes, we
observed a reference position near the galaxy for $10$--$15$ seconds. This was
chosen using the THINGS \hi\ maps to pick a position free of gas that lies
outside but near the optical disk of the galaxy.  Every $10$--$15$ minutes we
calibrated the intensity scale of the data using sky emission from the
reference position and the backend counts from loads at ambient and cold
(liquid nitrogen) temperatures.

In most cases the target area was the optical disk of the galaxy, defined by
the 25$^{\rm th}$ magnitude $B$-band isophote ($r_{25}$), though in several
galaxies we extended the map to probe obvious \hi\ peaks or filaments outside
the optical disk.

To construct a single map of a galaxy, we observed the target area using a
series of parallel scans. These ``long edge'' scans were usually aligned with
the major axis of the galaxy and were spaced to cover the whole target area.
We then immediately made a series of ``short edge'' scans covering the same
area, but oriented perpendicular to the original long edge scans.  The goal of
this cross--hatched observing strategy was to minimize artifacts in the final
data by ensuring that each area of the sky was observed by many different
receivers. Figure \ref{STRATEGY} illustrates this approach for NGC~3521, an
inclined Sb galaxy. The beam-print of the array is shown at the reference
position and arrows over the body of the galaxy indicate individual scans.
Here there are 3 ``long edge'' scans and 7 ``short edge'' scans. The second
and third columns in Table \ref{OBSTAB} give the orientation of the long edge
--- measured north through east --- and the number of scans along the short
and long edges ($7 \times 3$ for NGC~3521).

In order to fully sample the $11\arcsec$ (FWHM) beam of the 30--m, we used the
derotator to rotate the beam pattern of HERA by $9.5\arcdeg$ relative to the
direction of scanning\footnote{For a full description and illustration of this
  observing mode see the HERA User Manual (Version 2) by Schuster et al.
  (2006), available online at {\tt http://www.iram.fr/IRAMES}.}. This yields a
pixel spacing of $\approx 4\arcsec$ on the sky but leaves a gap between the
HERA pixels, which are separated by $28\arcsec$. We filled this gap by
offsetting the next scan by $11.9\arcsec$ perpendicular to the scan direction
and then repeating the original scan leg in reverse. This strategy yields a
fully sampled map with spectra spaced by $4\arcsec$ both along and
perpendicular to the scan direction.

We repeated the full set of long and short edge scans $15$--$20$ times for
each galaxy. This yielded the equivalent of $2$--$3$ minutes of integration
per independent beam. The fifth column in Table \ref{OBSTAB} lists the number
of individual spectra used in the final data cube for each galaxy (in units of
$10^4$). Each spectra corresponds to a 0.5 second integration with one
receiver, so that the full 18-element array produces 36 spectra each second of
on-source integration and an hour on source produces $13 \times 10^4$
spectra. Note that these numbers come after rejection of high-RMS spectra and
removal of bad pixels (\S \ref{REDUCE}) and that this time estimate takes no
account of reference observations, calibrations, slewing, tuning, or other
overheads.

\section{Reduction}
\label{REDUCE}

\subsection{Basic Calibration and Reference Subtraction }

We carried out the basic data reduction using the Multichannel Imaging and
Calibration Software for Receiver Arrays (MIRA)\footnote{{\tt
    http://www.iram.fr/IRAMFR/GILDAS/doc/html/mira-html/mira.html}.}, which is
part of the Grenoble Image and Line Data Analysis Software (GILDAS)
package\footnote{{\tt http://www.iram.fr/IRAMFR/GILDAS}}. This consisted of
combining each scan with the nearest reference measurement and calibration
observation via

\begin{equation}
  T_{\rm A}^* = T_{\rm cal} \times \frac{{\rm {\it On}}-{\rm {\it Off}}}{{\rm
      {\it Hot}}-{\rm {\it Off}}}~.
\end{equation}

\noindent Here {\it On} refers to the backend count rate from the on-source
measurement, {\it Off} to the count rate from the reference (empty sky)
measurements before and after the on-source measurement, $Hot$ to the count
rate from the ambient load during the calibration observation, and $T_{\rm
  cal}$ to a calibration factor determined from the calibration
observation\footnote{For a full discussion of the ``chopper wheel''
  calibration applied to the IRAM 30-m telescope, see ``Calibration of
  Spectral Line Data at the IRAM 30-m Telescope'' by Kramer (1997) available
  online at {\tt http://www.iram.fr/IRAMES}.}. The resulting antenna
temperature, $T_{\rm A}^*$, is defined to be the brightness temperature of a
source filling the entire $2\pi$ steradians in front of the telescope and
outside the atmosphere.

The data, in units of $T_{\rm A}^*$, were written out for further reduction
using the Continuum and Line Analysis Single-dish Software (CLASS) package,
which is also part of GILDAS.

\subsection{Fitting Baselines Based on \hi\ Velocity}
\label{BASEFIT}

\begin{figure*}  
  \epsscale{1.2}
  \plotone{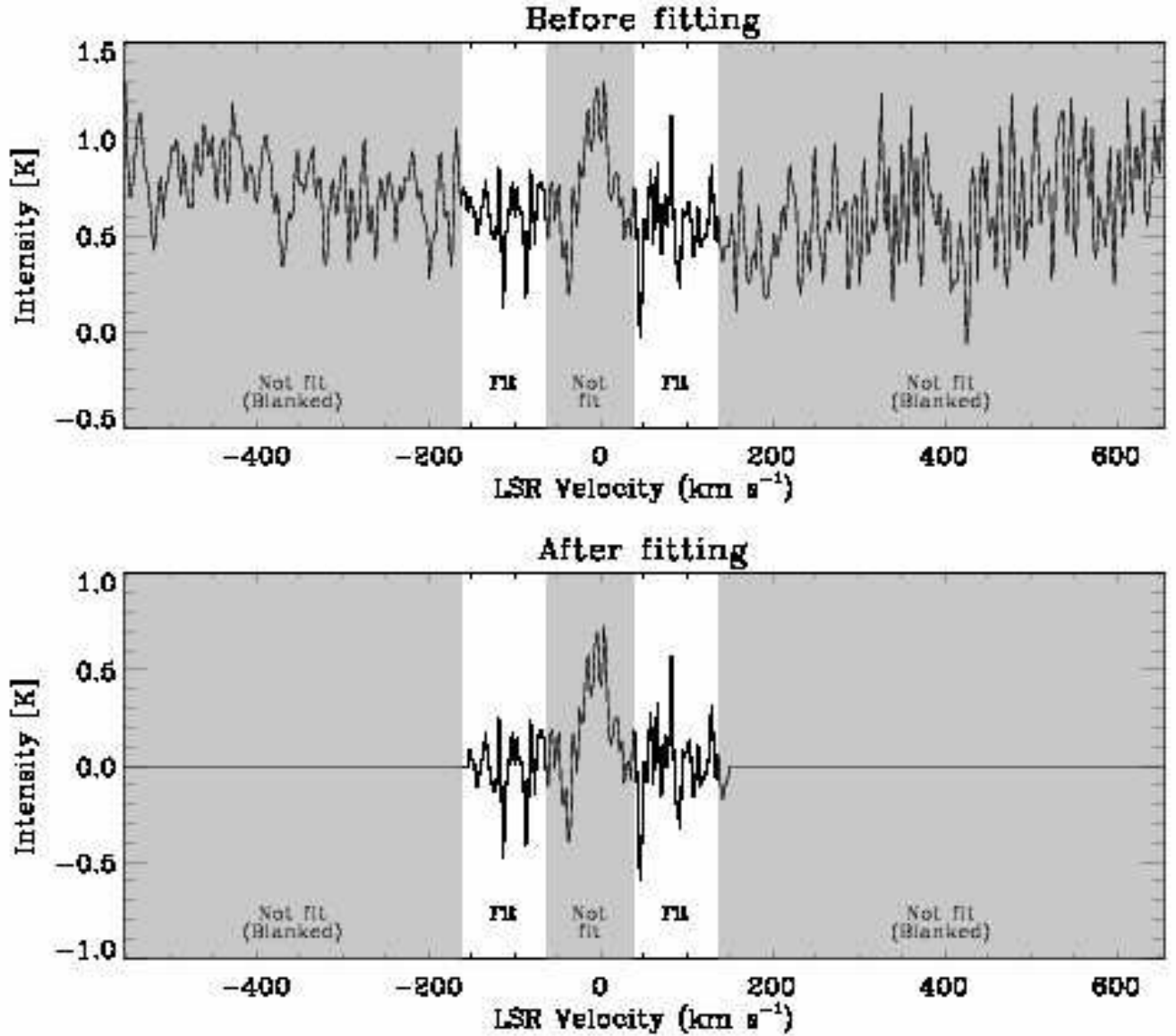}
  \caption{\label{FITTING} Our approach to baseline fitting, illustrated for a
    single (0.5 second) spectrum of NGC~6946. We use the mean
    \hi\ velocity to predict the likely velocity of CO emission. From this, we
    define three regions: one near the line that is likely to contain signal
    and two that are close to the line but displaced enough not to contain
    signal. We use the data in these two adjacent regions to fit a first-order
    baseline (while avoiding the central region). We subtract this baseline
    from the whole spectrum and discard data outside the fitting regions.}
\end{figure*}

After this basic reduction, total power variations in both the receiver and
the atmosphere and nonlinearities in the backend and receiver still made it
necessary to fit baselines to individual spectra before combining them into
data cubes. Because the data total $> 10^7$ spectra, this needed to be done in
an automated manner. 

The CO line typically covers a small but variable part of the bandpass, so
fitting a linear baseline to a single wide window did not yield satisfactory
results. Instead, we used the THINGS \hi\ data cubes to define the region of
the spectrum likely to contain the CO line and to fit a linear baseline to a
restricted part of the spectrum. This yielded good results and was
straightforward to automate. The underlying assumption is that \hi\ and
\htwo\ (traced by CO) are reasonably well-mixed, so that the mean velocity of
CO emission is similar to the mean \hi\ velocity. This is a reasonable
expectation: the two lines both trace dissipative gas moving in the same
potential well and \htwo\ is believed to form out of \hi . In \S \ref{HICOMP}
we use our data to verify this assumption, indeed finding an excellent match
between the mean velocities of \hi\ and CO.

The exact approach, illustrated in Figure \ref{FITTING}, was:

\begin{enumerate}
\item We defined a window over which the CO line was likely to appear. This
  window was centered on the mean \hi\ velocity and $60$--$300$~km~s$^{-1}$
  wide, depending on the galaxy. For many galaxies, we defined two windows:
  one used near the center of the galaxy and another used over the rest of the
  disk with the fitting region varying smoothly between the two regimes. We
  selected the window widths for each galaxy based on a preliminary reduction.
\item Adjacent to this window, we defined two regions of the spectrum that we
  used to fit a linear baseline. These had the same width as the central
  window, which was not included in the fit. The fit is subtracted from the
  whole spectrum.
\item We blank all data outside the fitting windows, so that only the fitting
  windows and the central region --- the likely location of signal --- are
  left.
\end{enumerate}

In this way, we fit baselines to all spectra in each galaxy. We reduced each
data cube three times. First, we made a crude reduction fitting baselines to a
fairly wide area around the \hi\ line.  Based on the results, we refined the
fitting region (white areas in Figure \ref{FITTING}, step 1 above), defining a
central window that was as narrow as possible while still including all of the
CO emission seen in the preliminary reduction. With the third reduction, we
iterated the process, using the second reduction to further refine our fitting
regions and more carefully identifying and removing pathological data (\S
\ref{REJECT}).

\subsection{Rejecting Remaining Pathological Spectra}
\label{REJECT}

A small fraction of spectra are still not well--fit by our approach. If
included in the final data cube, these introduce artifacts and obscure signal.
To remove them in a straightforward, systematic way, we discard the $10\%$ of
the data with the highest RMS residuals about the baseline fit from each day
of observing (note that these residuals are determined from the fitting
region, which is chosen to be signal free). Removing $10\%$ of the data causes
a negligible decrease in signal--to--noise but removing the pathological
spectra yields a noticeable increase in the quality of the data cubes. In a
few cases where a clear population of high RMS spectra survive the baseline
fitting, we adjust the rejection threshold to reject more data (these
observations are indicated in Table \ref{OBSTAB}). Typical values for the
$90^{\rm th}$ percentile are ${\rm RMS} \left(T_{\rm A}^*\right) \sim
0.35$--$0.7$~K for 2.6~km~s$^{-1}$ wide channels. 

\subsection{Constructing a Data Cube}

We combine the reduced spectra into a table and use the CLASS gridding routine
{\tt xy\_map} to construct a data cube with $2\arcsec$ pixel size and
$2.6$~km~s$^{-1}$ channel width. This routine convolves the irregularly
gridded OTF data with a Gaussian kernel with a full with width of $\sim 1/3$
the FWHM beam size, yielding a final angular resolution approximately
$11\arcsec$ and noise correlated on scales of $4 \arcsec$.

Because the noise is correlated on scales that are small compared to the beam
(response to astronomical signal), smoothing the data slightly offers a
significant gain in sensitivity while minimally degrading the beam size
\citep[e.g.,][]{MANGUM07}. In the rest of this paper, we show data that have
been convolved with a $7\arcsec$ (FWHM) Gaussian, yielding a final resolution
of $13\arcsec$.

We converted the units of the data cube from antenna temperature, $T_{\rm
  A}^*$, to main beam temperature, $T_{\rm MB}$. Main beam temperature is the
temperature of a source filling only the main beam of the telescope (recall
that $T_{\rm A}^*$ refers to a source filling the full forward $2\pi$ of the
sky). To convert from $T_{\rm A}^*$ to $T_{\rm MB}$, we replaced the forward
efficiency, $F_{\rm eff}=0.91$, with the main beam efficiency, $B_{\rm
  eff}=0.52$, yielding $T_{\rm MB} = 0.91 / 0.52~T_A^* = 1.75~T_A^*$
(efficiencies were adopted from the 30-m online documentation). For the
remainder of the paper, we work in units of $T_{\rm MB}$.

\subsection{Uncertainties}
\label{UNCERTAINTIES}

Table \ref{RESULTTAB} lists the RMS noise for each galaxy (column 2). We
estimate this from signal-free parts of the cube, which has $13\arcsec$
angular resolution and $2.6$~km~s$^{-1}$ channel width. Typical values are in
the range 20 -- 25~mK. Mostly this noise averages in the expected manner, but
on very large scales our method of calibration introduces an additional
consideration. Because we subtract the same reference measurements from all
spectra along an individual scan leg (Section \ref{OBSERVE}), low level
correlated noise is present in the maps. As a result, the RMS noise of spectra
derived from integrating entire data cubes is $2$--$3$ times higher than
expected from averaging independent, normal noise; this ratio is consistent
with the ratio of observing time spent on the reference to that on source.

We calculated how the integrated intensity of regions with very high ($\gtrsim
50$) signal--to--noise ratio (SNR) varied when measured with different
polarizations and on different days. We identified such regions in NGC~2903,
3351, 3521, 5055, and 6946 and then reduced the data separately for each day
and polarization. The RMS day-to-day scatter in the integrated intensity from
these high SNR regions is $\sim 20\%$. This is an approximate, but by no means
rigorous, measure of the uncertainty in the calibration of the telescope
(e.g., it includes the effect of pointing errors but not the uncertainty in
the assumed efficiencies of the telescope).

We tested the uncertainty associated with our method of reduction and baseline
fitting by reducing all of the data for one galaxy independently. We flagged
bad data and identified baseline fitting regions by eye. Over regions of
significant emission ($I_{\rm CO} >2\sigma$ in 3 successive channels at
$30\arcsec$ resolution) the two reductions agree very well.  The best-fit line
relating our automated reduction to the ``by-hand'' one in these regions has a
slope of $1.01$ and an intercept of $0.06$~K~km~s$^{-1}$. Along lines of sight
with strong CO emission ($I_{\rm CO} > 2.6$~K~km~s$^{-1}$), the ratio of the
two reductions scatters by $10\%$ (RMS). This check offers an estimate of the
uncertainty associated with how we carry out the data reduction and verifies
that our automated approach (which is less subjective and much simpler to
apply) agrees well with a ``by-hand'' reduction.

A final important uncertainty is related to our observing strategy.  In some
cases, pixel-to-pixel variations in gain, system temperature, and bandpass
shape leave the imprint of our cross-hatch observing strategy in the data.
This may be seen in the channel and integrated intensity maps present (\S
\ref{RESULTS}) as striping along the two scan directions (i.e., it is possible
to see the scanning path of individual pixels in the noise). The magnitude of
this striping is usually comparable to or below that of the statistical noise
and mostly these artifacts do not obscure or mimic signal. The exceptions are
four highly inclined spirals --- NGC~2841, 2903, 3521, and 7331 --- in which
significant striping is visible in individual channel maps. The likely cause
is a breakdown in our baseline fitting: because the CO line is often wide in
these galaxies, we are forced to use relatively broad fitting windows.

\section{The Distribution of CO $J=\jtwo$ Emission}
\label{RESULTS}

\begin{figure*}
  \plotone{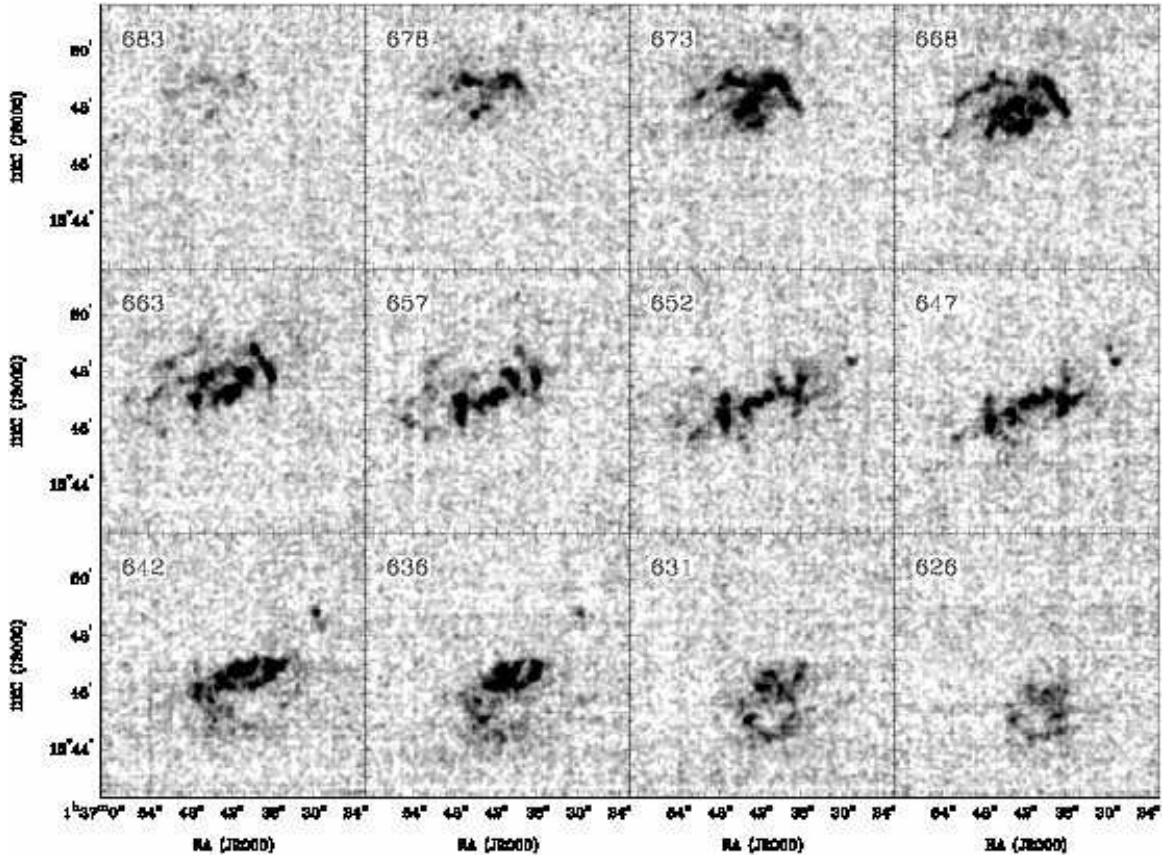} 

  \caption{ \label{NGC0628CHAN}Channel maps for NGC 628 at $13\arcsec$
  resolution. The grayscale runs from -0.025 to 0.15 K. Each channel is 5.2 km
  s$^{-1}$ wide and the RMS intensity in one channel is $19$~mK. The central
  LSR velocity for each channel (in km s$^{-1}$) appears in the top left
  corner of the map.}
\end{figure*}

\begin{figure*}
  \plotone{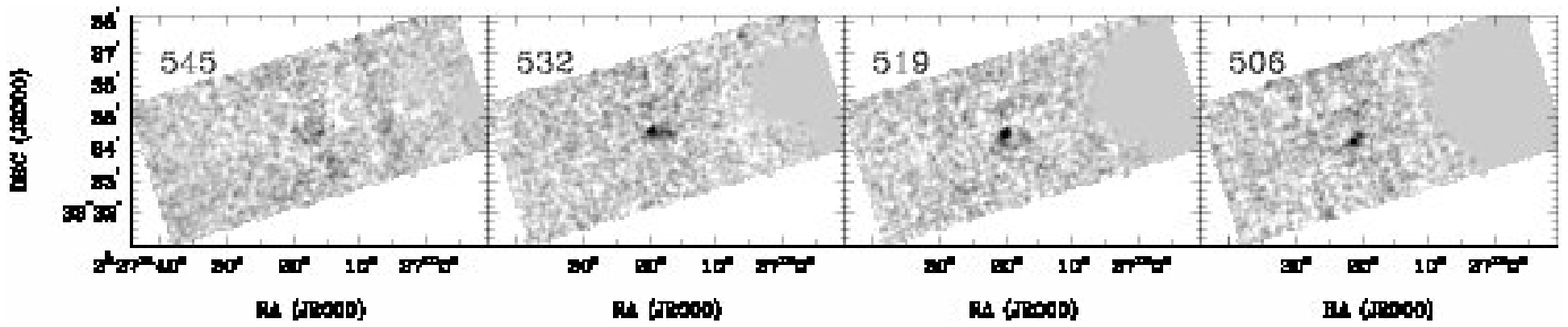}
  \label{NGC0925CHAN}
  
  \caption{Channel maps for NGC 925 at $13\arcsec$ resolution. The grayscale
    runs from -0.025 to 0.10 K. Each channel is 13 km s$^{-1}$ wide and the
    RMS intensity in one channel is $12$~mK. The smooth gray area indicates
    data at velocities outside the baseline fitting window, which are blanked
    during the reduction (\S \ref{BASEFIT}).}
\end{figure*}

\begin{figure*}
  \plotone{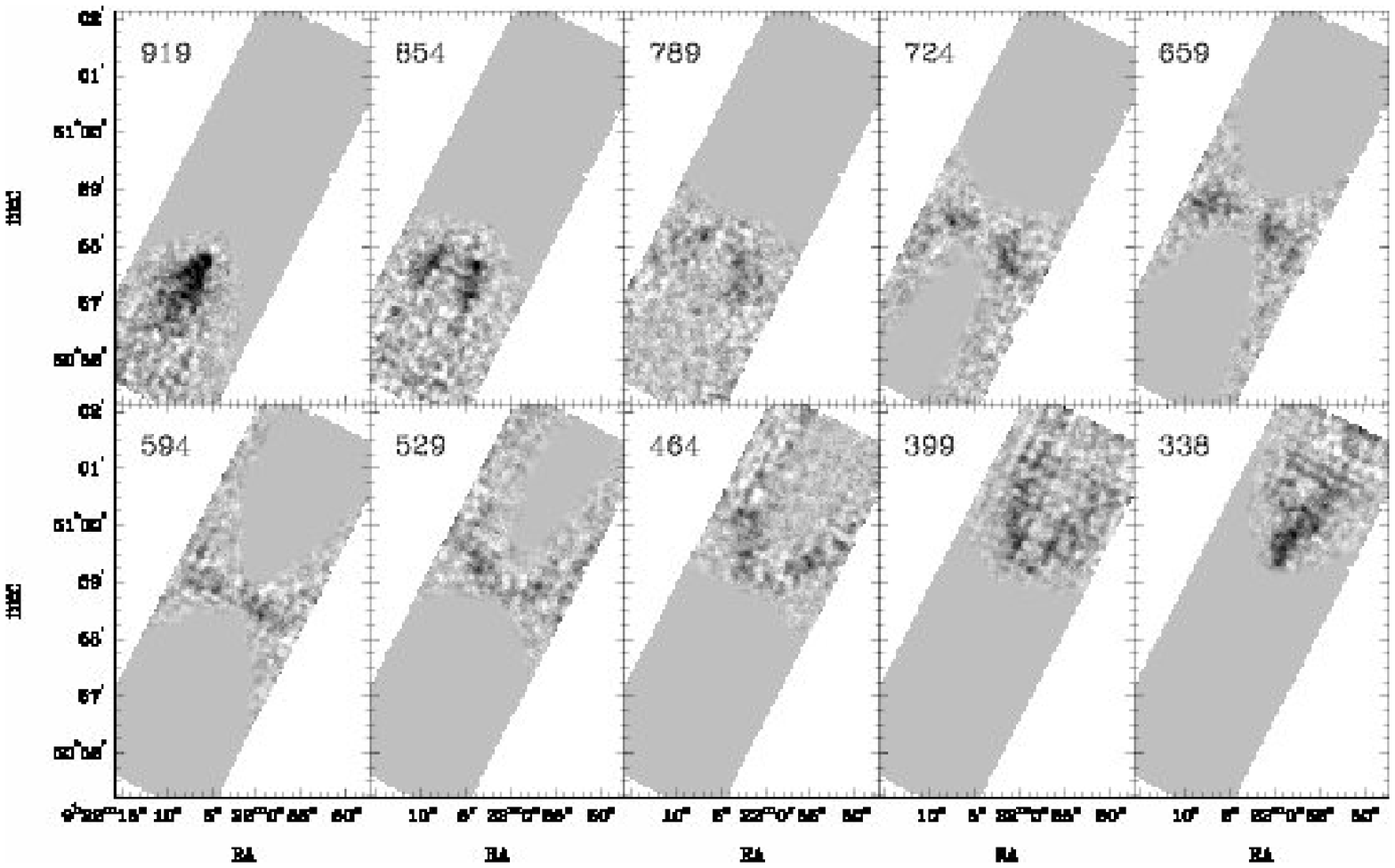}
  \label{NGC2841CHAN} 

\caption{Channel maps for NGC 2841 at $13\arcsec$
  resolution. The grayscale runs from -0.025 to 0.075 K. Each channel is 65 km
  s$^{-1}$ wide and the RMS intensity in one channel is $13$~mK. The smooth
  gray area indicates data at velocities outside the baseline fitting window,
  which are blanked during the reduction (\S \ref{BASEFIT}).}
\end{figure*}

\begin{figure*}
  \plotone{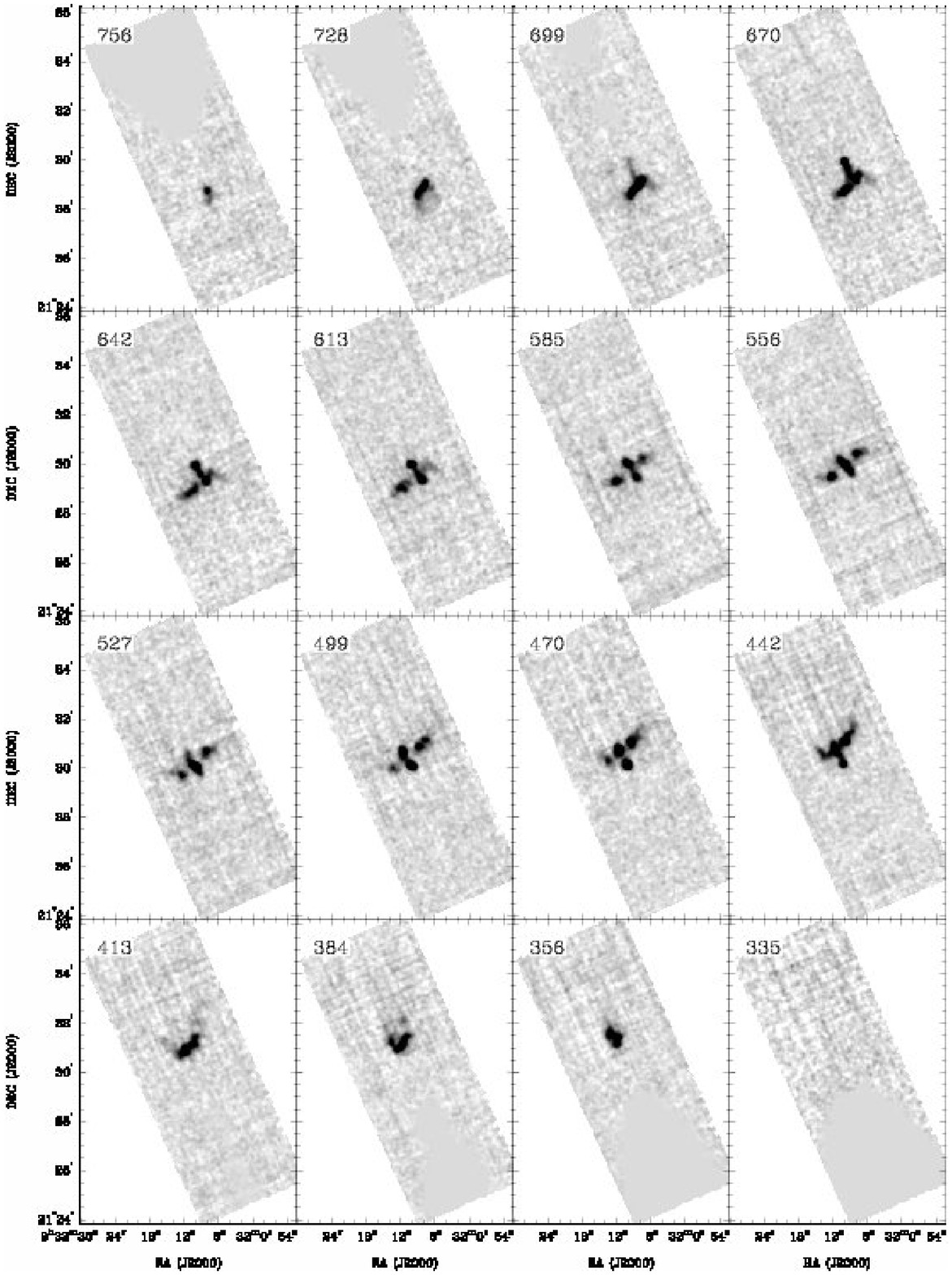} 

\caption{Channel maps for NGC 2903 at $13\arcsec$
  resolution. The grayscale runs from -0.025 to 0.15 K. Each channel is 28.6
  km s$^{-1}$ wide and the RMS intensity in one channel is $10$~mK.  The
  smooth gray area indicates data at velocities outside the baseline fitting
  window, which are blanked during the reduction (\S \ref{BASEFIT}).}
\end{figure*}

Our basic result is the distribution of CO $J=\jtwo$ emission in our
$18$ targets. In $14$ galaxies, extended emission is clearly detected. Here,
we show that emission using channel maps, peak intensity maps, integrated
intensity maps, and radial profiles. We make several basic measurements,
summarized in Table \ref{RESULTTAB}, including the integrated flux,
luminosity, and exponential scale length.

\begin{deluxetable*}{l c c c c c c c c}
  \tabletypesize{\small} \tablewidth{0pt} \tablecolumns{9}
  \tablecaption{\label{RESULTTAB} Basic Results} 
  \tablehead{
    \colhead{Galaxy} 
    & \colhead{$\sigma_{\rm rms}$} 
    & \colhead{$F_{\rm CO}$}
    & \colhead{$L_{\rm CO}$} 
    & \colhead{$l_{\rm CO}$}
    & \colhead{$d_{\rm max}$}
    & \colhead{$M_{\rm H2} / M_{\rm HI}$ }
    & \colhead{$M_{\rm H2} / M_*$} 
    & \colhead{$\left( M_{\rm H2} + M_{\rm HI} \right) / M_*$} 
    \\
    & (mK) & ($10^5$~K~km~s$^{-1}$~arcsec$^{2}$) &
    ($10^7$~K~km~s$^{-1}$~pc$^{2}$) 
    & (kpc) & ($r_{25}$) & & & }
  \startdata
  NGC~628 & 22 & 1.8 & 23  & 2.4 & 0.72 & 0.24 & 0.10 & 0.52 \\
  NGC~925 & 21 & 0.25 & 4.9 & 3.2 & 0.17 & 0.04 & 0.04 & 0.90 \\
  Holmberg~II & 36 & $<0.10$\tablenotemark{a} & $<0.33$\tablenotemark{a} 
  & \nodata & \nodata & \nodata & \nodata & $>3.6$ \\
  NGC~2841 & 44 & 0.61 & 28  & \nodata & 0.65 & 0.13 & 0.03 & 0.22 \\
  NGC~2903 & 23 & 2.9 & 53  & 1.6 & 0.47 & 0.50 & 0.25 & 0.77 \\
  Holmberg~I & 24 & $<0.34$\tablenotemark{a} & $<0.94$\tablenotemark{a} 
  & \nodata & \nodata & \nodata & \nodata & $> 7.2$ \\
  NGC~2976 & 21 & 0.40 & 1.2 & 1.2 & 0.53 & 0.36 & 0.05 & 0.18 \\
  NGC~3184 & 20 & 1.0 & 30 & 2.9 & 0.55 & 0.39 & 0.09 & 0.33 \\
  NGC~3198 & 17 & 0.25 & 11 & 2.7 & 0.43 & 0.04 & 0.04 & 1.03 \\
  IC~2574 & 33 & $<0.74$\tablenotemark{a} & $<2.8$\tablenotemark{a} 
  & \nodata & \nodata & \nodata & \nodata & $>3.8$ \\
  NGC~3351 & 20 & 0.78 & 19 & 2.5 & 0.41 & 0.63 & 0.05 & 0.12 \\
  NGC~3521 & 22 & 2.8 & 76 & 2.2 & 0.95 & 0.38 & 0.08 & 0.29 \\
  NGC~4214 & 18 & 0.09 & 0.17 & \nodata & 0.33 & 0.02 & 0.01 & 0.80 \\
  NGC~4736 & 23 & 2.2 & 11 & 0.8 & 1.33 & 1.13 & 0.03 & 0.06 \\
  DDO~154 & 18 & $<0.05$\tablenotemark{a} & $<0.21$\tablenotemark{a} 
  & \nodata & \nodata & \nodata & \nodata & $>34.8$ \\
  NGC~5055 & 24 & 4.1 & 98 & 3.1 & 0.67 & 0.43 & 0.08 & 0.26 \\
  NGC~6946 & 25 & 10.8 & 88 & 1.9 & 0.86 & 0.86 & 0.16 & 0.35 \\
  NGC~7331 & 20 & 2.2 & 113 & 3.1 & 0.61 & 0.50 & 0.07 & 0.21 \\
\enddata

\tablecomments{{\em Column 1:} galaxy; {\em Column 2:} RMS intensity n a
  2.6~km~s$^{-1}$-wide channel map with $13\arcsec$ (FWHM) angular resolution;
  {\em Column 3:} integrated flux of CO emission (see Equation \ref{FCOTOJY}
  for conversion to Jy km~s$^{-1}$); {\em Column 4:} integrated luminosity of
  CO emission (see Equation \ref{H2MASSEQ} for conversion to H$_2$ mass); {\em
    Column 5:} exponential scale length of CO emission; {\em Column 6:} radius
  of most extended high SNR detection (\S \ref{RADIAL}); {\em Column 8:} ratio
  of \htwo\ to \hi\ mass; {\em Column 9:} ratio of \htwo\ to stellar mass;
  {\em Column 10:} ratio of total gas ($\htwo + \hi$) to stellar mass.}
\tablenotetext{a}{$5\sigma$ upper limits.}
\end{deluxetable*}

\subsection{Channel, Integrated, and Peak Intensity Maps}

\begin{figure*}
  \plotone{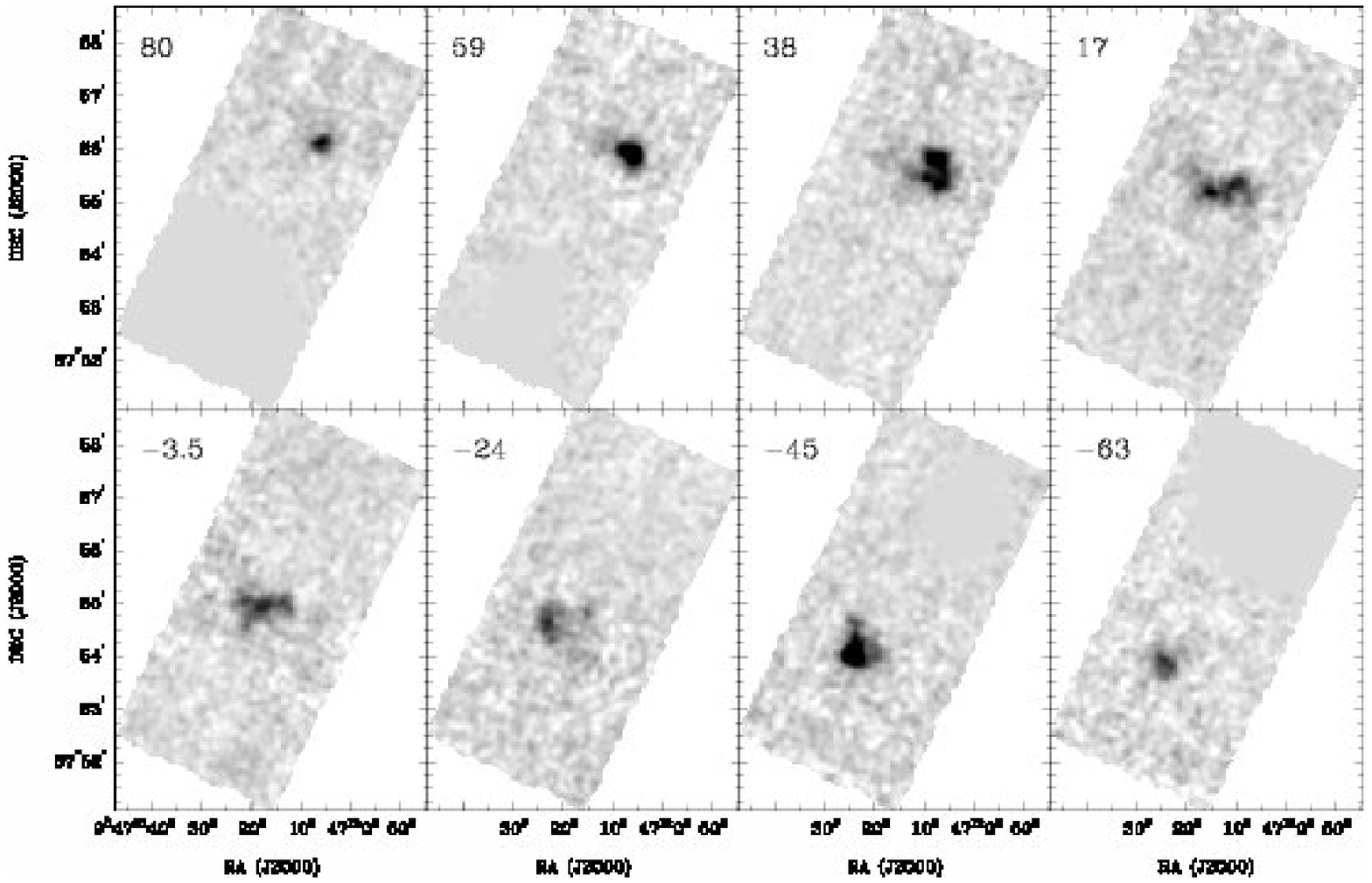}
  \label{NGC2976CHAN}

  \caption{Channel maps for NGC 2976 at $13\arcsec$ resolution. The grayscale
    runs from -0.025 to 0.15 K. Each channel is 20.8 km s$^{-1}$ wide and the
    RMS intensity in one channel is $9$~mK. The smooth gray area indicates data
    at velocities outside the baseline fitting window, which are blanked
    during the reduction (\S \ref{BASEFIT}).}
\end{figure*}

\begin{figure*}
  \plotone{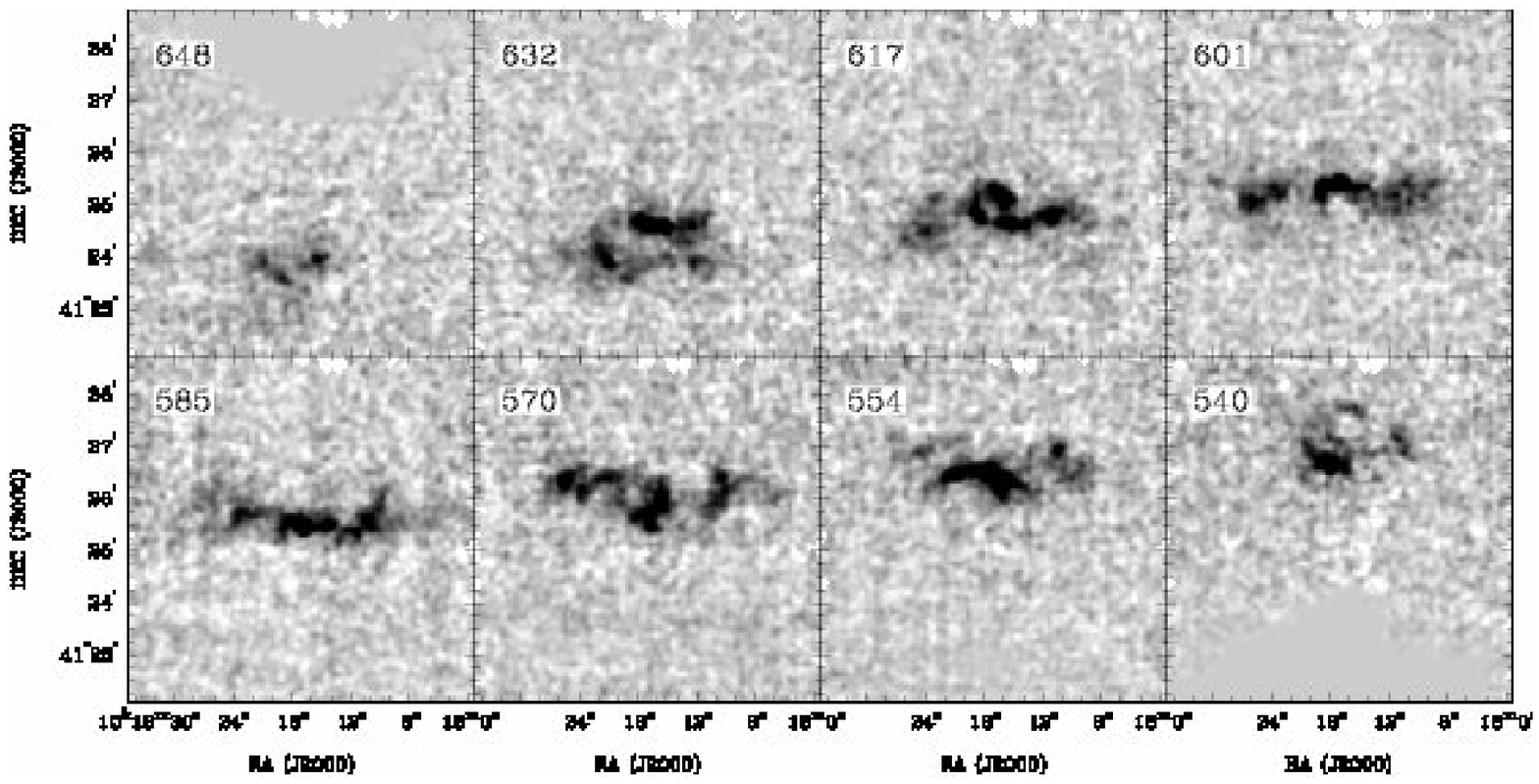} \label{NGC3184CHAN} 

\caption{Channel maps for NGC
  3184 at $13\arcsec$ resolution. The grayscale runs from -0.025 to 0.10 K.
  Each channel is 15.6 km s$^{-1}$ wide and the RMS intensity in one channel
  is $11$~mK. The smooth gray area indicates data at velocities outside the
  baseline fitting window, which are blanked during the reduction
  (\S \ref{BASEFIT}).}
\end{figure*}

\begin{figure*}
  \plotone{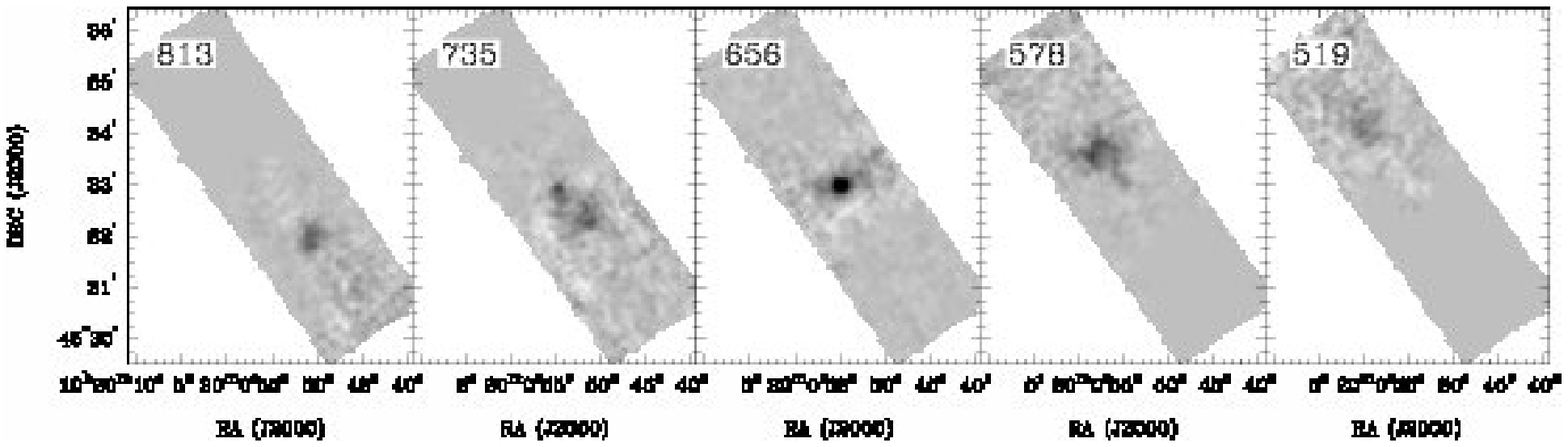}
  \label{NGC3198CHAN}

  \caption{Channel maps for NGC 3198 at $13\arcsec$ resolution. The grayscale
    runs from -0.025 to 0.10 K. Each channel is 78.0 km s$^{-1}$ wide and the
    RMS intensity in one channel is $4$~mK.  The smooth gray area indicates
    data at velocities outside the baseline fitting window, which are blanked
    during the reduction (\S \ref{BASEFIT}).}
\end{figure*}

\begin{figure*}
  \plotone{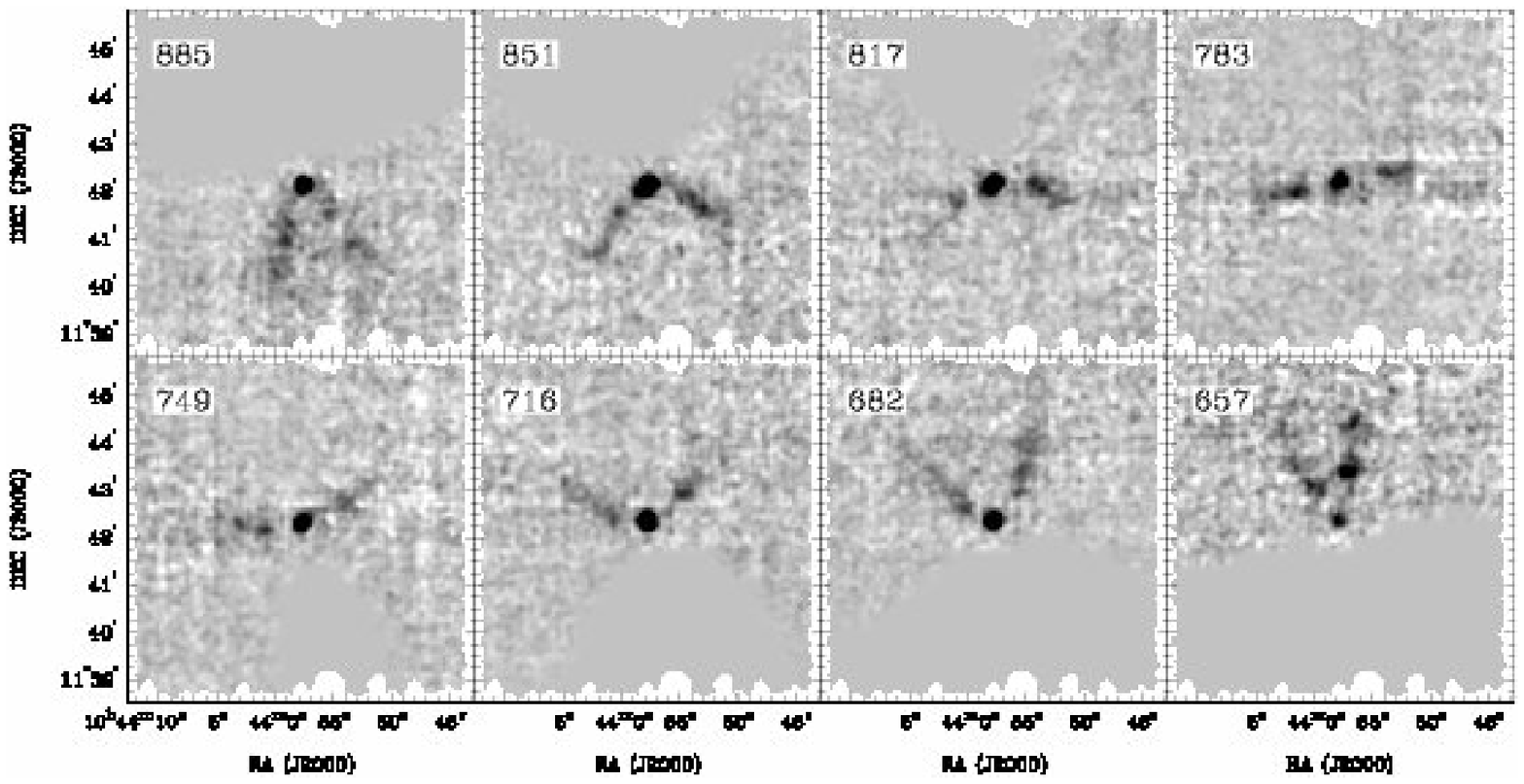} \label{NGC3351CHAN} 

\caption{Channel maps for NGC
  3351 at $13\arcsec$ resolution. The grayscale runs from -0.025 to 0.08 K.
  Each channel is 33.8 km s$^{-1}$ wide and the RMS intensity in one channel
  is $7$~mK. The smooth gray area indicates data at velocities outside the
  baseline fitting window, which are blanked during the reduction
  (\S \ref{BASEFIT}).}
\end{figure*}

\begin{figure*}
\plotone{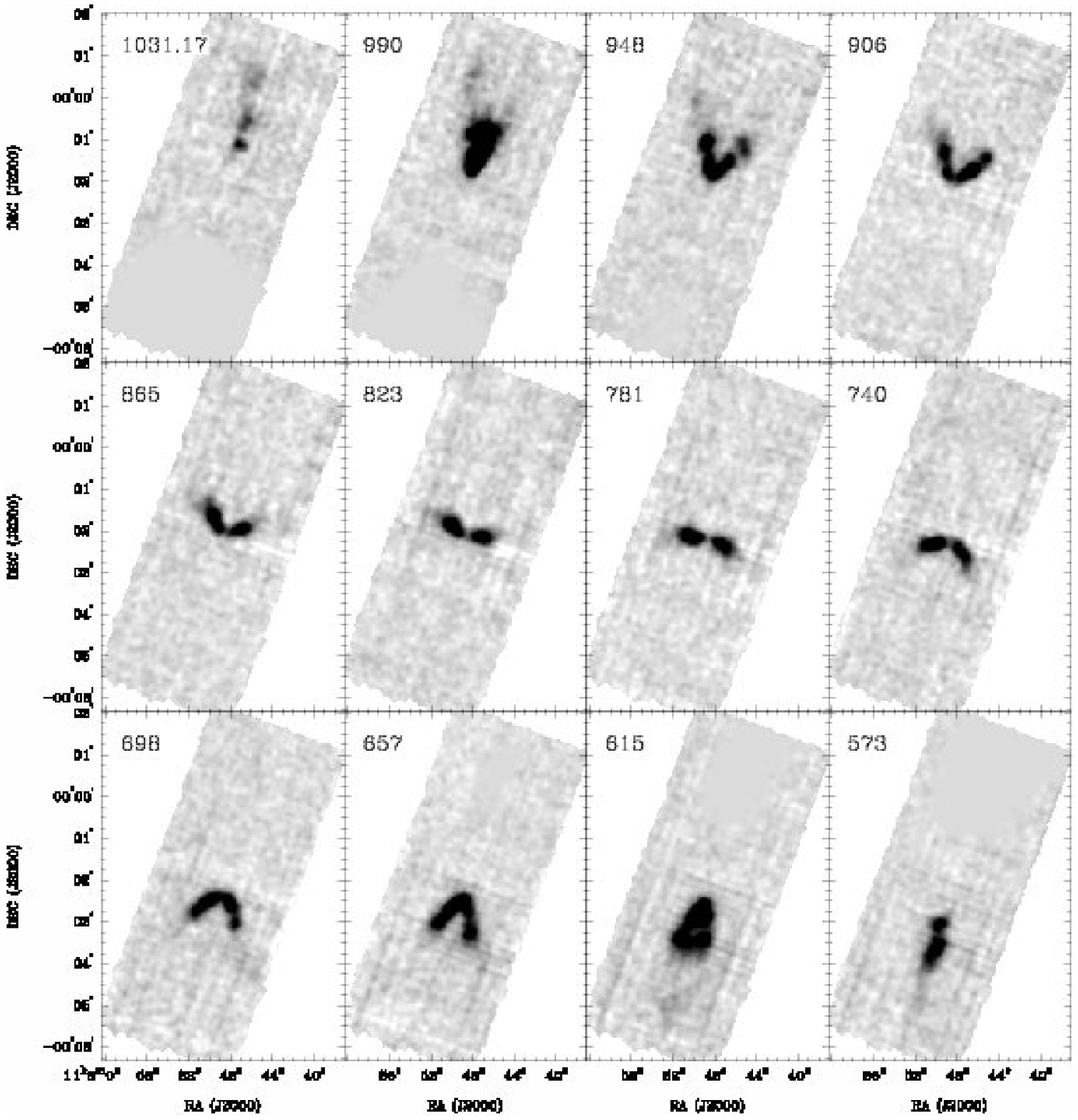} \label{NGC3521CHAN} 

\caption{Channel maps for NGC
  3521 at $13\arcsec$ resolution. The grayscale runs from -0.025 to 0.15 K.
  Each channel is 41.6 km s$^{-1}$ wide and the RMS intensity in one channel
  is $9$~mK. The smooth gray area indicates data at velocities outside the
  baseline fitting window, which are blanked during the reduction
  (\S \ref{BASEFIT}).}
\end{figure*}

\begin{figure*}
  \plotone{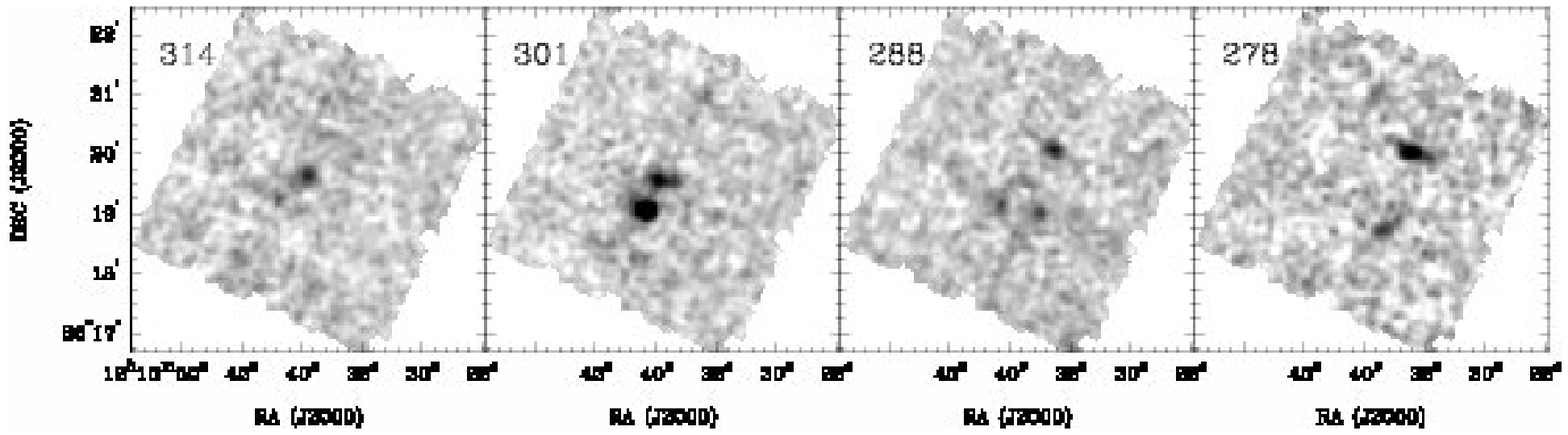}
  \label{NGC4214CHAN}

  \caption{Channel maps for NGC 4214 at $13\arcsec$ resolution. The grayscale
    runs from -0.025 to 0.1 K. Each channel is 13.0 km s$^{-1}$ wide and the
    RMS intensity in one channel is $10$~mK. The smooth gray area indicates data
    at velocities outside the baseline fitting window, which are blanked
    during the reduction (\S \ref{BASEFIT}).}
\end{figure*}

Figures \ref{NGC0628CHAN} -- \ref{NGC7331CHAN} show channel maps (i.e.,
intensity integrated over a succession of velocity ranges) of each clearly
detected galaxy.  These span approximately the velocity range over which we
find CO emission and show the area of uniform sensitivity (i.e., the inner
rectangle in Figure \ref{STRATEGY}). The gray scale and final channel width
vary from galaxy to galaxy and are indicated in each caption. Data at
velocities outside the baseline fitting region (\S \ref{BASEFIT}) are blanked
in the reduction and appear as zeroes (smooth gray regions) in the channel
maps.

The left column in Figures \ref{NGC0628MOM0} -- \ref{NGC7331MOM0} shows
integrated intensity for the same $14$ galaxies. A bar next to each figure
indicates the color stretch, which varies from galaxy to galaxy. Contours in
the integrated intensity maps run from 1 K~km~s$^{-1}$ (gray), increasing by
factors of 2 each step. The first black contour, $I_{\rm CO} =
2$~K~km~s$^{-1}$, corresponds to $\Sigma_{\rm H2} \sim
11$~M$_{\odot}$~pc$^{-2}$ in a face-on spiral galaxy (Equation
\ref{H2SURFEQ}). This approximately indicates where the ISM is dominated by
H$_2$. The first white contour, $I_{\rm CO} = 32$~K~km~s$^{-1}$, corresponds
to $\sim 176$~M$_{\odot}$~pc$^{-2}$, roughly the surface density of an
individual Galactic GMC \citep{SOLOMON87}. Regions with such high surface
densities over a large area are rare but not unknown in our survey (e.g., the
bar of NGC~2903 and the centers of NGC~2903, 3351, 4736, 5055, 6946).

We created the integrated intensity maps by summing along the velocity axis
over regions that contain signal. To identify these regions, we first smooth
the data to $30\arcsec$ angular resolution and bin to reach a velocity
resolution equal to half the width of the channel maps (this varies with
galaxy, see Figures \ref{NGC0628CHAN} -- \ref{NGC7331CHAN}). All regions with
$I_{\rm CO}>3\sigma$ in two consecutive velocity channels in these low
resolution cubes were labeled ``signal,'' and we integrated the original cube
over these regions to yield the maps in Figures \ref{NGC0628MOM0} --
\ref{NGC7331MOM0}. This procedure has a low false positive rate ($\lesssim 1$
expected per cube) so that most emission seen here is real with high
confidence, though in a few cases (NGC~2841, 2903, 7331) artifacts from
baseline fitting persist. Because the condition to identify signal is fairly
restrictive these maps are not ideal indicators of low brightness, extended
emission.

The peak intensity maps in the right column of Figures \ref{NGC0628MOM0} --
\ref{NGC7331MOM0} give a clearer view of the area mapped, extent of low
brightness emission, and morphology. These are created by measuring the peak
brightness along each line of sight at 2.6~km~s$^{-1}$ velocity
resolution. This procedure suppresses the worst artifacts in our data, which
tend to persist at a low level over many consecutive channels. It also lowers
the contrast between bright and faint regions, which arise partially from
differences in the line width. As a result, several interesting low-lying
features are visible in these maps: e.g., arms extending out from the
molecular rings in NGC~3521 and NGC~7331; faint filamentary structure in the
disk of NGC~4736; and a faint, previously unidentified molecular complex in
the southwest of the dwarf starburst NGC~4214 \citep[c.f.,][]{WALTER01}.

Peak intensity maps of inclined systems tend to show radial structure on small
scales. In order for emission to add constructively in such a map, it must be
at the same velocity.  When a galaxy with a rapidly changing velocity field is
observed with finite angular resolution and then collapsed to a peak intensity
map, signal will tend to be smoothed along isovelocity contours, leaving them
faintly visible in the final image.

The channel, integrated, and peak intensity maps show extended molecular
structures covering the inner part of the disk in most of our galaxies. The
detailed morphologies are fairly varied. Spiral structure is particularly
evident in the Sc galaxies NGC~628, NGC3184, and NGC~6946.  NGC~2903 is
dominated by a central, bright bar. The Sb galaxies NGC~3351 and NGC~7331 both
show well-defined molecular rings and similar structures are suggested by the
channel maps for NGC~2841 and NGC~3198.  Molecular gas completely covers the
inner parts of NGC~3521, NGC~4736, and NGC~5055. Over this area, NGC~3521 and
NGC~4736 show ring-like structure and outside this central region, NGC~3521,
NGC~4736, NGC~5055, and NGC~7331 show spiral structure. The late-type galaxies
NGC~925 and NGC~4214 both show only faint CO emission, which is common for
late-type galaxies. The other low mass galaxy that we detect, NGC~2976, defies
this trend, showing bright CO dominated by two large star--forming complexes.

\subsection{Radial Profiles}
\label{RADIAL}

\begin{figure*}
  \plotone{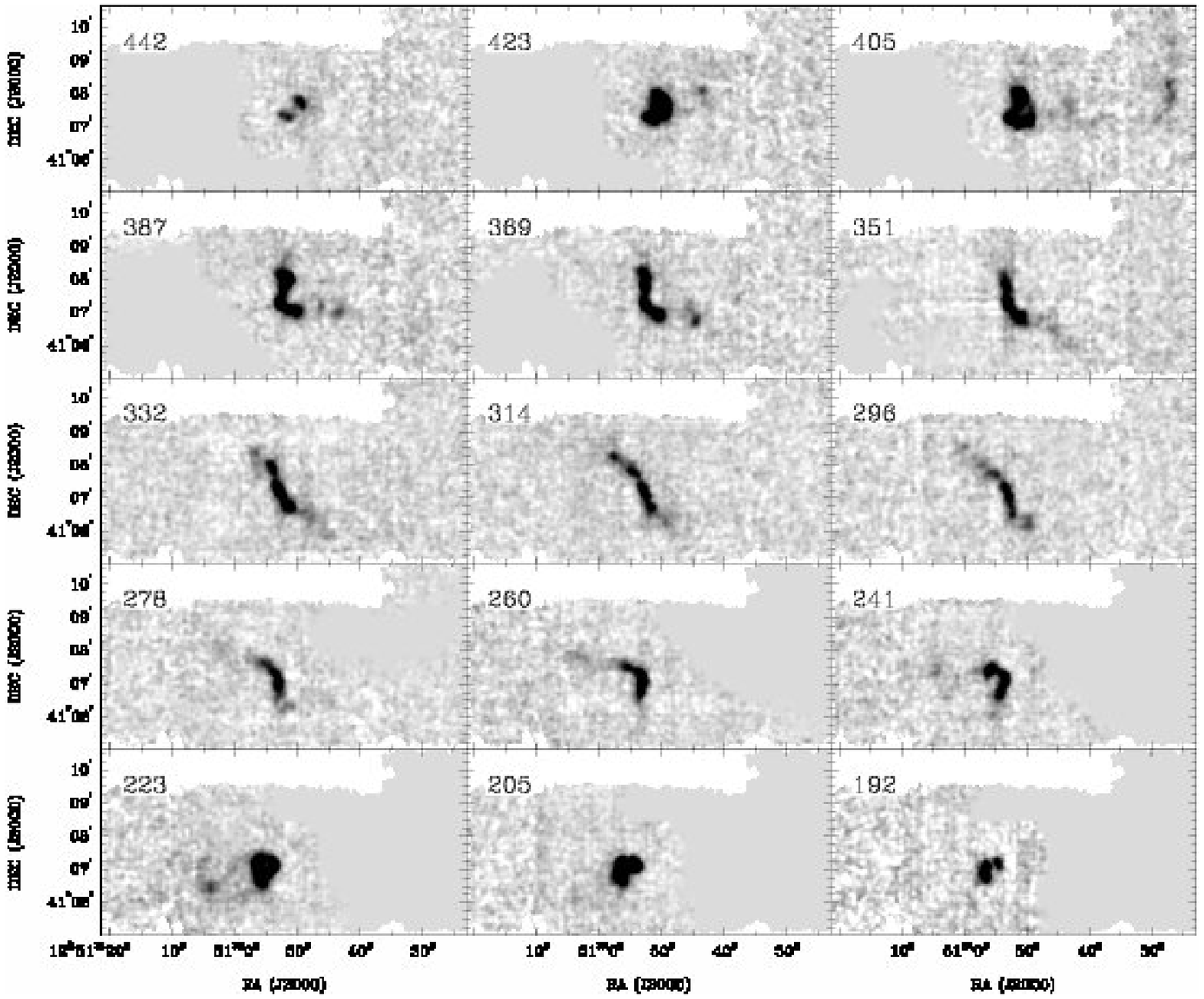} 
  \label{NGC4736CHAN} 

  \caption{Channel maps for NGC 4736 at $13\arcsec$ resolution. The grayscale
runs from -0.025 to 0.15 K. Each channel is 18.2 km s$^{-1}$ wide and the RMS
intensity in one channel is $12$~mK.  The smooth gray area indicates data at
velocities outside the baseline fitting window, which are blanked during the
reduction (\S \ref{BASEFIT}).}
\end{figure*}

\begin{figure*}
  \plotone{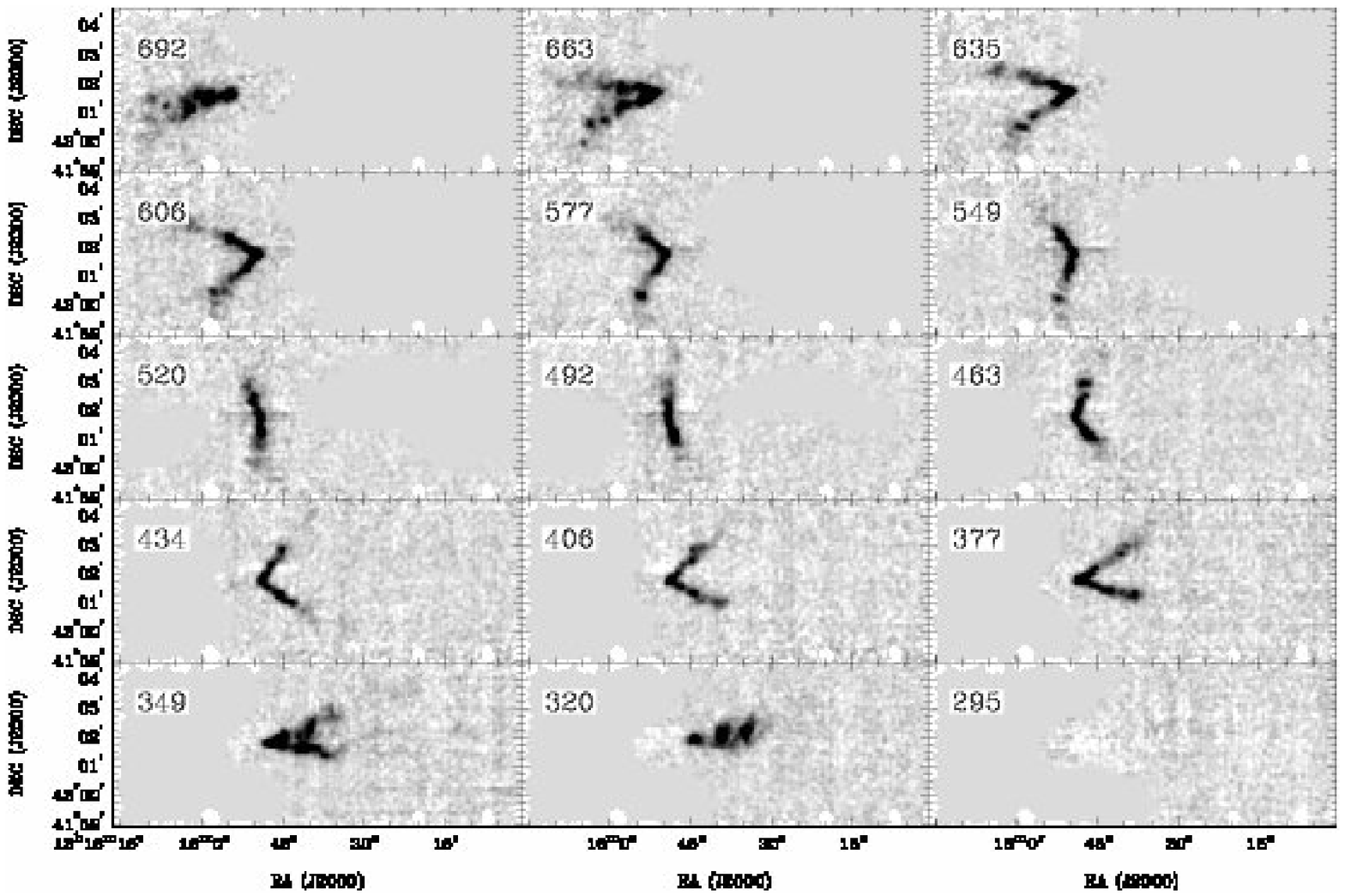}
  \label{NGC5055CHAN}

  \caption{Channel maps for NGC 5055 at $13\arcsec$ resolution. The grayscale
    runs from -0.025 to 0.15 K. Each channel is 28.6 km s$^{-1}$ wide and the
    RMS intensity in one channel is $10$~mK.  The smooth gray area indicates
    data at velocities outside the baseline fitting window, which are blanked
    during the reduction (\S \ref{BASEFIT}).}
\end{figure*}

\begin{figure*}
  \plotone{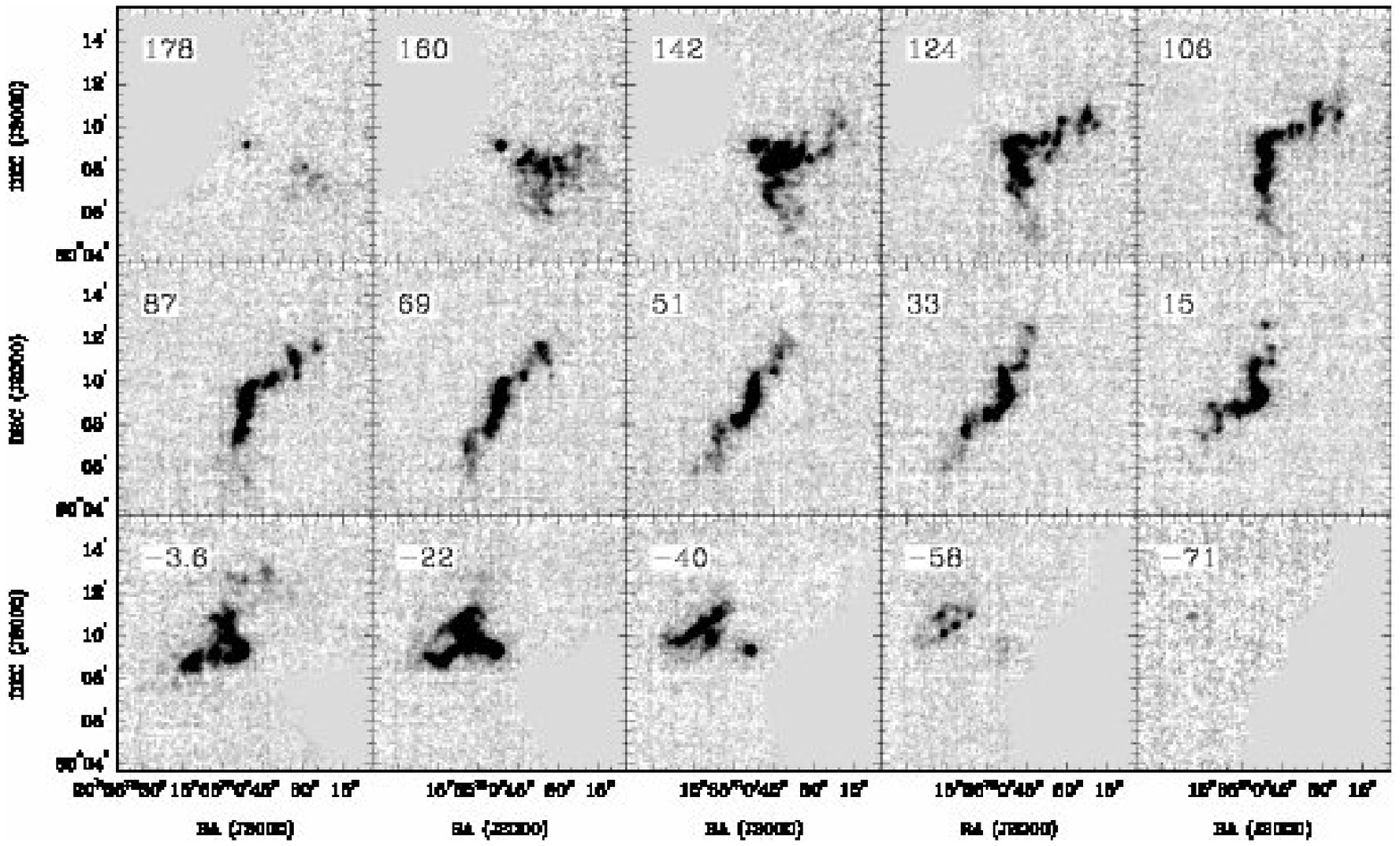}
  \label{NGC6946CHAN}

  \caption{Channel maps for NGC 6946 at $13\arcsec$ resolution. The grayscale
    runs from -0.025 to 0.15 K. Each channel is 18.2 km s$^{-1}$ wide and the
    RMS intensity in one channel is $13$~mK.  The smooth gray area indicates
    data at velocities outside the baseline fitting window, which are blanked
    during the reduction (\S \ref{BASEFIT}).}
\end{figure*}

\begin{figure*}
  \plotone{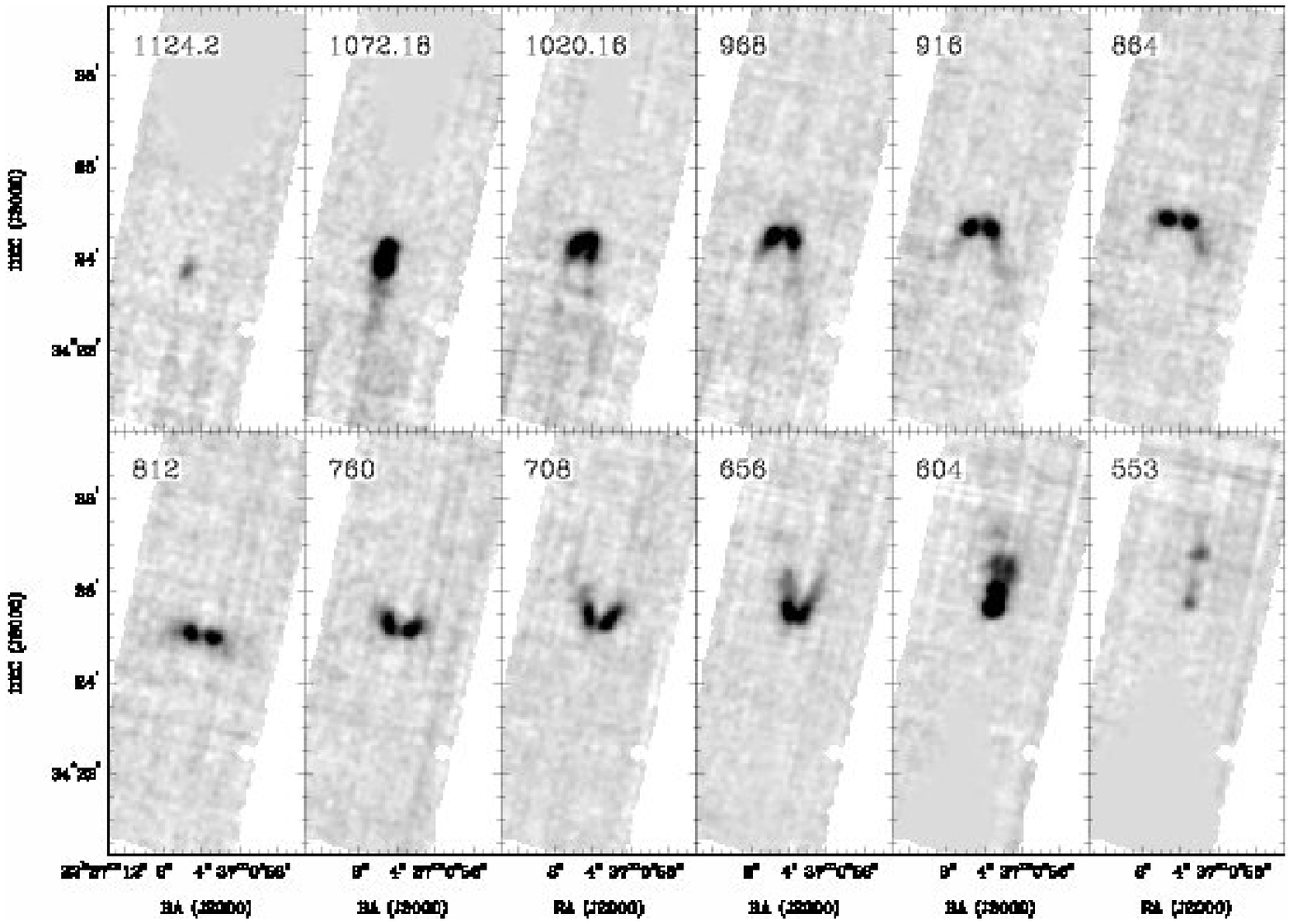}

  \caption{ \label{NGC7331CHAN}Channel maps for NGC 7331 at $13\arcsec$
    resolution. The grayscale runs from -0.025 to 0.15 K. Each channel is 52.0
    km s$^{-1}$ wide and the RMS intensity in one channel is $8$~mK. The
    smooth gray area indicates data at velocities outside the baseline fitting
    window, which are blanked during the reduction (\S \ref{BASEFIT}).}
\end{figure*}

\begin{figure*}
  \plottwo{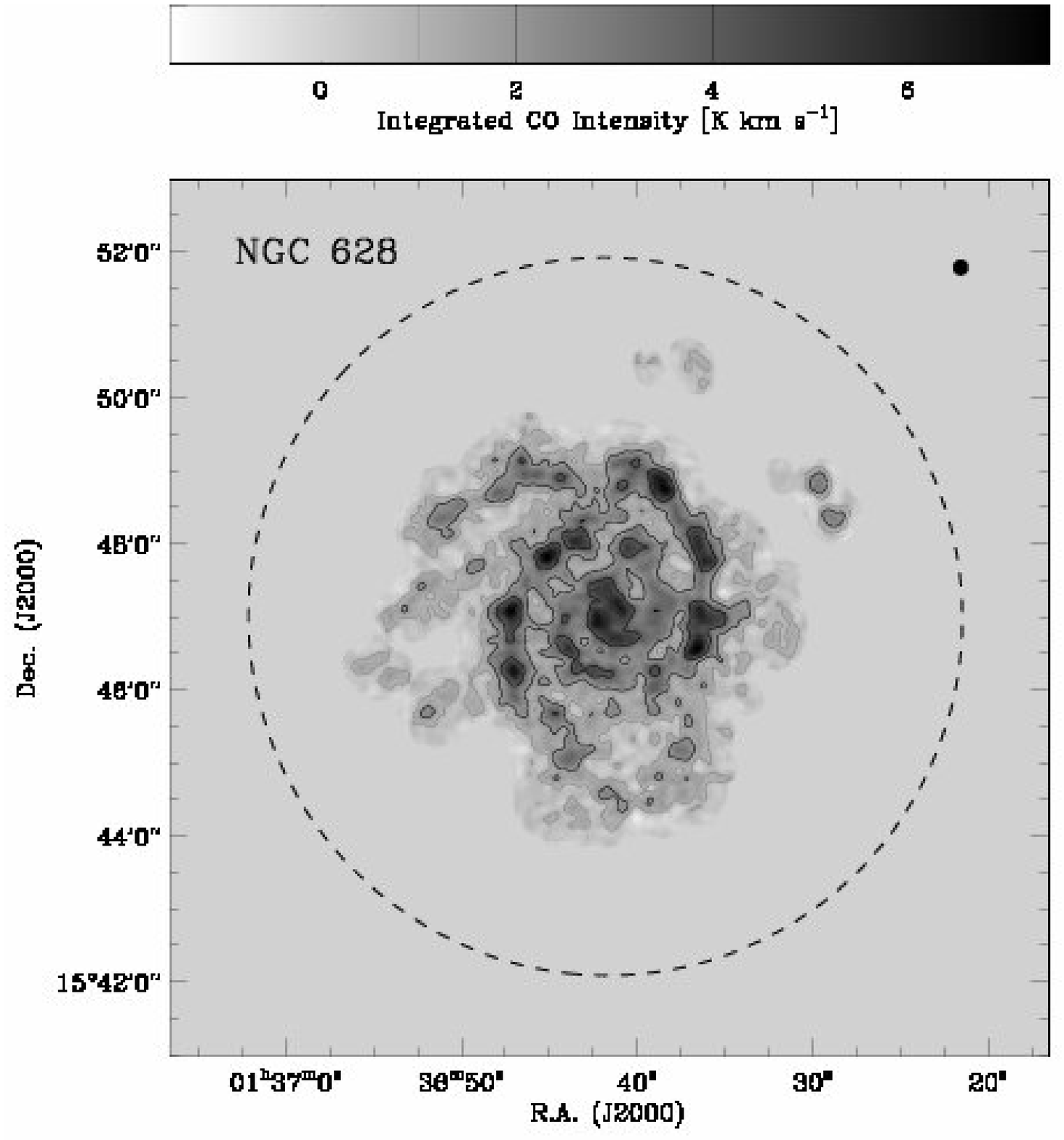}{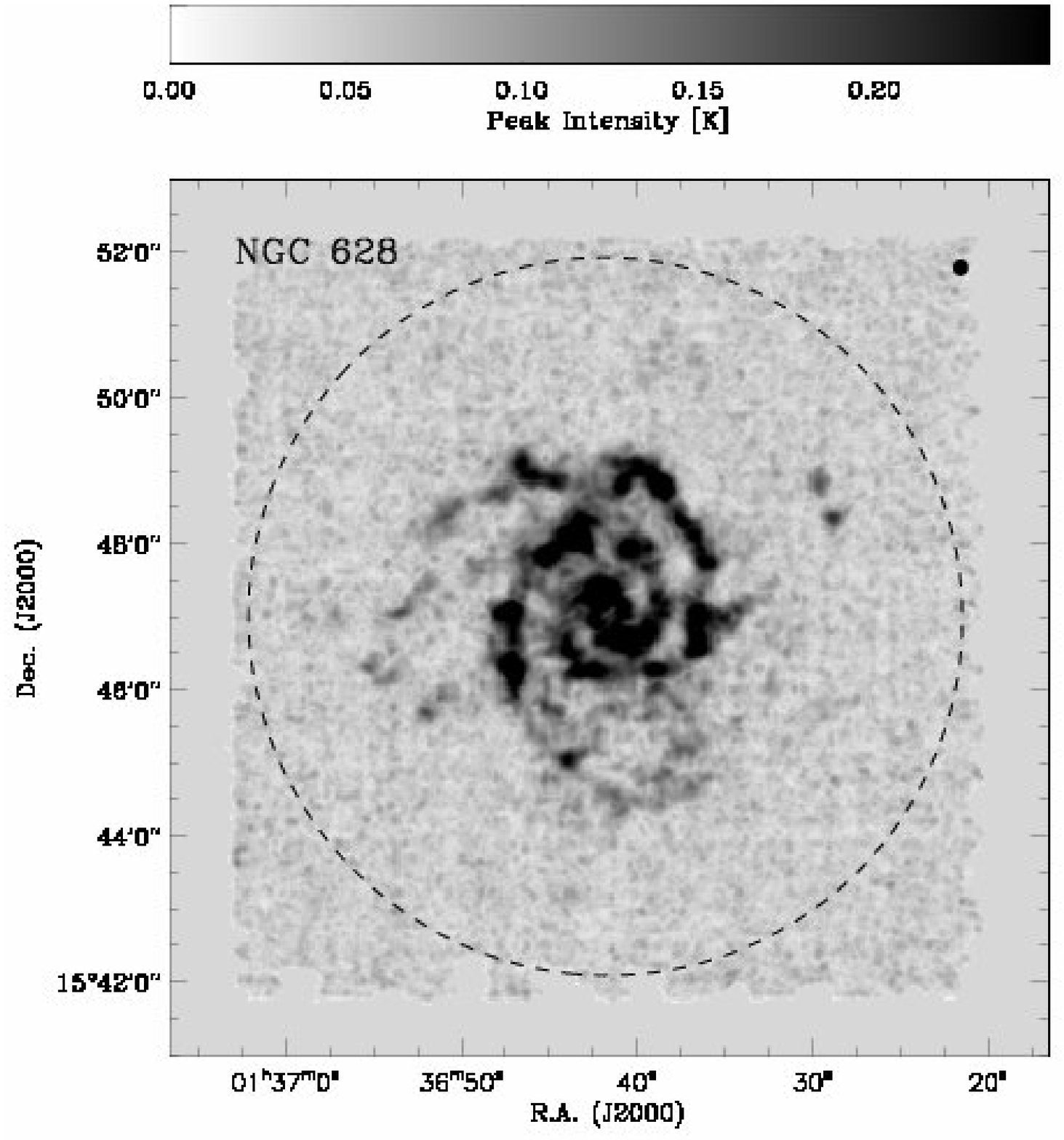}
\caption{\label{NGC0628MOM0} ({\em left}) Integrated intensity CO J$=2\rightarrow1$ intensity along each line
  of sight, summing over regions where $I_{\rm CO} > 2\sigma$ in consecutive
  channel maps (see text). Contours begin at 1 K~km~s$^{-1}$ and increase by a
  factor of two each step. The first black contour indicates $2$~K~km~s$^{-1}$
  and the first white contour shows $32$~K~km~s$^{-1}$, which correspond
  approximately to surface densities of $11$ and $176$~M$_\odot$~pc$^{-2}$. In
  both figures a dashed ellipse shows $r_{25}$, the $25^{\rm th}$ magnitude
  isophote, and a solid black circle indicates the $13\arcsec$ angular
  resolution of the data. ({\em right}) Peak CO J$=2\rightarrow1$ intensity
  along each line of sight for NGC 628.}
\end{figure*}

Despite different morphologies, our targets show a common behavior on large
scales: azimuthally averaged CO intensity declines steadily as a function of
radius in most galaxies. This is clear from Figure \ref{PROFILES1}, in which
we plot integrated intensity as a function of radius for each detected galaxy.
To make these profiles, we collapse the data cubes without masking into
integrated intensity maps and then average these over a series of concentric
tilted rings. These rings assume the position angle and inclinations in Table
\ref{SAMPLETAB} and are each $10\arcsec$ in width, about 1 resolution
element. Note that these profiles do not include any correction for
inclination (i.e., they correspond to the observed values). Error bars show
the uncertainty in the mean integrated intensity, derived by measuring the RMS
scatter within the tilted ring and dividing by $\sqrt{N}$, where $N$ is the
number of independent data points in the ring. Measurements with $>3\sigma$
significance appear as black circles, those with less as gray triangles.

Thick gray lines show exponential fits to the profiles, carried out over the
region bracketed by the arrows at the top of each plot. The corresponding
scale lengths, given in Table \ref{RESULTTAB}, span from $0.8$ to
$3.2$~kpc. In most galaxies, an exponential decline captures the large-scale
behavior well, though we cannot reliably fit NGC~2841 or NGC~4214. Studying
BIMA SONG, which overlaps our sample, \citet[][]{REGAN01} found the same
variations in detailed morphology averaging into a common exponential
decline. They discuss this topic in more detail. Both \citet[][]{REGAN01} and
\citet[][]{HELFER03} noted deviations from these fits in galaxy centers, which
we also observe in the form of depressions (e.g., NGC~3521, NGC~7331) and
excesses (e.g., NGC~3351, NGC~6946).

The exponential decline results from the combination of a decreasing filling
fraction of CO emission and decreased intensity along individual lines of
sight. To illustrate the relative contributions of these two effects, Figure
\ref{PROFILES1} also shows the maximum integrated intensity inside each
ring. Filled diamonds indicate rings where this maximum value is $>4\sigma$
and significantly ($>2\sigma$) larger than the absolute value of the minimum
integrated intensity (roughly accounting for artifacts). Open diamonds show
where the maximum intensity is below this value and we are correspondingly
uncertain that the measurement is not biased by noise and artifacts. Comparing
the two profiles, we observe a fairly close correspondence in the central
parts of galaxies and then a shallow decline in maximum intensity compared to
average intensity at larger radii. The difference corresponds to the
decreasing area subtended by bright CO emission.

Although \hi\ typically dominates the ISM outside $\sim 0.5~r_{25}$, we detect
individual regions in the outskirts of several galaxies at high
significance. In Figure \ref{PROFILES1} and Table \ref{RESULTTAB}, we report
the largest radius at which we detect a ``high signal spectrum.'' We define
this as a region with three consecutive velocity channels at $>3\sigma$
significance over 10 spatially contiguous pixels. This conservative definition
generates no false positives in the whole survey when applied to the negative
portions of the data. Figure \ref{PROFILES1} clearly shows that more extended,
but lower significance, signal is present in many maps.

For the most part, these detections are simply associated with the outer part
of the star forming disk and can be clearly identified with structures seen in
\hi , UV, or IR maps that extend continuously from the inner to outer
galaxy. This is the case for the detections at $\sim 0.7$--$1.0~r_{25}$ in
NGC~628, NGC~3521, NGC~5055, and NGC~6946 \citep[the outer region in the
  latter corresponds to positions P10, 11, and 12 observed
  by][]{BRAINE07}. The notable exception is the bright CO emission at
$1.33~r_{25}$ in NGC~4736. This relatively narrow-line width feature is
clearly visible in the channel map centered at 405~km~s$^{-1}$ in Figure
\ref{NGC4736CHAN}. The emission is $\sim 5\arcmin$ (projected distance of
$\sim 7$~kpc) from the center of the galaxy and well outside the main region
of active star formation \citep[e.g., see maps in][]{LEROY08}. This feature
matches the velocity and position of an \hi\ filament outside the main body of
the galaxy; it also shows UV and mid-IR (24$\mu$m) emission, indications of
recent massive star formation.

\subsection{Flux, Luminosity, and H$_2$ Mass}

Columns 3 and 4 of Table \ref{RESULTTAB} report the integrated CO flux and
luminosity for each target, measured by summing the same integrated intensity
maps used to make the radial profiles. In all cases, the statistical
uncertainty from Gaussian noise alone is small, meaning that the choice of
area to integrate over, artifacts, and calibration of the telescope are the
dominant sources of uncertainty. We estimate the uncertainty in the
calibration at $\sim 20\%$ (Section \ref{UNCERTAINTIES}) and find that
applying masking (e.g., as in Figures \ref{NGC0628MOM0} -- \ref{NGC7331MOM0})
alters the flux by $\lesssim 15\%$ (slightly more in the cases of NGC~925,
NGC~4214, and NGC~2841).

In four low metallicity dwarf irregulars (Holmberg~I, Holmberg~II, IC~2574,
and DDO~154) we do not detect an extended distribution of CO.  Instead, Table
\ref{RESULTTAB} gives upper limits on their CO content. To derive these, we
create an integrated spectrum for each cube and estimate the RMS noise at
velocities offset from the \hi\ line, regions expected to be
signal-free. Combining this RMS noise with the \hi\ line width \citep[$W_{20}$
  from][]{WALTER08} and area of the map yields the $5\sigma$ upper limits on
$F_{\rm CO}$ in Table \ref{RESULTTAB}. This approach takes into account the
low-level correlated noise discussed in Section \ref{UNCERTAINTIES}.

We quote CO flux, $F_{\rm CO}$, in units of K~km~s$^{-1}$~arcsec$^{2}$
($T_{\rm MB}$). This may be converted to Jy~km~s$^{-1}$ via

\begin{equation}
\label{FCOTOJY}
  F_{\rm CO} \left[ {\rm Jy~km~s}^{-1} \right] = 0.036~F_{\rm CO} \left[ {\rm K~km~s}^{-1}~{\rm arcsec}^{2} \right]~,
\end{equation}

\noindent We report CO luminosity, $L_{\rm CO}$, in units of
K~km~s$^{-1}$~pc$^{2}$ ($T_{\rm MB}$). This may be combined with a
CO-to-\htwo\ conversion factor to estimate the mass of molecular gas via

\begin{equation}
\label{H2MASSEQ}
M_{\rm H2} \left[ M_\odot \right] = 5.5~X_{2}~\frac{R_{\rm 21}}{0.8}~L_{\rm
  CO} \left[ {\rm K~km~s}^{-1}~{\rm pc}^{2} \right]~.
\end{equation}

\noindent Here $X_{2}$ is the CO $J=\jone$-to-\htwo\ conversion factor
normalized to $2\times 10^{20}$~\xcounits , approximately the Solar
Neighborhood value \citep[][]{STRONG96,DAME01}. $R_{\rm 21}$ is the CO
$J=\jtwo$ to $J=\jone$ ratio; $0.8$ is a typical value that we
find comparing HERACLES to other surveys (\S \ref{LINERAT}).  Equation
\ref{H2MASSEQ} includes a factor of 1.36 to account for helium.

\subsection{GMC Filling Fraction and Point Source Sensitivity}

\begin{figure*}
  \plottwo{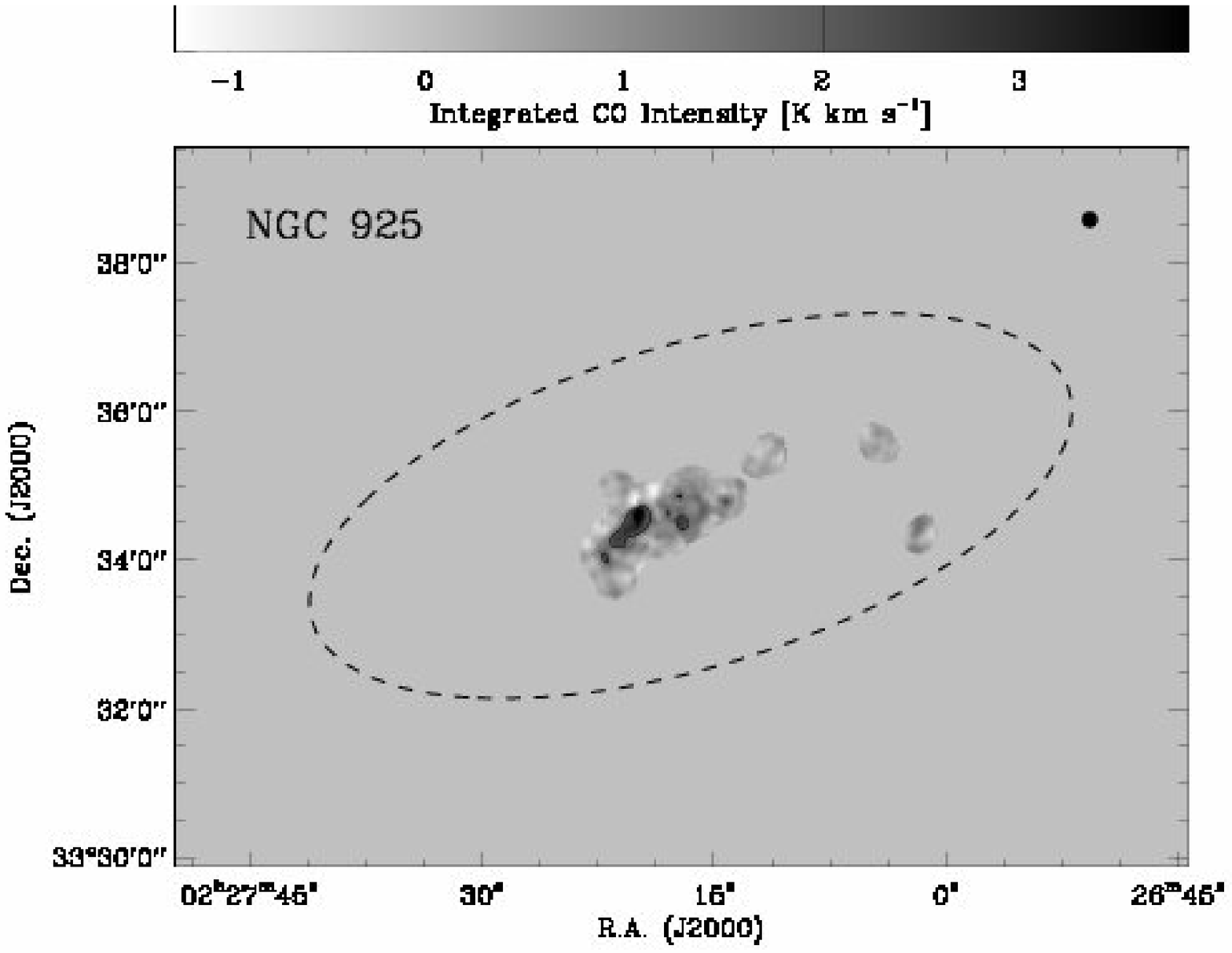}{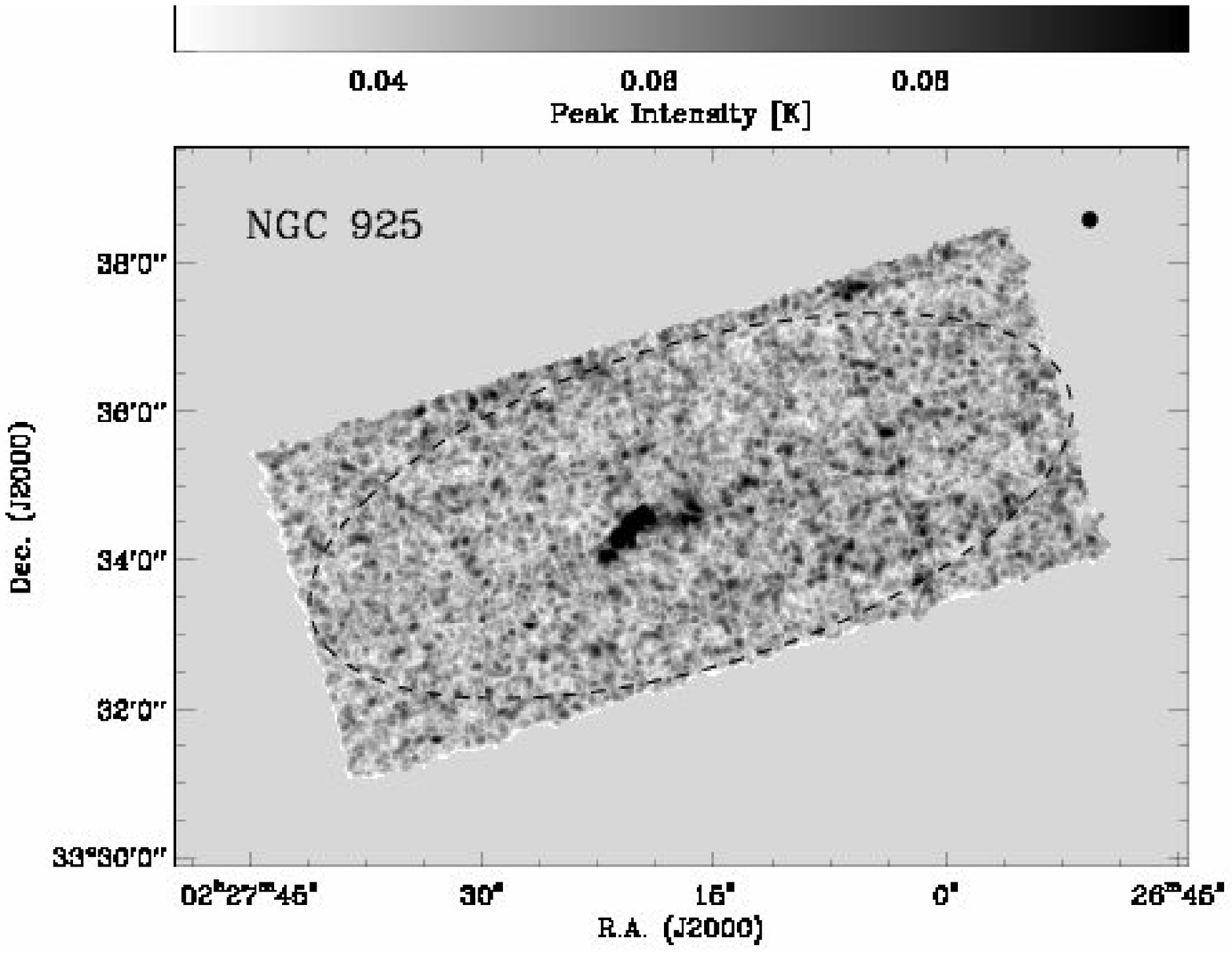}
\caption{\label{NGC0925MOM0}
  As Figure \ref{NGC0628MOM0} but for NGC 925.}
\end{figure*}

\begin{figure*}
  \epsscale{0.8} 
\plottwo{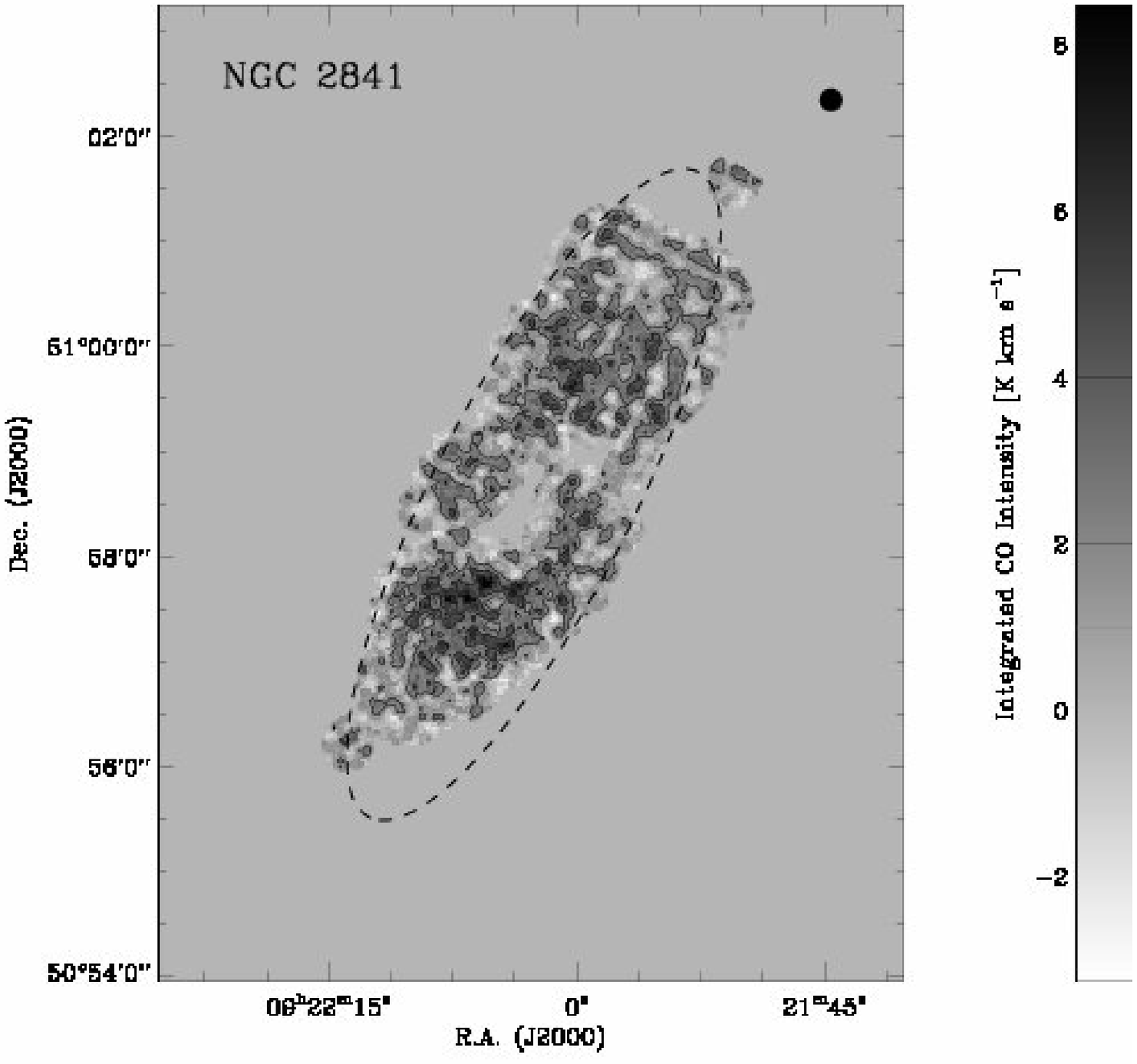}{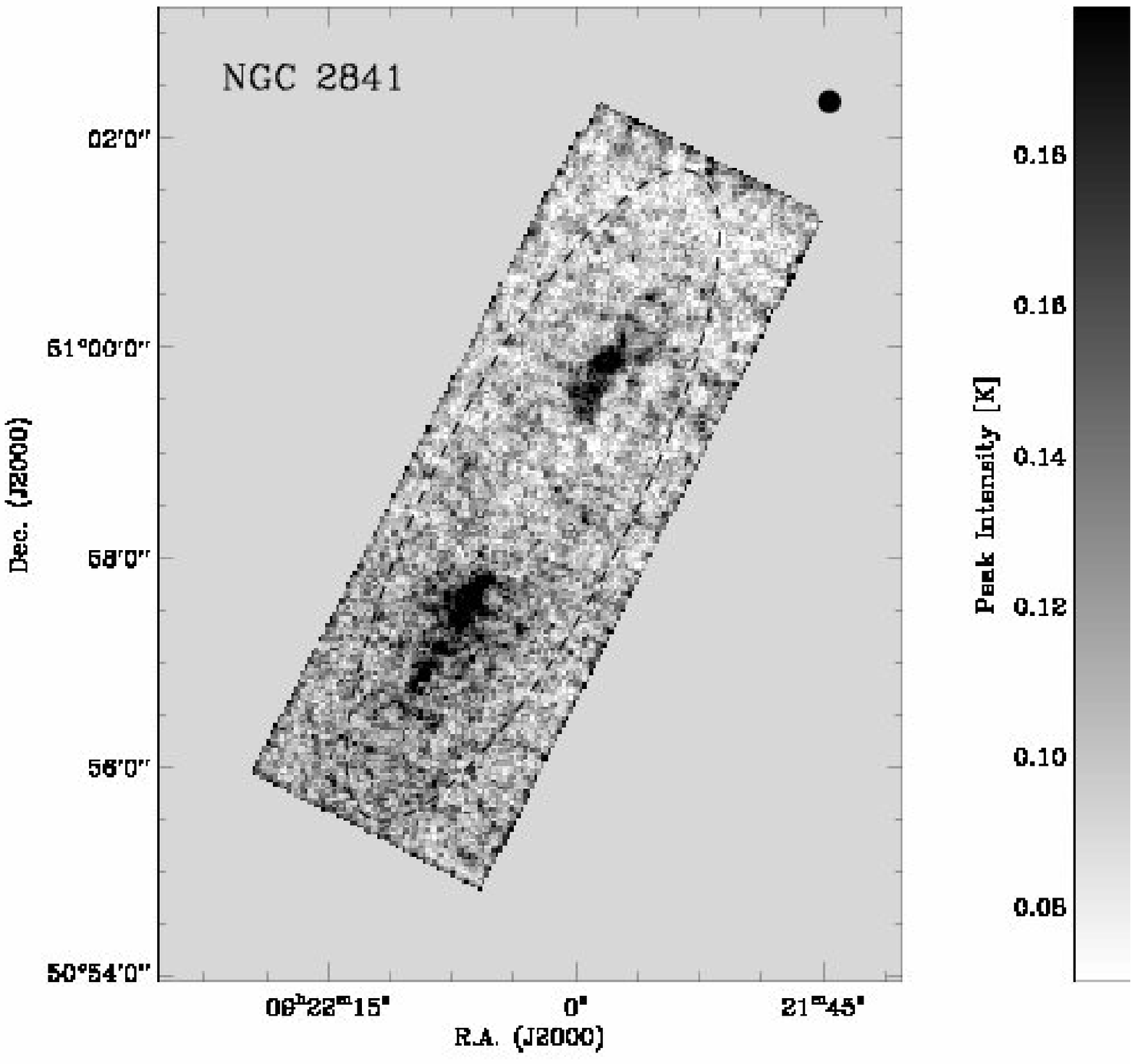} 
\caption{\label{NGC2841MOM0}
  As Figure \ref{NGC0628MOM0} but for NGC 2841.}
\end{figure*}

\begin{figure*}
  \plottwo{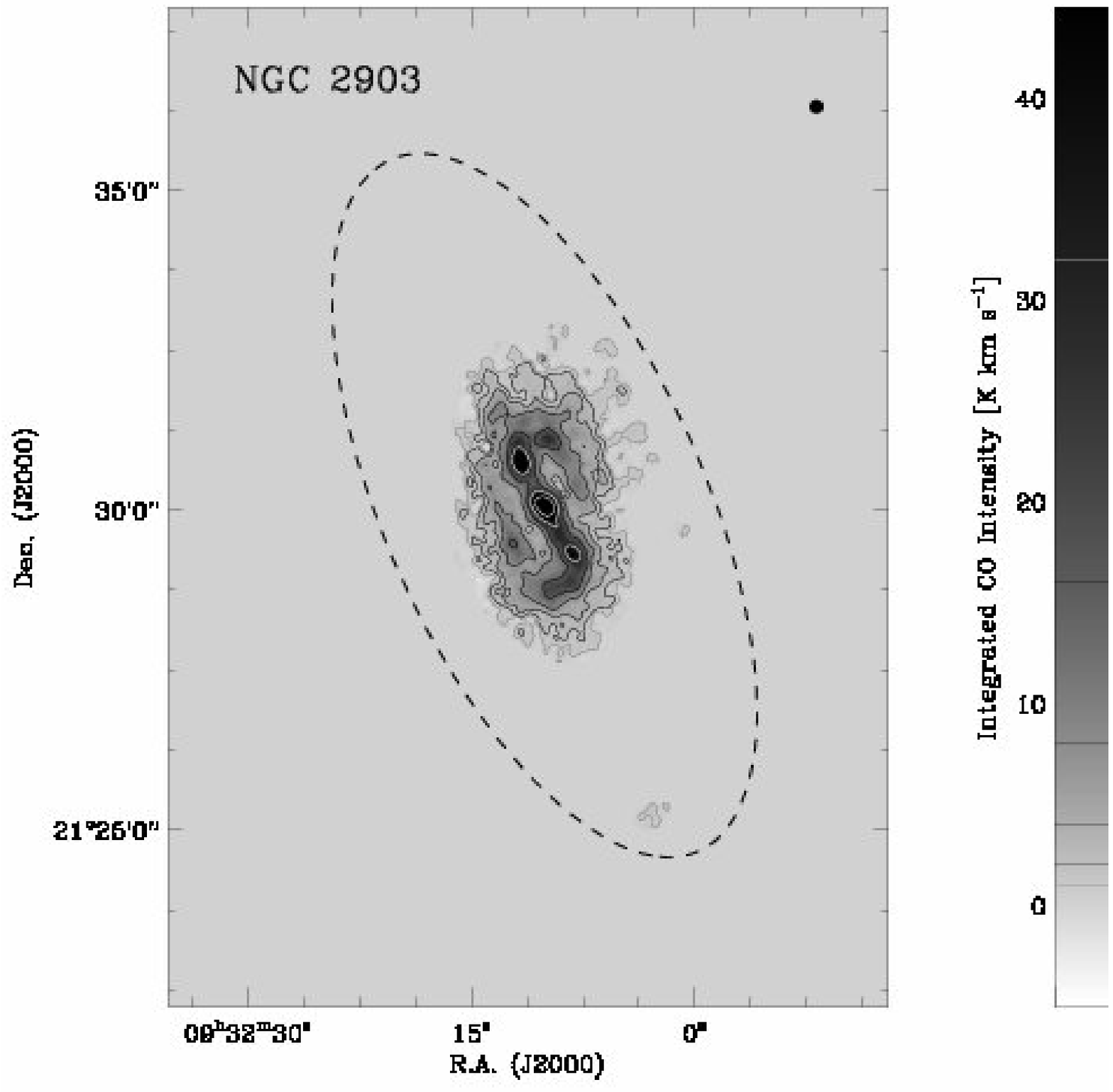}{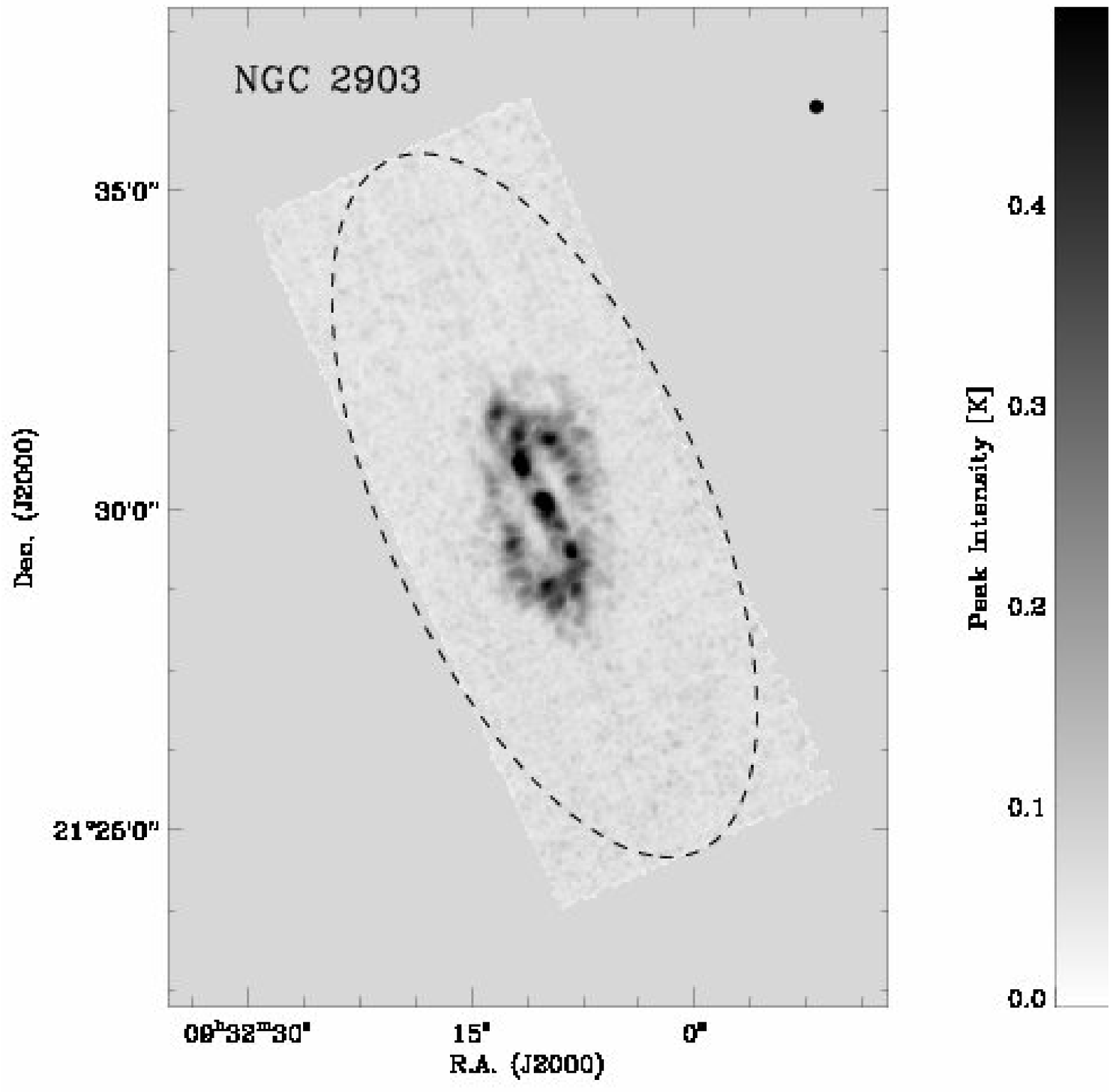}
  \caption{\label{NGC2903MOM0} As Figure \ref{NGC0628MOM0} but for NGC 2903.}
\end{figure*}

\begin{figure*}
  \plottwo{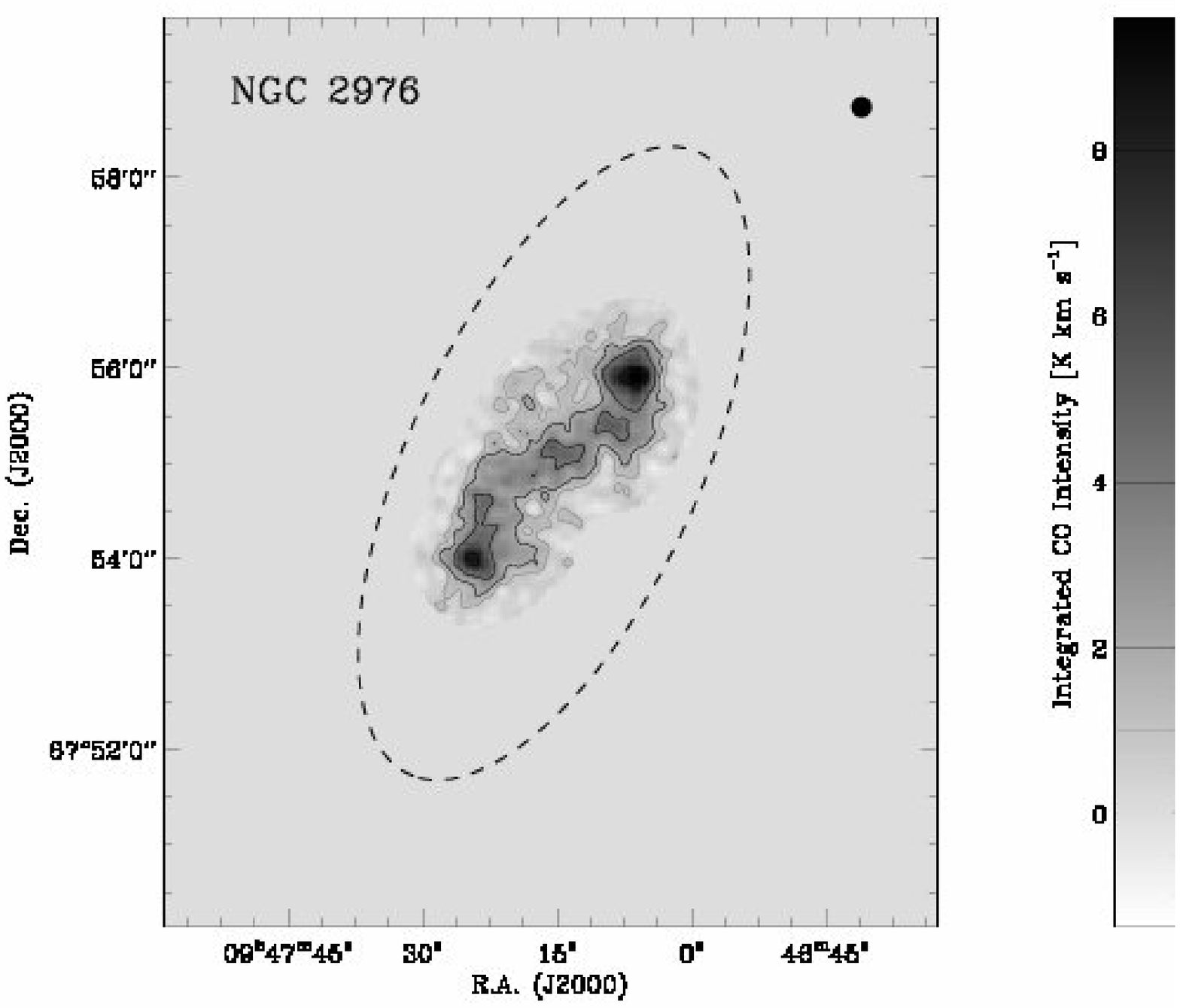}{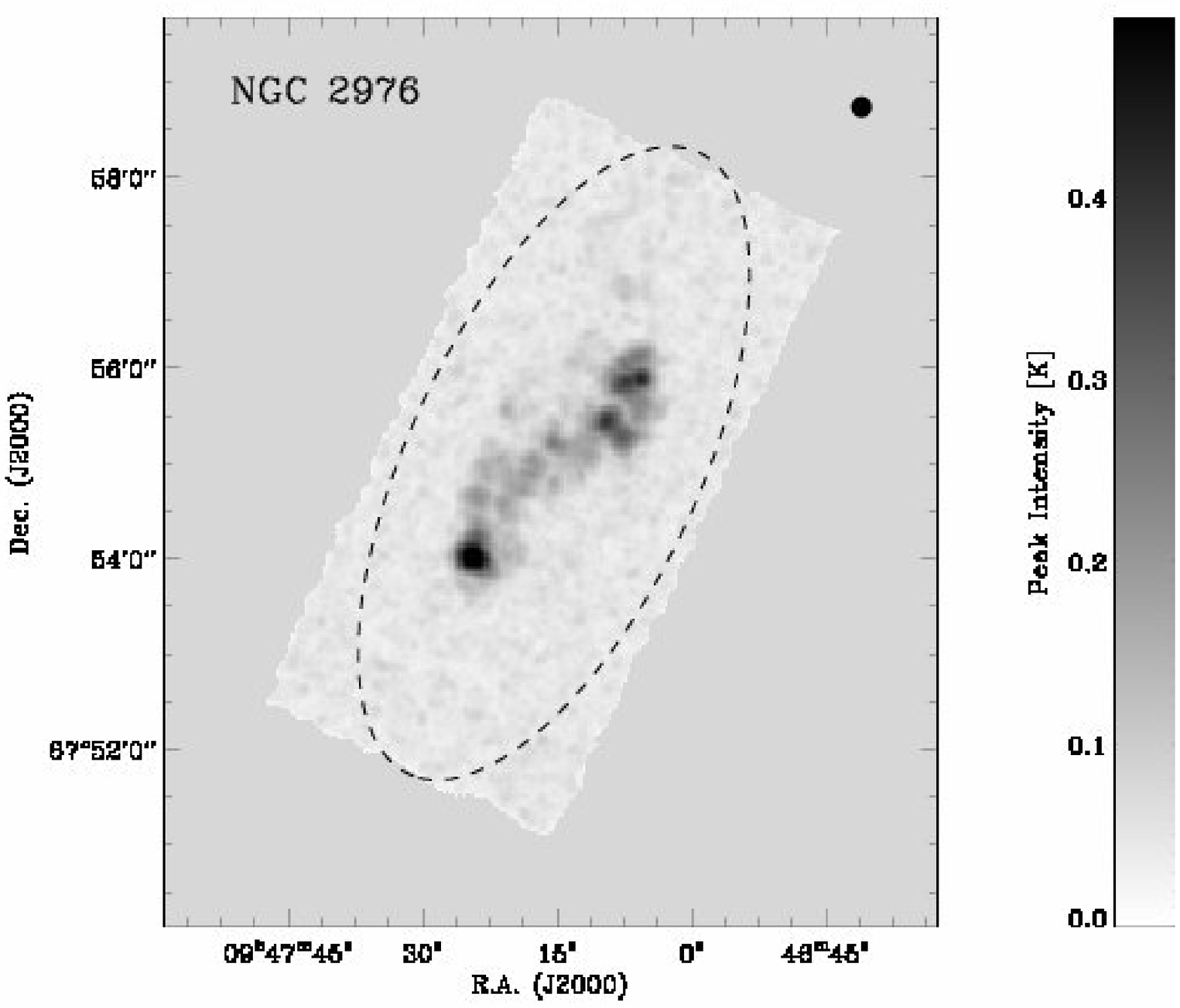}
  \caption{\label{NGC2976MOM0} As Figure \ref{NGC0628MOM0} but for NGC 2976.}
\end{figure*}

In the Milky Way and nearby disk galaxies, CO emission is observed to come
mostly from GMCs. These clouds have approximately constant surface densities,
$\Sigma_{\rm H2} \sim 170$~M$_{\odot}$~pc$^{-2}$, and brightness temperatures
$\gtrsim 4$~K \citep{SOLOMON87}.  We expect that these clouds similarly
dominate the ISM across the disks of our sample.

The contrast between our measured surface densities and those of Galactic GMCs
may give an idea of the filling factor of GMCs inside a resolution element. To
translate CO intensity, $I_{\rm CO}$, into molecular gas surface density,
$\Sigma_{\rm H2}$, one can apply an analog of Equation \ref{H2MASSEQ},

\begin{equation}
\label{H2SURFEQ}
\Sigma_{\rm H2} \left[{\rm M}_{\odot}~{\rm pc}^2\right] =
5.5~X_{2}~~\frac{R_{\rm 21}}{0.8}~I_{\rm CO}~\left[ {\rm K~km~s}^{-1}\right]~.
\end{equation}

\noindent The contours in Figures \ref{NGC0628MOM0} -- \ref{NGC7331MOM0} thus
correspond to $\Sigma_{\rm H2} \approx 5.5$ (gray), $11$ (first black), $22$,
$44$, $88$, $176$ (first white) M$_{\odot}$~pc$^{-2}$, and so on, values that
can be explained if $\sim 3$, $6$, $13$, $25$, $50$, and $100\%$ of the area
inside the beam is covered by Galactic GMCs. Most area where we see CO
emission has relatively low filling factors, $\sim 5$--$15\%$.

The peak intensity maps allow a similar comparison. If $\sim 4$~K is the
typical brightness for a Galactic cloud, $T_{\rm peak} = 0.1$ and $0.5$~K ---
values that bracket most of the observed peak temperatures --- correspond to
filling factors of $\sim 3$ and $13\%$.

The point mass sensitivity of our maps corresponds to a collection of several
Galactic GMCs. For a source with full line width of $\sim 10.4$~km~s$^{-1}$
\citep[typical for Galactic GMCs,][]{SOLOMON87}, our typical RMS sensitivity
translates to a $3\sigma$ point source sensitivity of $4.8 \times
10^5~d_{10}$~K~km~s$^{-1}$~pc$^2$, where $d_{\rm 10}$ is the distance to the
source divided by 10~Mpc. From Equation \ref{H2SURFEQ}, this corresponds to an
H$_2$ mass of $2.7 \times 10^6~d_{\rm 10}$~M$_{\odot}$, which is near the high
end found for Galactic GMCs \citep[clouds with masses
  $10^5$--$10^6$~M$_{\odot}$ account for most of the Galactic H$_2$,
  e.g.,][]{BLITZ93}. For reference, the most massive Milky Way GMC, the most
massive cloud detected in M33, and the Orion molecular complex have masses of
$\sim 6 \times 10^6$~M$_{\odot}$, $\sim 7 \times 10^5$~M$_\odot$, and $\sim 5
\times 10^5$~M$_{\odot}$, respectively
\citep{SOLOMON87,ENGARGIOLA03,WILSON05}.

\section{Comparison With Other Data}

Previous observations of our targets at wavelengths from radio to UV allow a
range of comparisons. Here we use these data to calculate the
$H_2$-to-\hi\ and $H_2$-to-stellar mass ratio, to test our assumption that
\hi\ and CO exhibit the same mean velocity, to check our observations against
previous CO measurements, and to estimate the CO $J=\jtwo /
\jone$ line ratio. More detailed comparisons to \hi , IR, and UV data
may be found in \citet{LEROY08} and \citet{BIGIEL08}.

\subsection{Relative Masses of H$_2$, \hi , and Stars}

Columns 7, 8, and 9 in Table \ref{RESULTTAB} list the ratios of \htwo\ mass to
\hi\ mass ($M_{\rm H2} /M_{\rm HI}$), \htwo\ mass to stellar mass ($M_{\rm H2}
/ M_*$), and gas to stellar mass ($\left( M_{\rm H2} + M_{\rm HI} \right) /
M_*$) for each galaxy. We adopt $M_{\rm HI}$ from \citet{WALTER08}, scaling by
$1.36$ to account for helium. The stellar masses are taken from
\citet{LEROY08}, who estimate them from SINGS 3.6$\mu$m imaging assuming a
Kroupa IMF \citep[NGC~2903 comes from][]{DEBLOK08}. \citet{DEBLOK08}
investigate possible variations in the $3.6\mu$m mass-to-light ratio and find
a factor of $2$ plausible.

The range of $M_{\rm H2} / M_*$ is $0.01$--$0.25$, with most values between
$0.03$ and $0.1$. $M_{\rm H2} / M_{\rm HI}$ spans from $0.02$ to $1.13$, with
most targets in the range $\sim 0.2$--$0.6$ (in our sample only NGC~4736
appears to have more H$_2$ than \hi ). Combined, these ratios yield a wide
range of gas-richness, from $\left( M_{\rm HI} + M_{\rm H2} \right) / M_* \sim
0.2$ in relatively early type spirals to $\gtrsim 3$ in our late-type,
low-mass nondetections.

\subsection{Scale Lengths At Other Wavelengths}

\begin{figure*}
  \plottwo{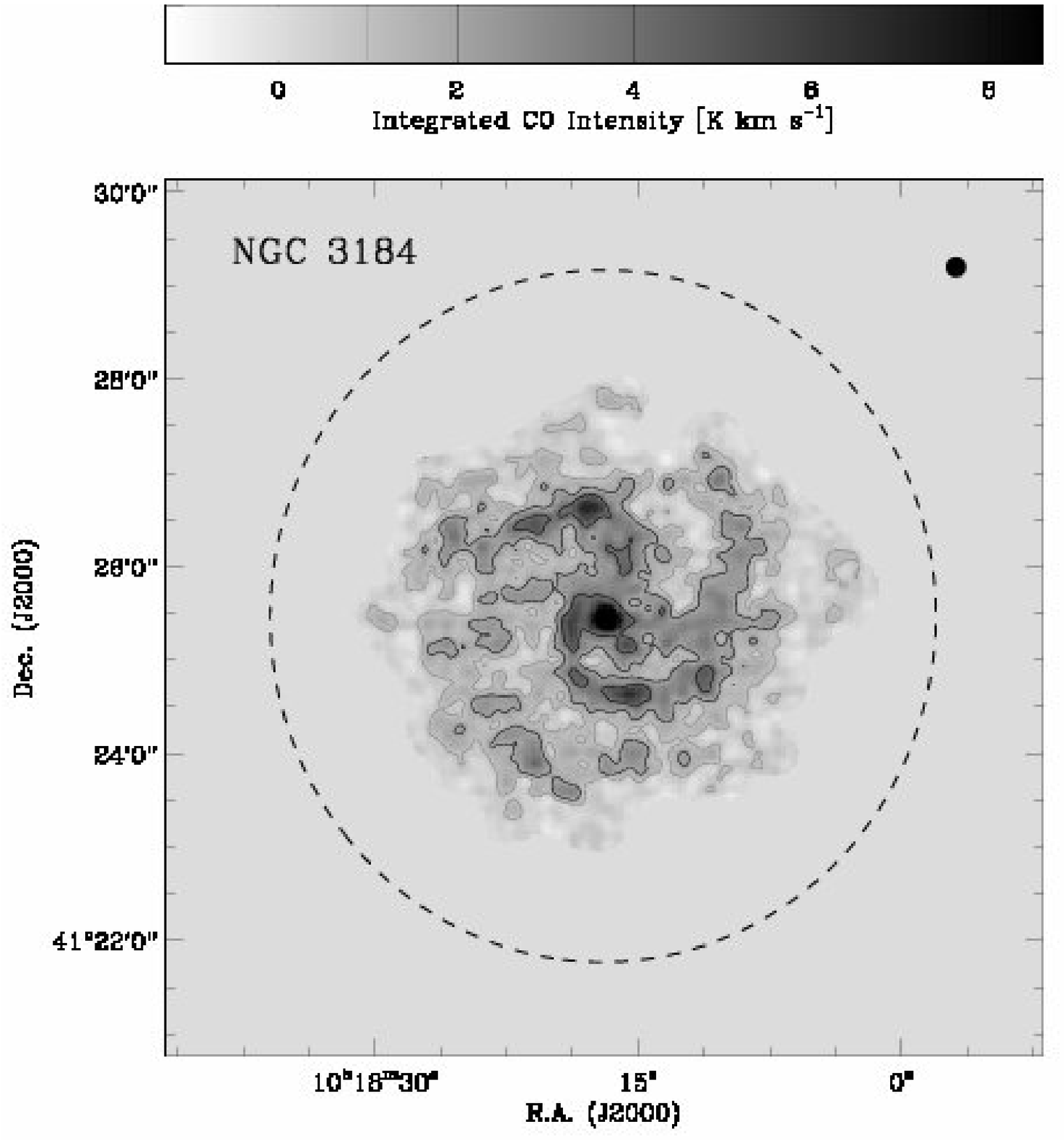}{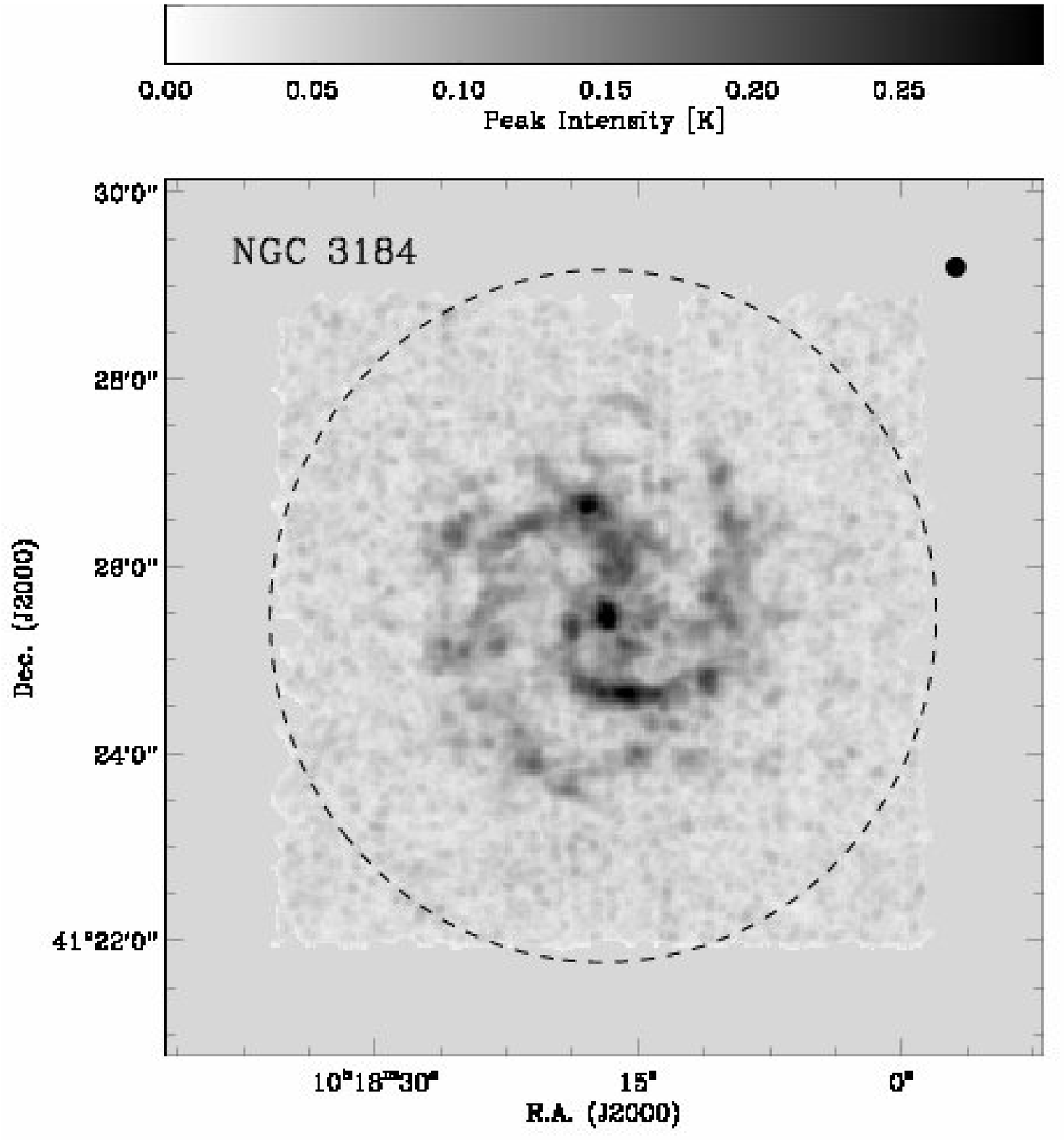}
  \caption{ \label{NGC3184MOM0} As Figure \ref{NGC0628MOM0} but for NGC 3184.}
\end{figure*}

\begin{figure*}
  \plottwo{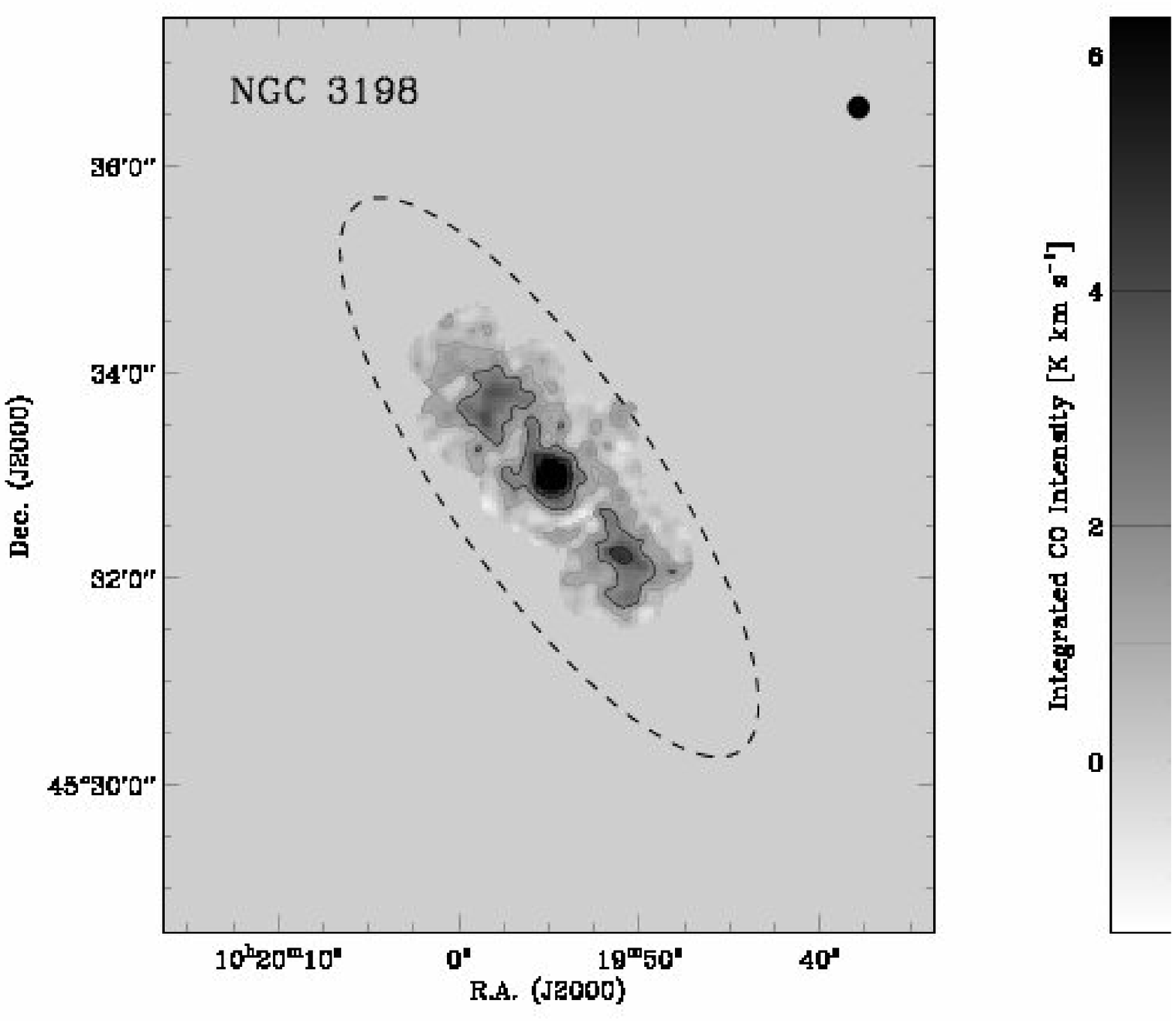}{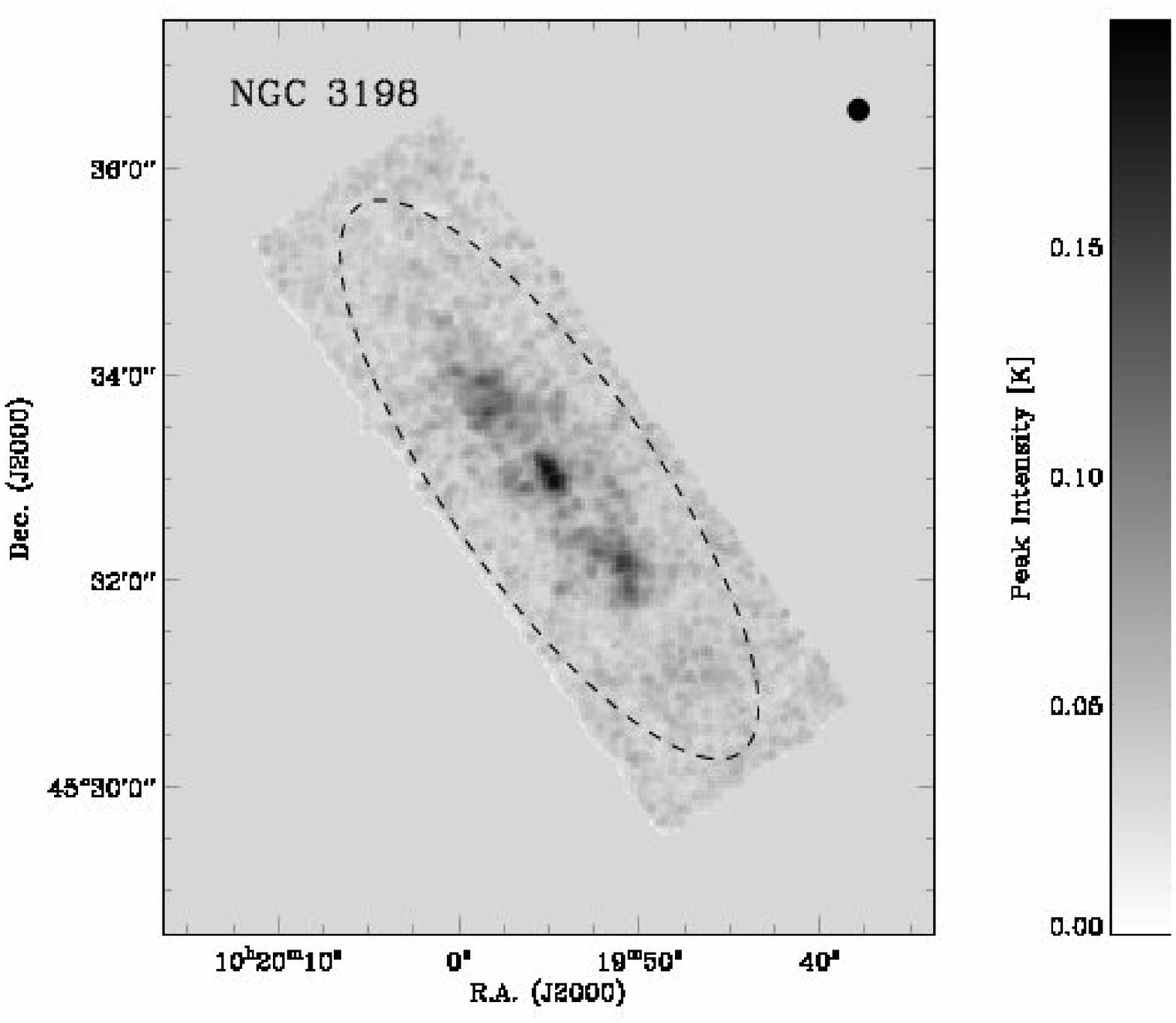}
  \caption{\label{NGC3198MOM0} As Figure \ref{NGC0628MOM0} but for NGC 3198.}
\end{figure*}

\begin{figure*}
  \plottwo{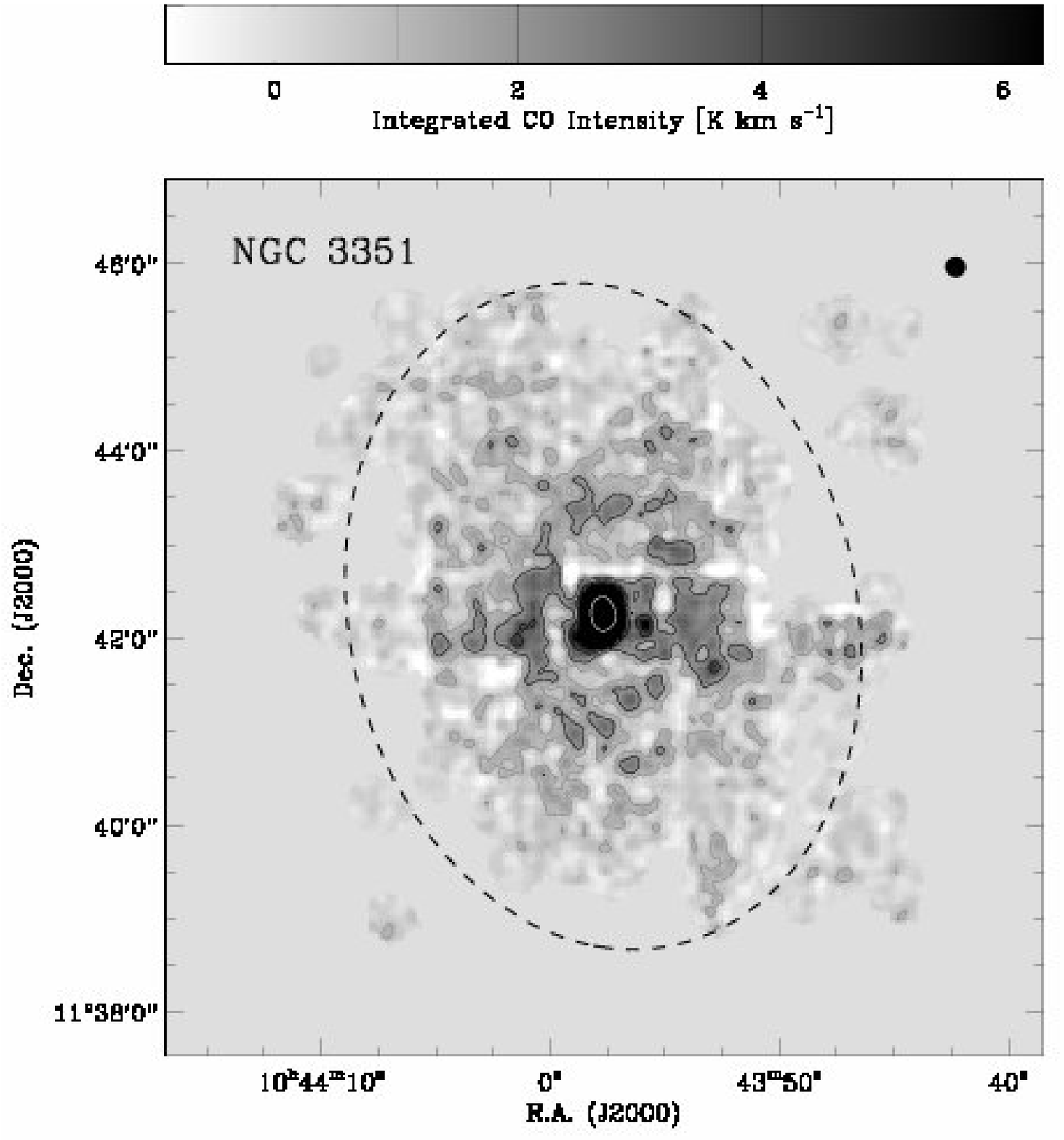}{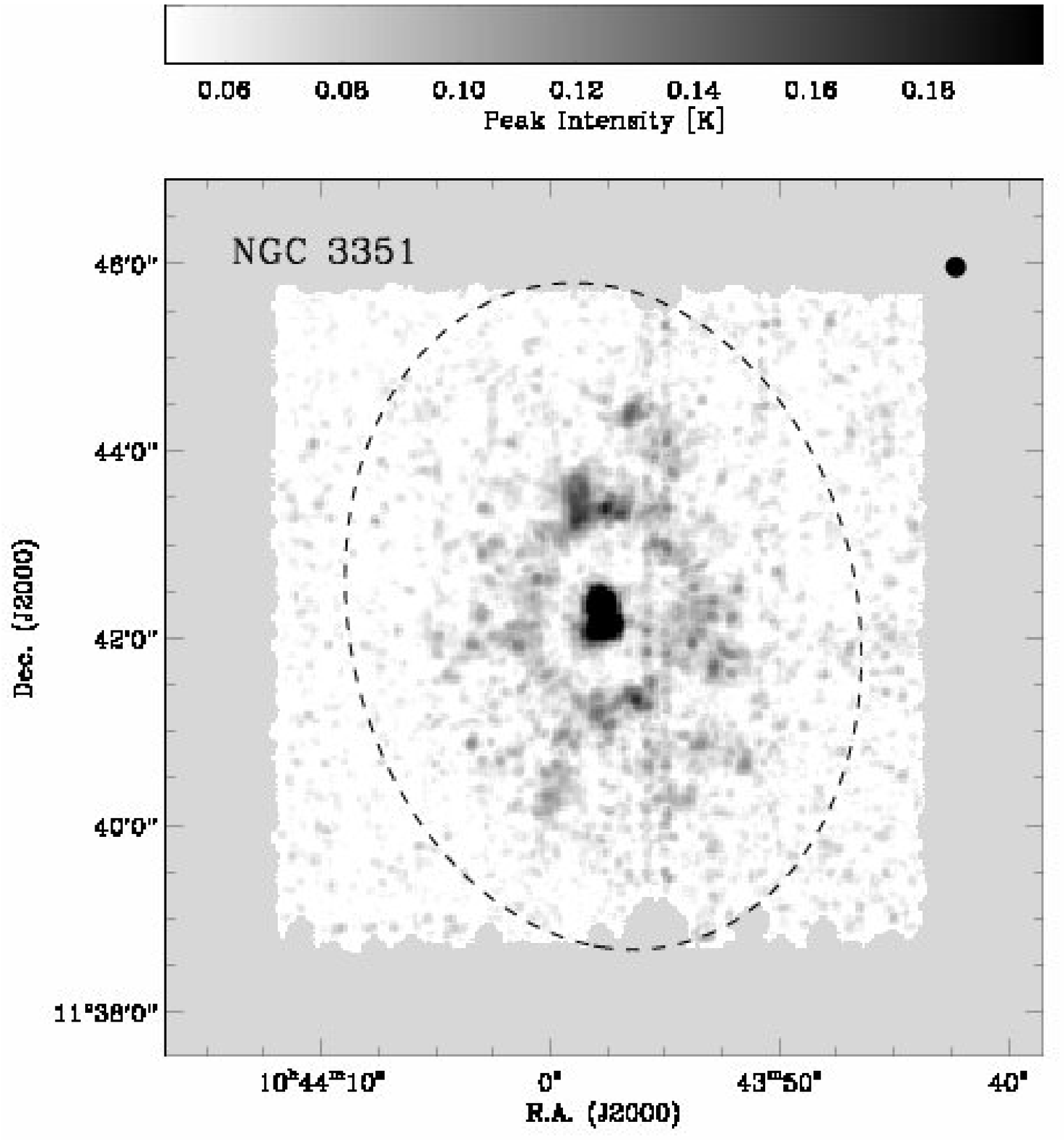}
  \caption{\label{NGC3351MOM0} As Figure \ref{NGC0628MOM0} but for NGC 3351.}
\end{figure*}

Figure \ref{SLSLPLOT} shows the size of our galaxies at other wavelengths as a
function of the exponential scale lengths that we measure from azimuthally
averaged CO profiles, $l_{\rm CO}$. Filled circles show $0.2~r_{25}$, which
\citet{YOUNG95} found to be a typical scale length of CO emission. We observe
the same here, $l_{\rm CO} = 0.2~r_{25}$ with a $1\sigma$ scatter of
$0.05$. Stars show the exponential scale length fit to median profiles of
$3.6\mu$m emission \citep[from][]{LEROY08}, a tracer of the distribution of
old stellar mass. Diamonds plot the exponential scale length fit to profiles
of star formation surface density, estimated from a combination of {\em GALEX}
FUV emission and {\em Spitzer} 24$\mu$m data \citep[from][]{LEROY08}. All
three quantities scatter about equality, with $l_{\rm CO}$ on average $\sim
5$--$10\%$ lower than the other two, not a significant difference. The plot
underscores the well-established close association between CO emission and
stellar light in disk galaxies \citep[e.g.,][]{YOUNG91,YOUNG95,REGAN01} and
the similar match between the distributions of CO and star formation
\citep[e.g.,][]{LEROY08}.

\subsection{HERACLES CO and THINGS {\sc Hi} Velocities}
\label{HICOMP}

\begin{figure*}
  \plottwo{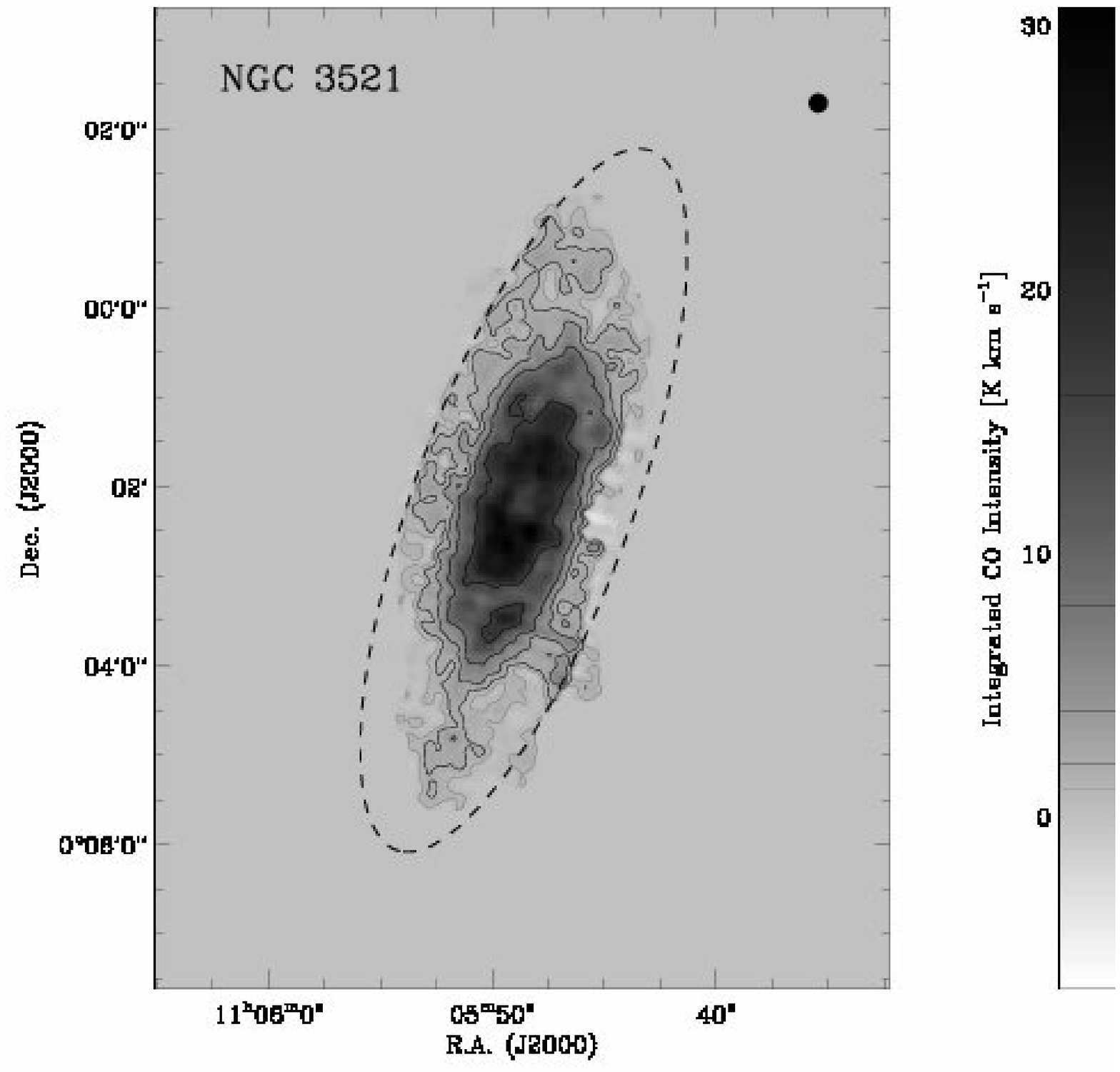}{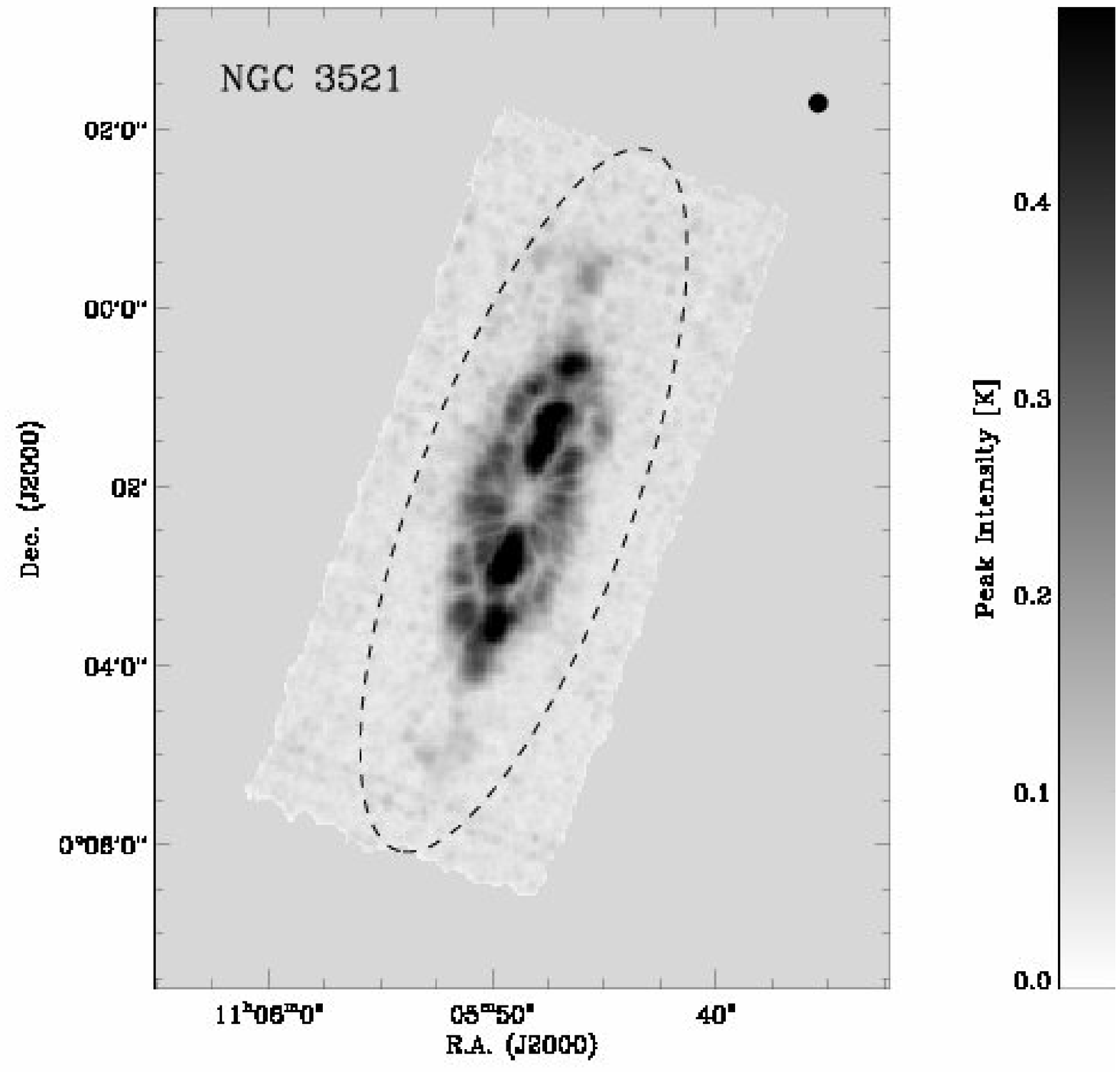}
  \caption{\label{NGC3521MOM0} As Figure \ref{NGC0628MOM0} but for NGC 3521.}
\end{figure*}

\begin{figure*}
  \plottwo{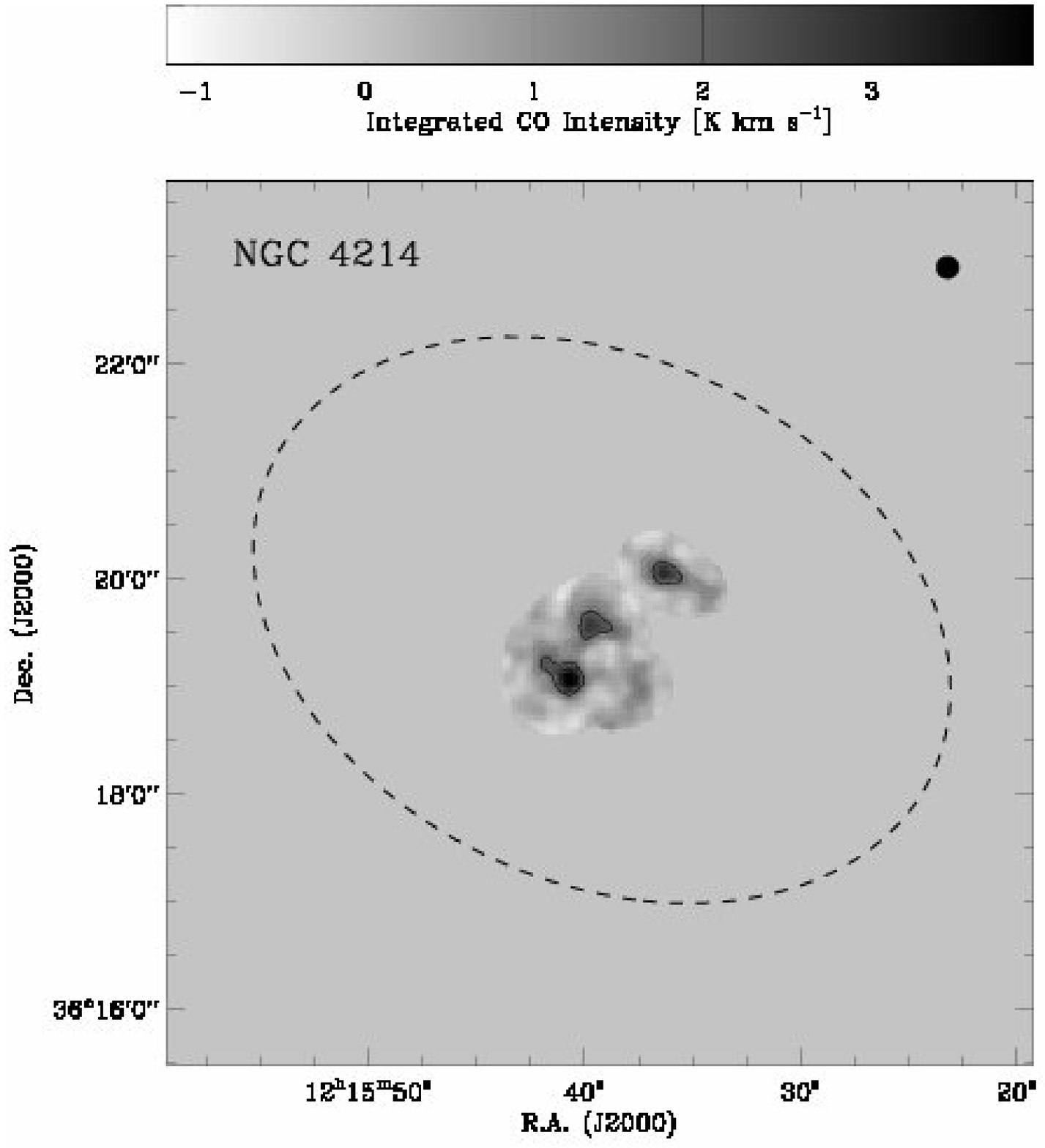}{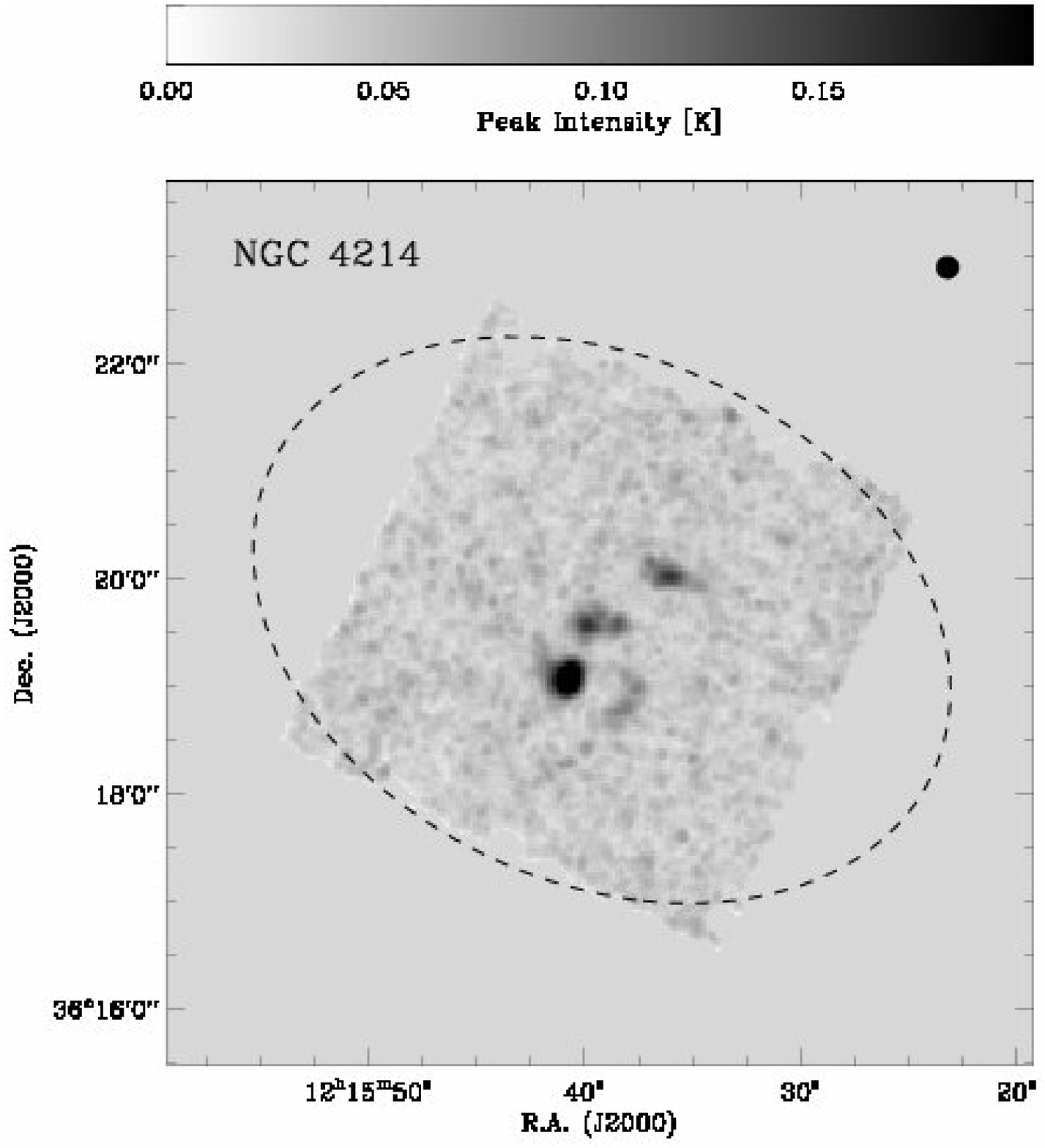}
  \caption{ \label{NGC4214MOM0} As Figure \ref{NGC0628MOM0} but for NGC 4214.}
\end{figure*}

\begin{figure*}
  \plottwo{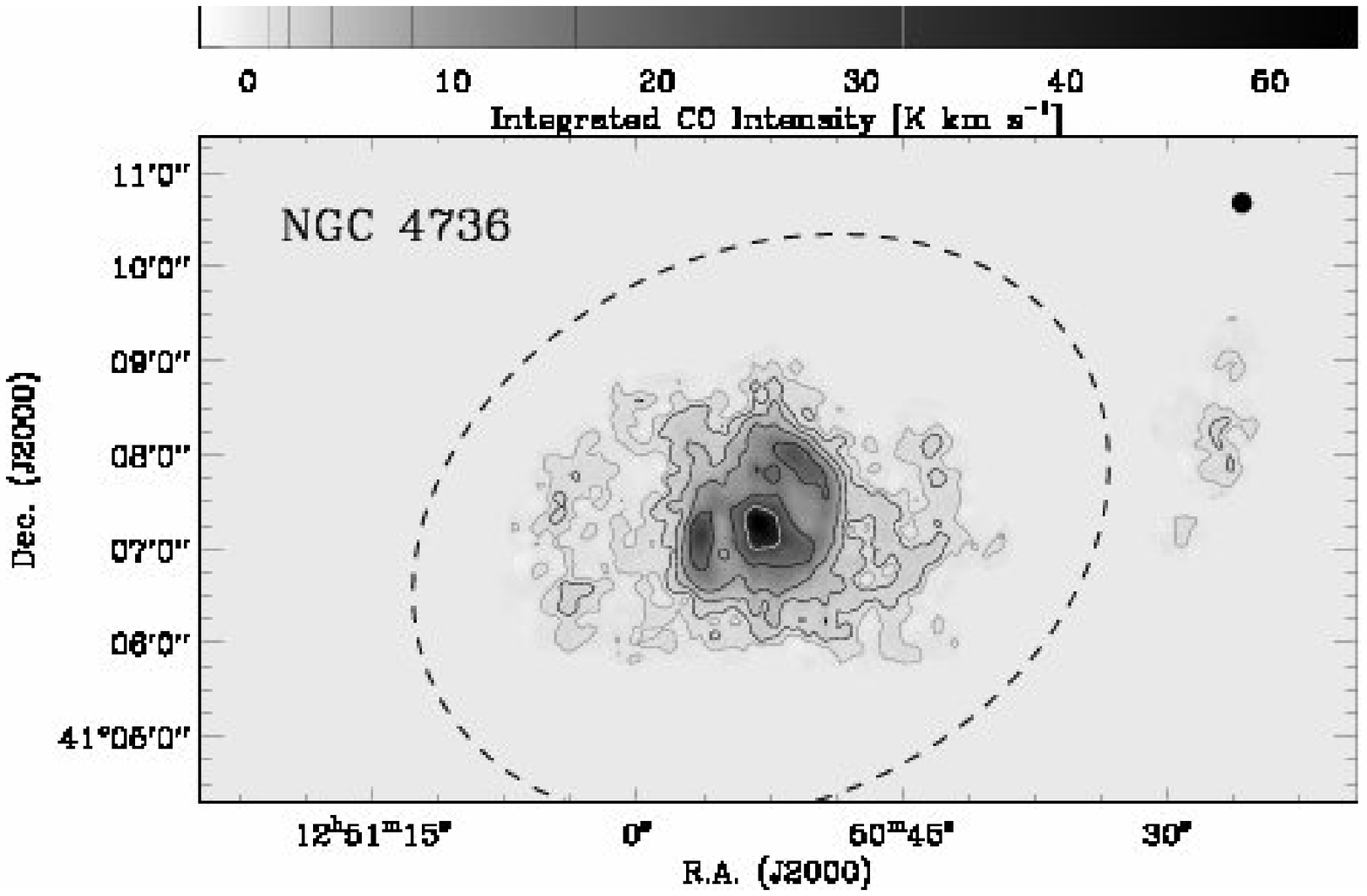}{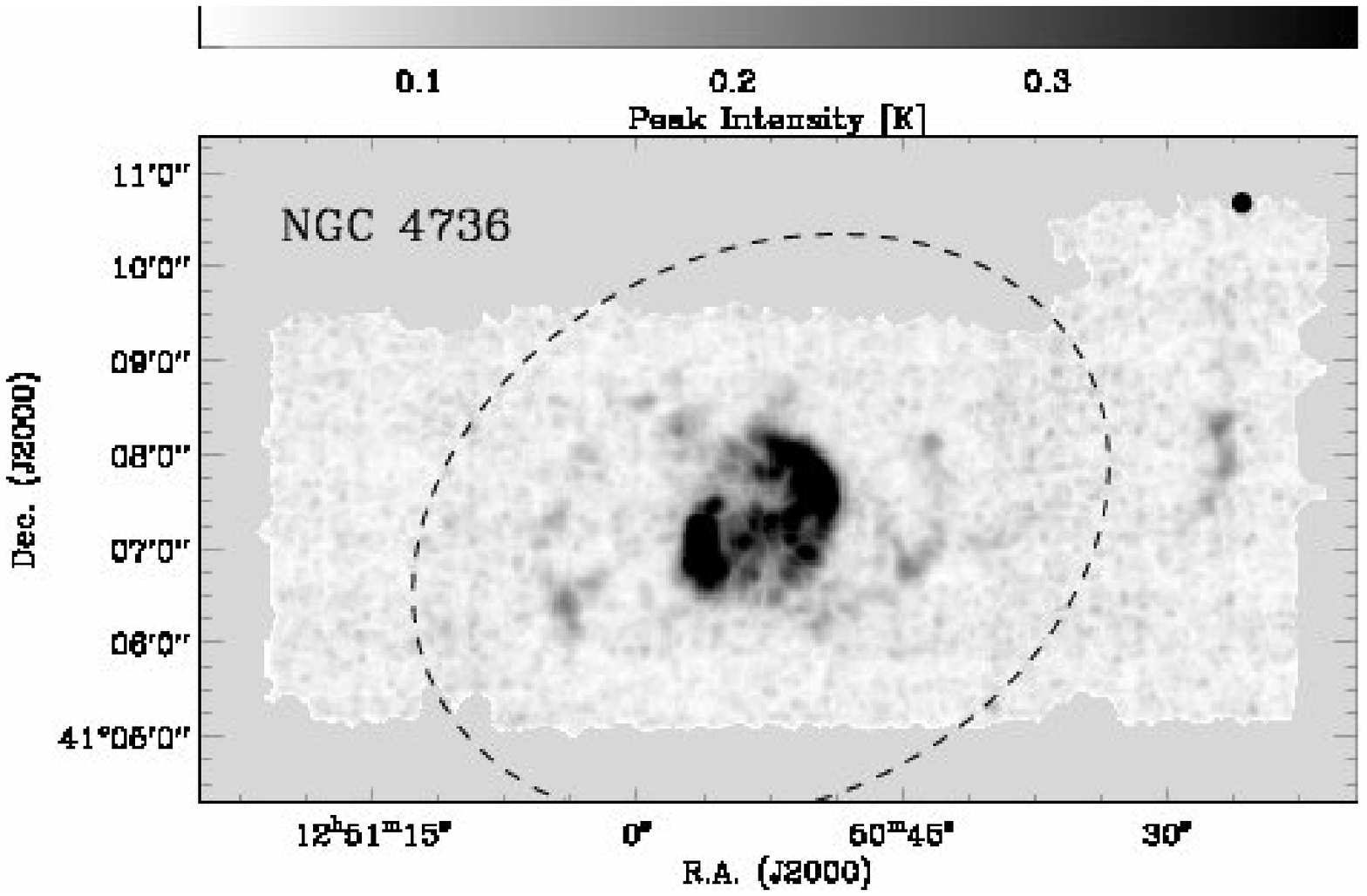}
  \caption{ \label{NGC4736MOM0} As Figure \ref{NGC0628MOM0} but for NGC 4736.}
\end{figure*}

The THINGS data allow us to test our assumption that CO and \hi\ have the same
mean velocity (\S \ref{BASEFIT}). To do so, we assemble a set of $\sim 1200$
spatially independent spectra, each with peak SNR greater than $5$. For each
spectrum, we measure the intensity-weighted mean velocity of CO
$J=\jtwo$ emission, $\langle v_{\rm CO}\rangle$, and extract the
intensity-weighted mean \hi\ velocity, $\langle v_{\rm HI}\rangle$, from the
THINGS first moment maps \citep{WALTER08}.

The gray histogram in Figure \ref{VELHIST} shows the distribution of $\langle
v_{\rm CO}\rangle - \langle v_{\rm HI}\rangle$ for all $1200$ spectra. The
mean velocities of CO and \hi\ appear closely matched for most spectra.  A
Gaussian fit to the distribution (thick dashed line) has $1\sigma$ width
$5.9$~km~s$^{-1}$ and a center consistent with zero offset
($-0.05$~km~s$^{-1}$). The scatter in $\langle v_{\rm CO}\rangle - \langle
v_{\rm HI}\rangle$ is much smaller than the $60$--$300$ km~s$^{-1}$ width of
our baseline fitting regions (\S \ref{BASEFIT}), so the agreement is not a
product of our reduction. The extended wings come almost entirely from a few
inclined, massive spirals (NGC~2841, 2903, 3521, 7331), galaxies that also
sometimes show multiple components in a single \hi\ spectrum. These outliers
are interesting, but here we emphasize the basic result that \hi\ and CO gas
show approximately the same mean velocity.

\subsection{Comparison to CO $J=\jone$ Observations}
\label{COMPARE}

\begin{figure*}
  \plottwo{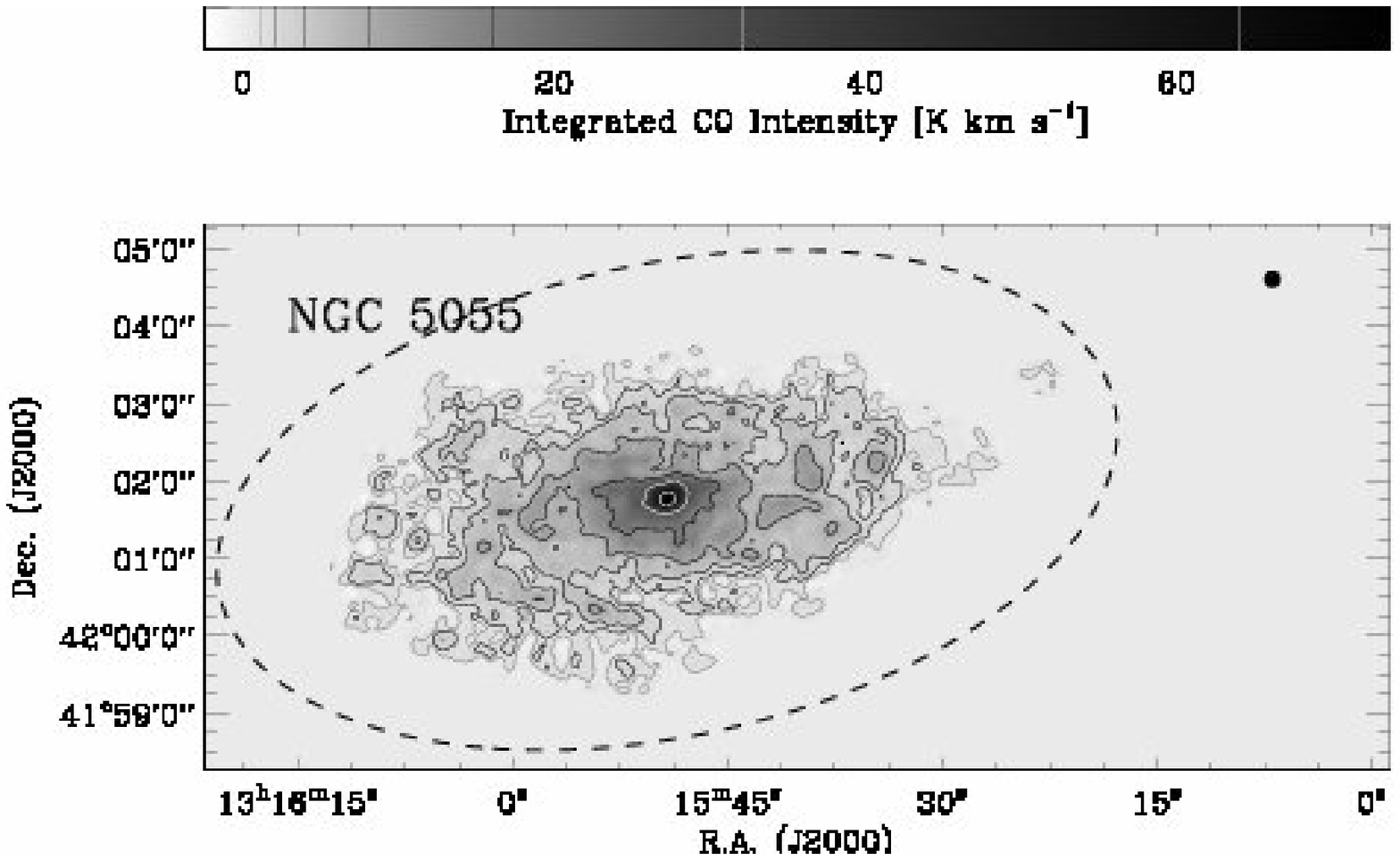}{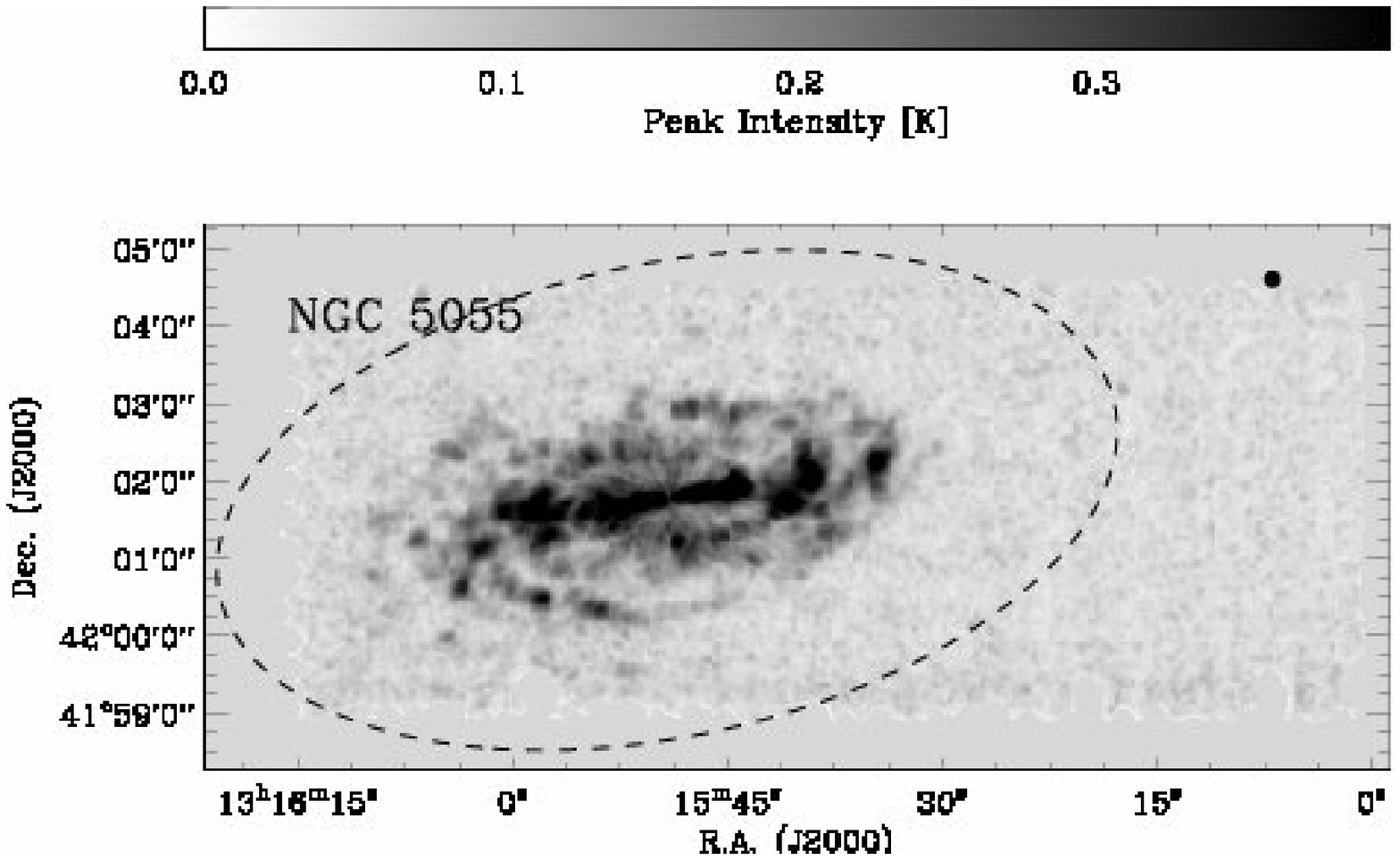}
  \caption{ \label{NGC5055MOM0} As Figure \ref{NGC0628MOM0} but for NGC 5055.}
\end{figure*}

\begin{figure*}
  \plottwo{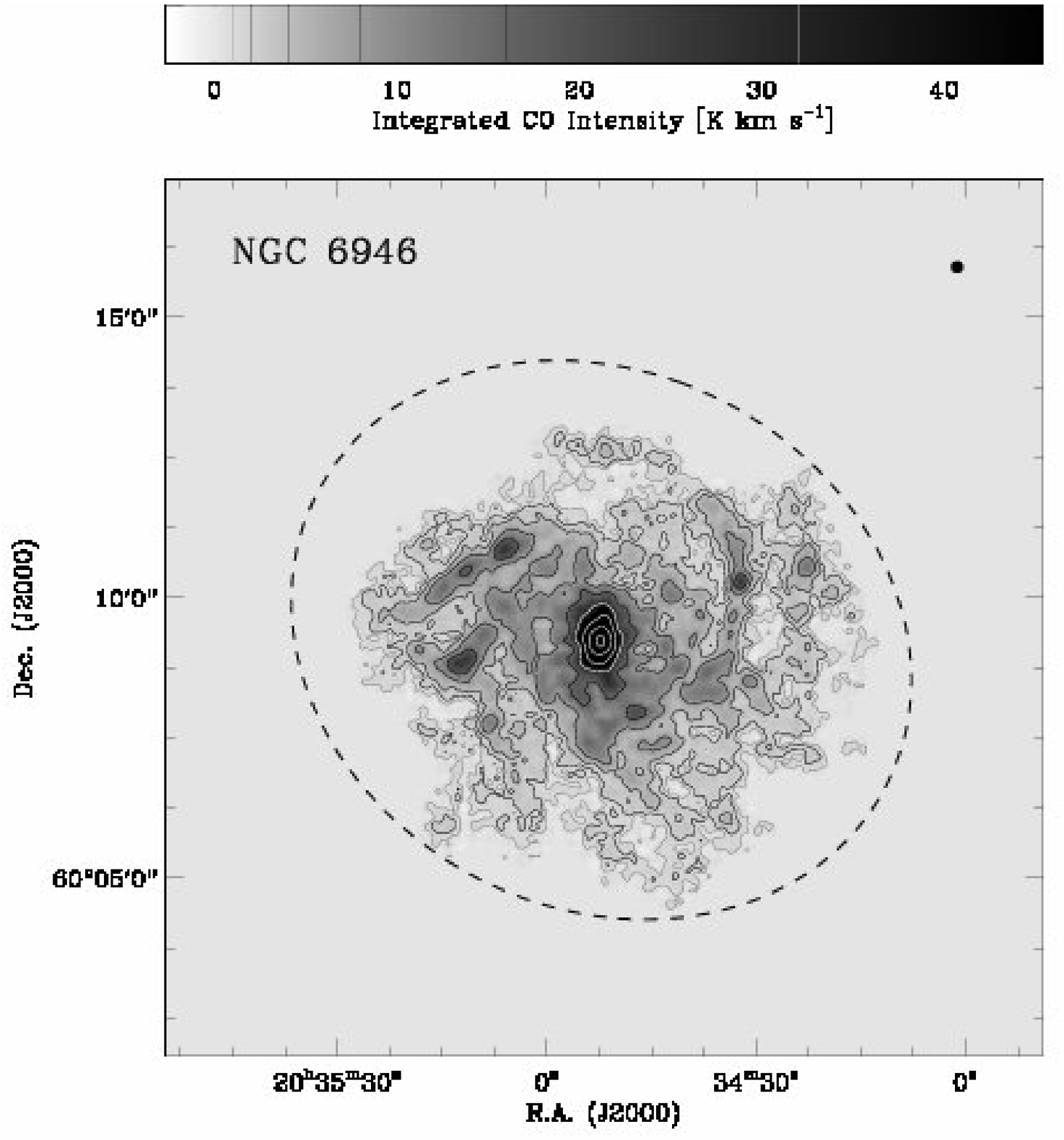}{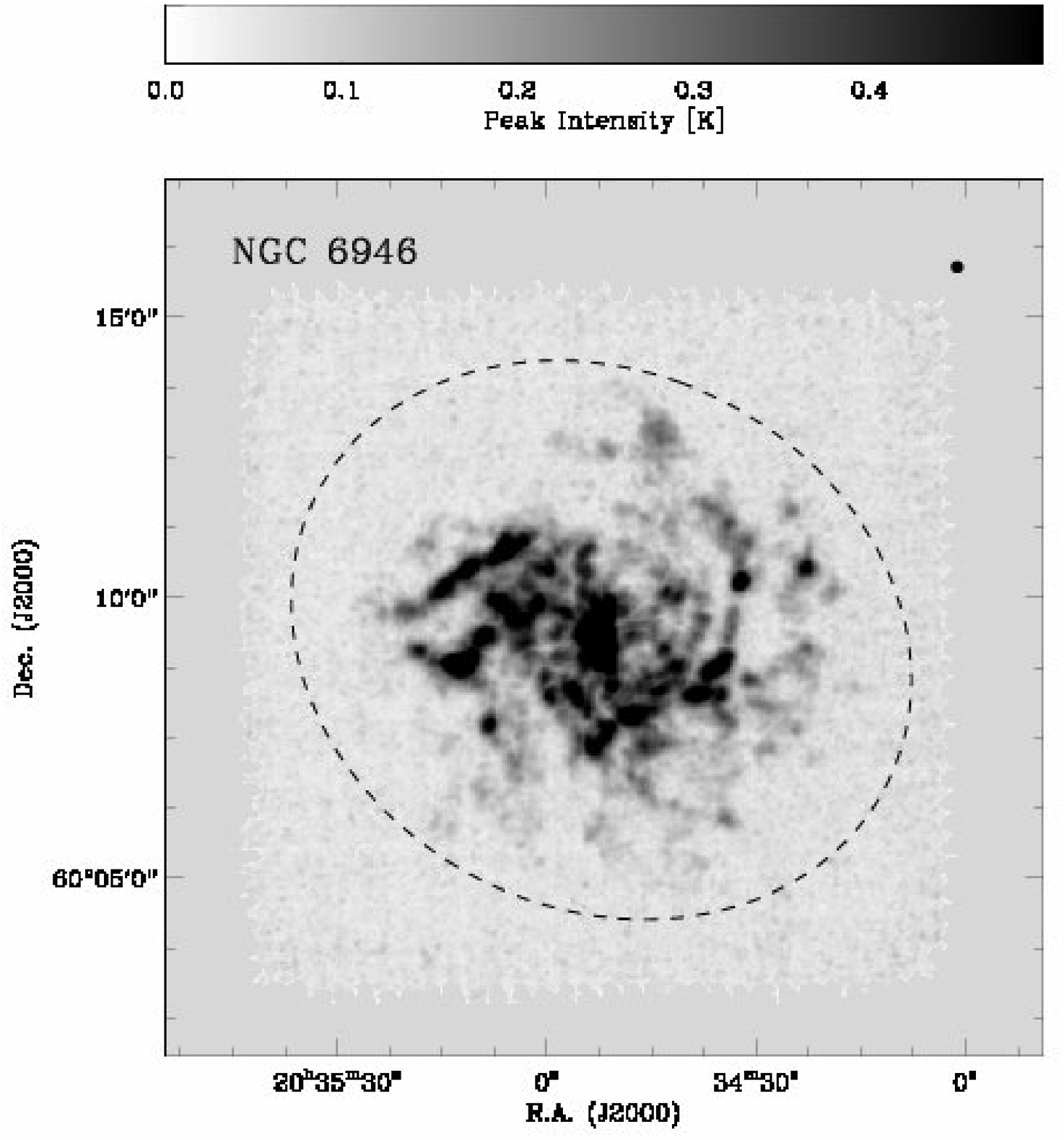}
  \caption{ \label{NGC6946MOM0} As Figure \ref{NGC0628MOM0} but for NGC 6946.}
\end{figure*}

\begin{figure*}
  \plottwo{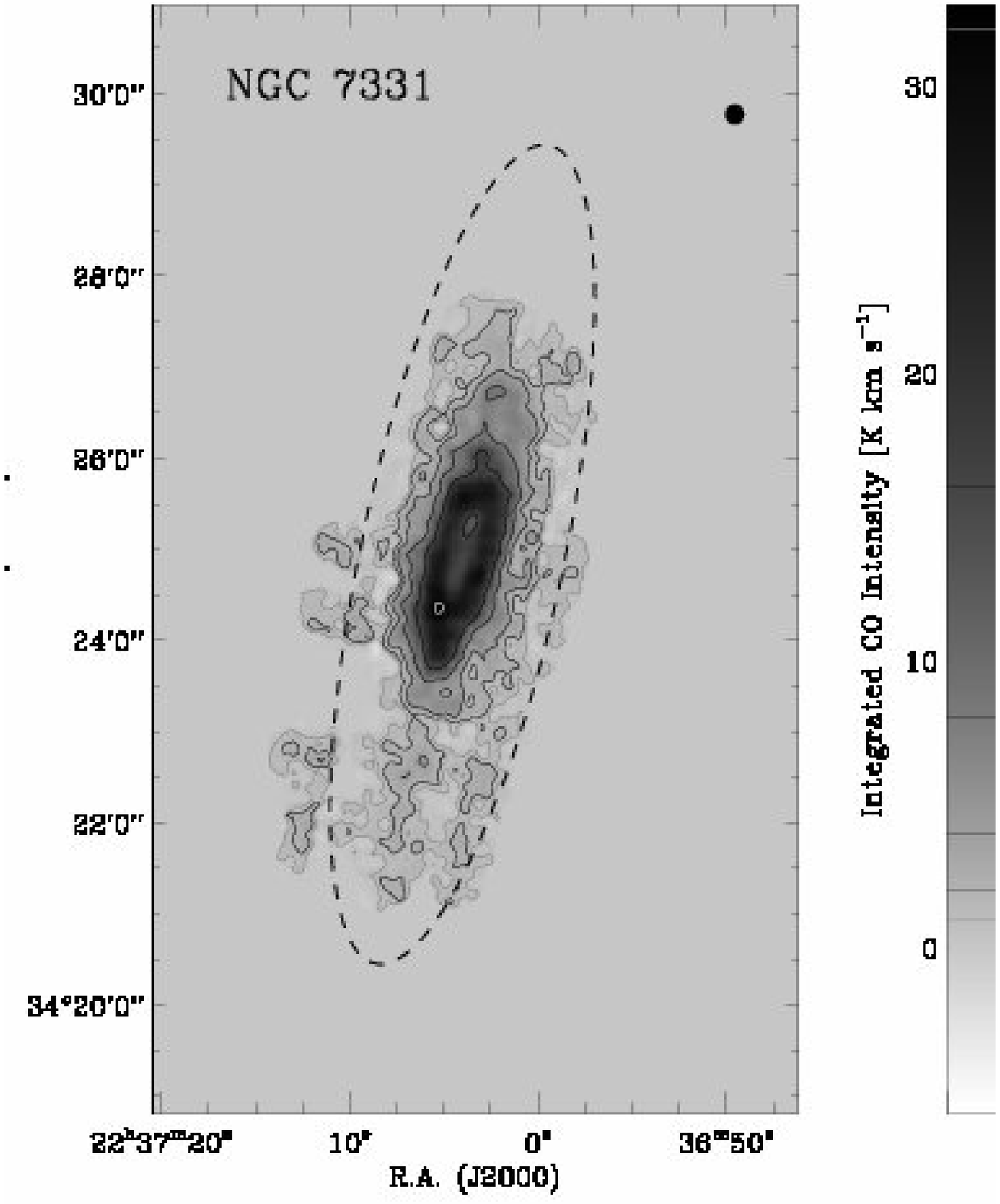}{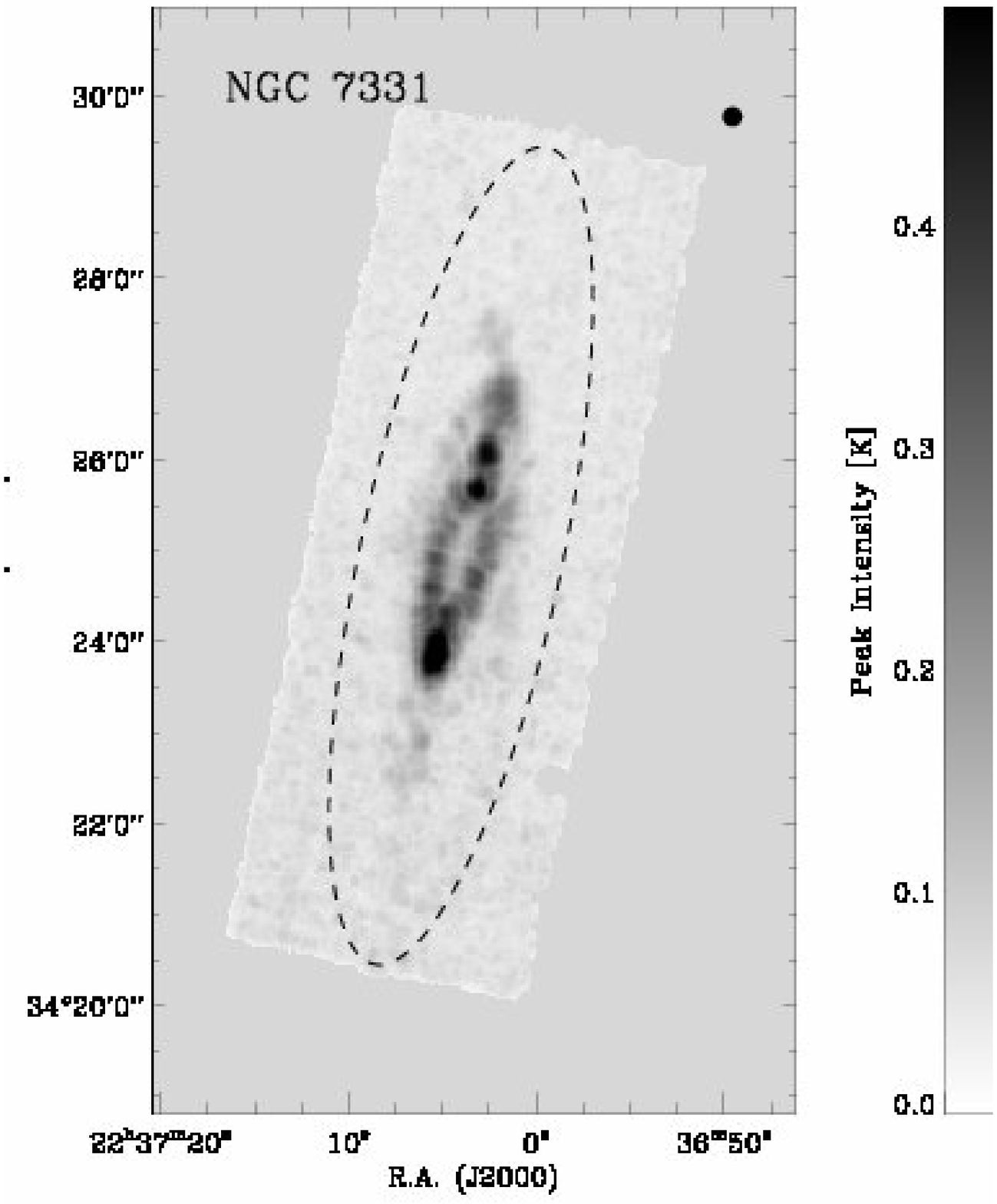}
  \caption{ \label{NGC7331MOM0} As Figure \ref{NGC0628MOM0} but for NGC 7331.}
\end{figure*}

CO $J=\jone$ emission has been mapped for many galaxies in our sample. Here we
compare HERACLES to the FCRAO survey \citep[][resolution
  $45\arcsec$]{YOUNG95}, BIMA SONG \citep[][]{HELFER03}, and the Nobeyama CO
Atlas of Nearby Spiral Galaxies \citep[][resolution $15\arcsec$]{KUNO07}. We
use two data sets drawn from BIMA SONG: on-the-fly maps made with the NRAO
12-m (resolution $55\arcsec$) and the combined BIMA + 12-m data (typical
resolution $\sim 6\arcsec$). We refer to the former as the ``NRAO 12-m'' and
the latter as BIMA SONG.

Our method of comparison is the following:

\begin{enumerate}
\item We convolve each plane of each HERACLES data cube with a series of
  Gaussian kernels to create data cubes with angular resolutions of $15$,
  $45$, and $55\arcsec$, appropriate for comparison with \citet{KUNO07}, the
  FCRAO survey, and the NRAO 12-m data. We convolve BIMA SONG to $15\arcsec$
  resolution in the same manner and compare it to HERACLES at this resolution.
\item For each galaxy, we extract spectra from each data cube at a series of
  independent pointings. When comparing to the FCRAO survey, these are the
  FCRAO pointings, which are usually spaced by a full ($45\arcsec$) beam width
  along the major axis. When comparing with \citet{KUNO07}, BIMA SONG, and the
  NRAO 12-m, the pointings are on a grid that covers most of the galaxy with
  sampling points spaced by a full beam width --- i.e., $15\arcsec$ and
  $55\arcsec$.
\item We smooth these spectra in velocity so that they all have comparable
  velocity resolution, $\sim15$~km~s$^{-1}$ (chosen to match the FCRAO
  survey).
\item We discard all spectra where the peak SNR is less than 5. When comparing
  with the FCRAO survey, we consider only pointings where \citet{YOUNG95}
  report a peak temperature, mean velocity, and line width.
\end{enumerate}

The result is a series of reasonably high SNR spectra at matched positions
with matching angular and velocity resolutions. For each spectrum, we measure
four parameters: the peak intensity of the line, $T_{\rm peak}$ (in K); the
intensity-weighted mean velocity of the line, $v_{\rm mean}$ (in km s$^{-1}$);
the full width at half maximum of the line, $v_{\rm FWHM}$ (in km s$^{-1}$);
and the integrated intensity of the line, $I_{\rm CO}$ (in K km s$^{-1}$).
Because the FCRAO survey data are not electronically available, we use the
values of these parameters reported by \citet{YOUNG95}.

\subsubsection{CO Velocity and Line Width}

\begin{figure*}
  \plotone{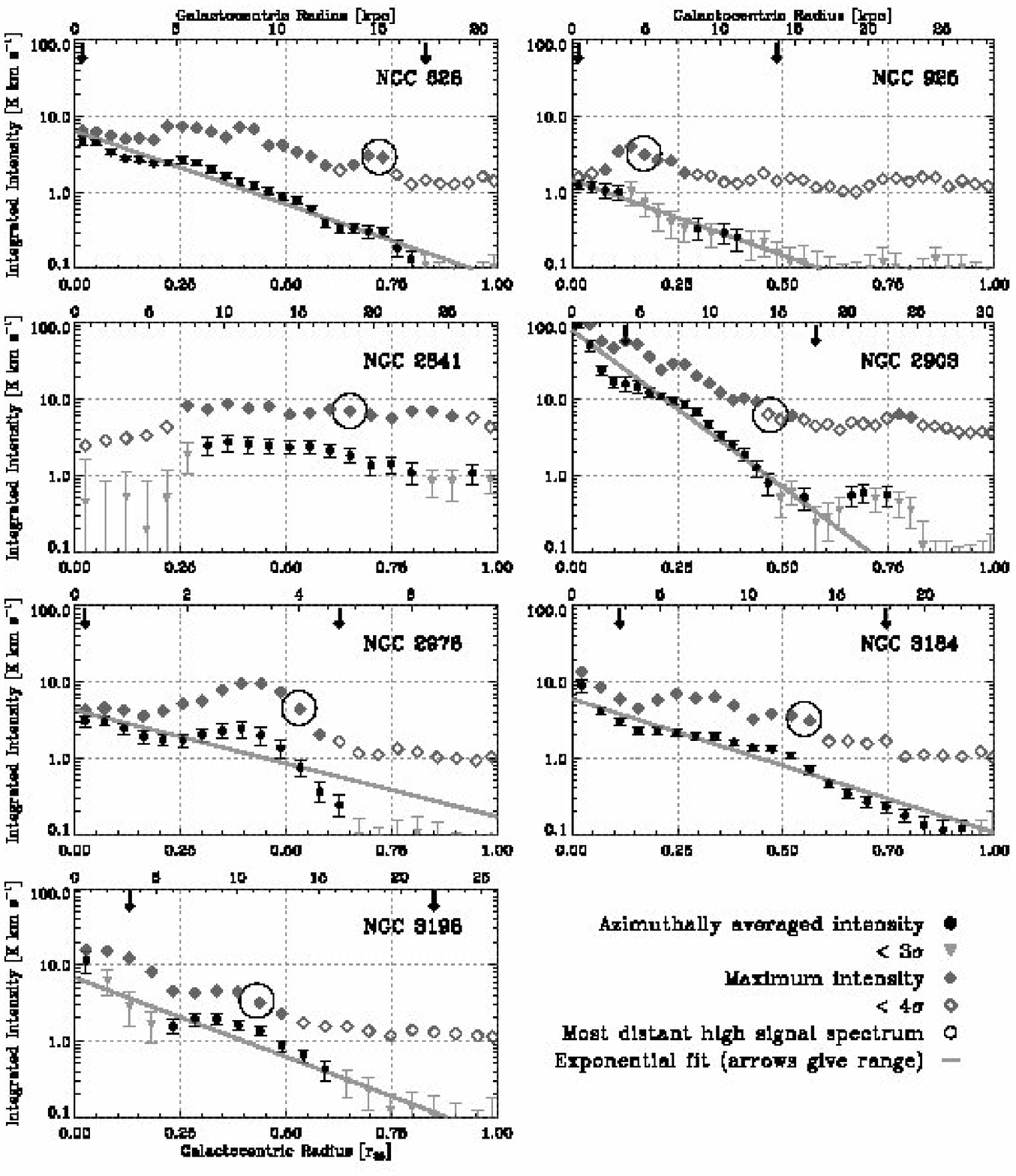}
  \caption{\label{PROFILES1} Integrated CO $J=\jtwo$ intensity, in
    K~km~s$^{-1}$, as a function of galactocentric radius in units of $r_{25}$
    (bottom) and kpc (top). Black circles and gray triangles show the
    intensity averaged over $10\arcsec$-wide tilted rings. Gray triangles
    indicate where the measurement divided by the RMS uncertainty is
    $<3$. Diamonds display the maximum intensity in each ring, with filled
    symbols showing where this value is clearly higher than expected from
    noise alone. The dark gray line shows the best fit scale length. Arrows at
    the top of the plot bracket the region over which we carry out this
    fit. Large open circles indicate the location of the most distance high
    signal spectrum (\S \ref{RADIAL}).}
\end{figure*}

\begin{figure*}
  \plotone{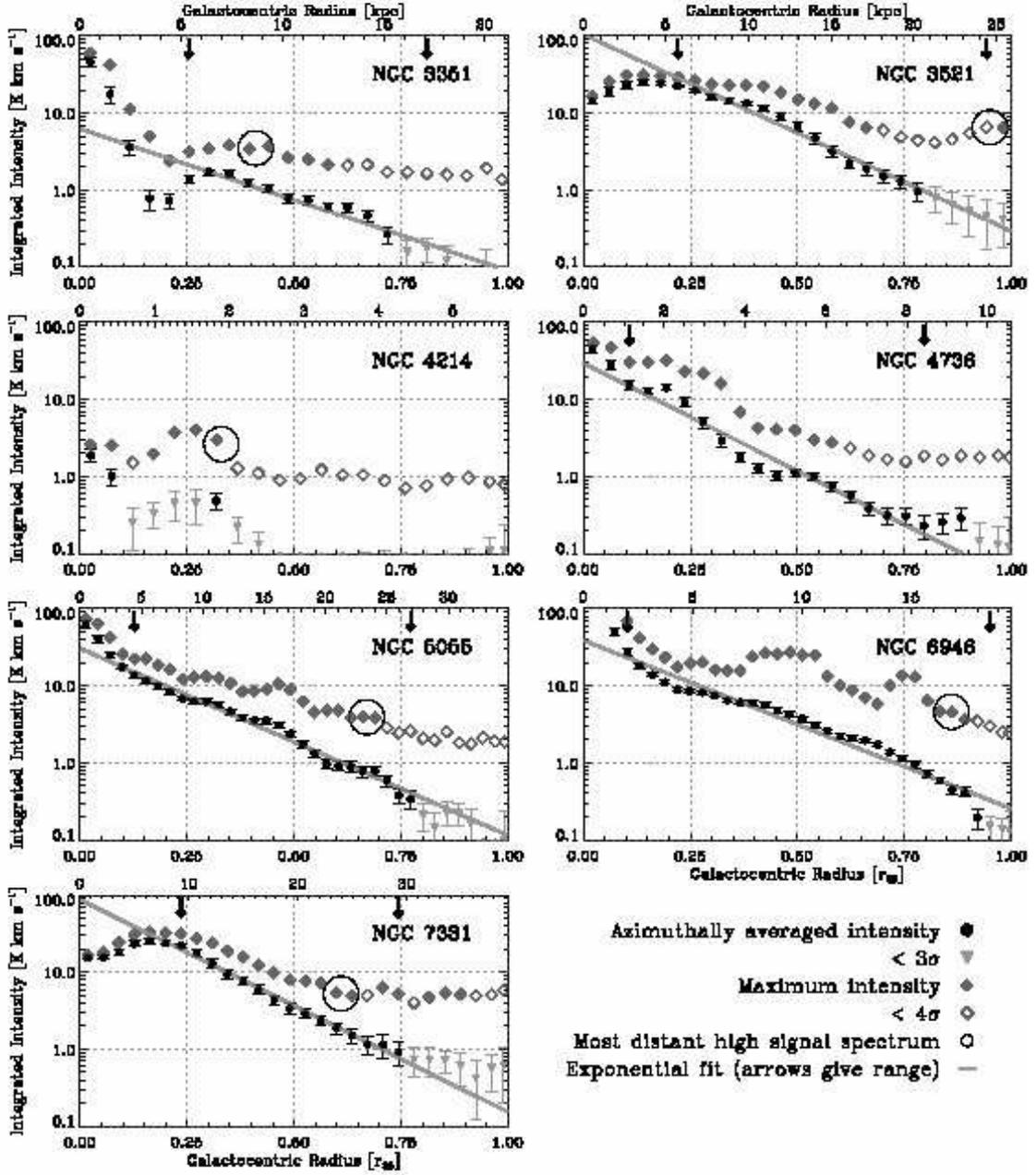}
  \caption{\figurenum{\ref{PROFILES1}} Integrated intensity as a function of
    galactocentric radius for the remaining galaxies.}
\end{figure*}

\begin{figure}
  \plotone{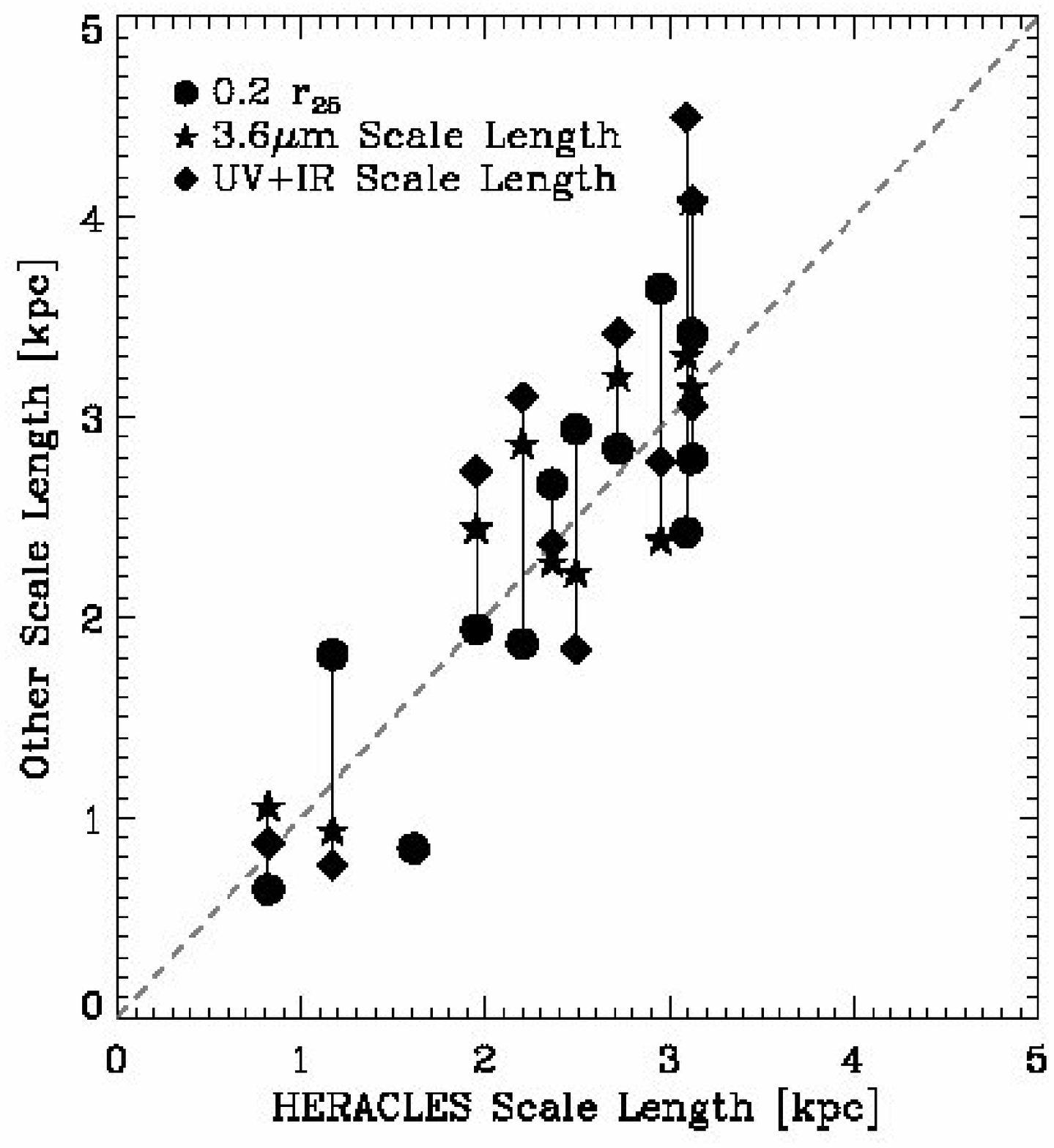}
  \caption{\label{SLSLPLOT} Several sizes for our targets ($y$-axis) as a
    function of the exponential scale lengths fit to the azimuthally average
    profiles ($x$-axis). Circles show $0.2~r_{25}$; stars show the 3.6$\mu$m
    scale length, a tracer of the old stellar distribution; and diamonds show
    the scale length {\em GALEX} FUV + {\em Spitzer} 24$\mu$m
    emission. Vertical lines connect points from the same galaxy. The dashed
    line shows equality.}
\end{figure}

\begin{figure}
  \plotone{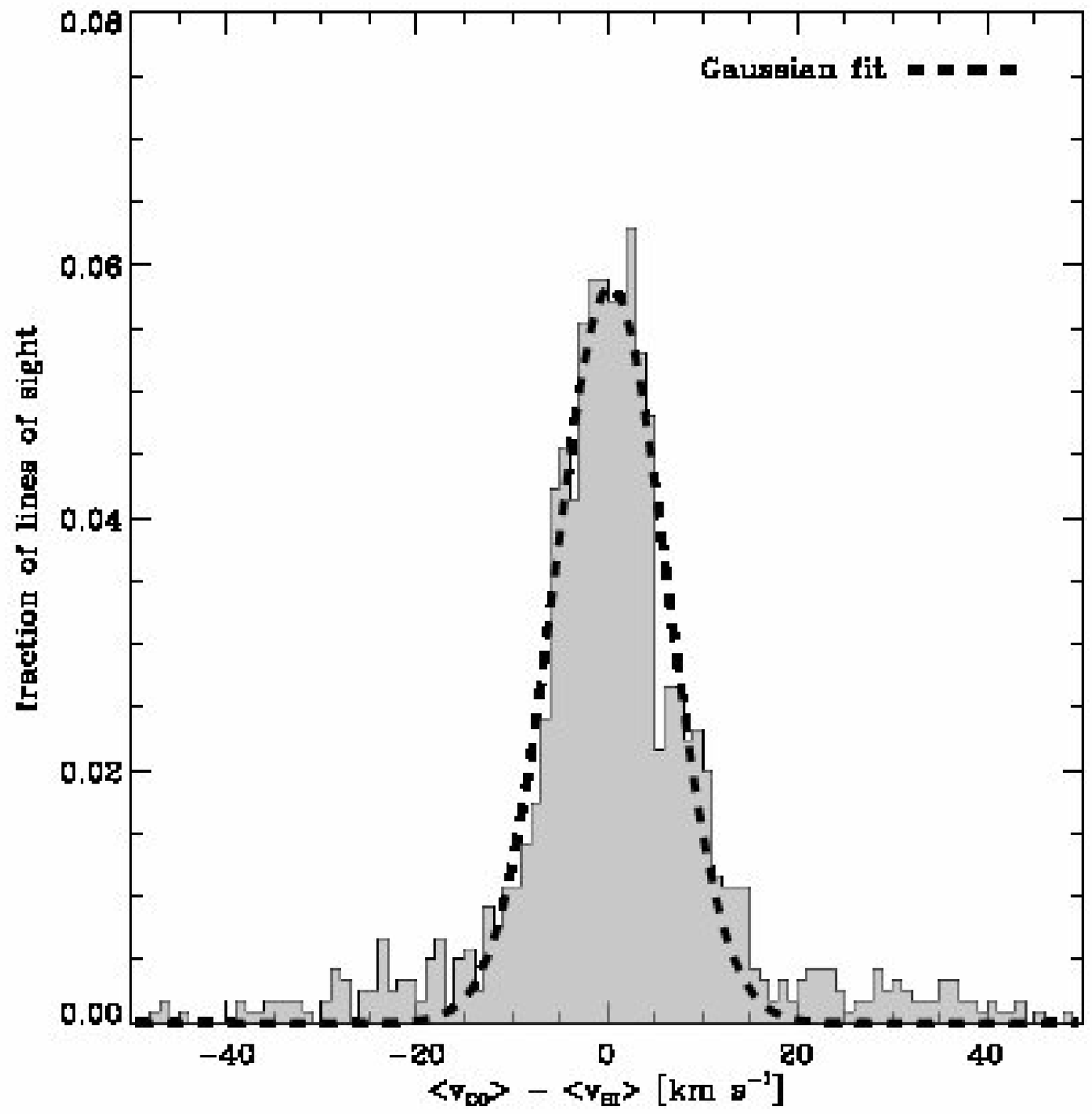}
  \caption{\label{VELHIST} Distribution of $\langle v_{\rm CO}\rangle -
    \langle v_{\rm HI}\rangle $, the difference between the intensity-weighted
    mean CO velocity (from HERACLES) and intensity-weighted mean \hi\ velocity
    (from THINGS), along lines of sight with good signal-to-noise.  The best
    fit Gaussian (dashed line) has $1\sigma$ width $5.9$~km~s$^{-1}$ and
    center $-0.1$~km~s$^{-1}$.}
\end{figure}

\begin{deluxetable*}{l c c c c}
  \tabletypesize{\small} \tablewidth{0pt} \tablecolumns{5}
  \tablecaption{\label{COMPTAB} Comparison of HERACLES and Other Surveys}
  \tablehead{ \colhead{Quantity} & \colhead{FCRAO Survey} & \colhead{NRAO 12-m} & \colhead{BIMA SONG} & \colhead{\citet{KUNO07}}}
  \startdata
  $\Delta v_{\rm mean}$ (km s$^{-1}$)\tablenotemark{a} & -2.9 & 3.6 & 2.4 & -0.4 \\
  ratio of $v_{\rm FWHM}$\tablenotemark{a} & 1.01 & 1.05 & 1.18 & 0.93 \\
  ratio of $T_{\rm peak}$\tablenotemark{a} & 0.76 & 0.87 & 0.67 & 0.81 \\
  ratio of $I_{\rm CO}$\tablenotemark{a} & 0.71 & 0.87 & 0.84 & 0.74 \\
  \enddata
  \tablenotetext{a}{Values are mean difference (HERACLES - other survey) or
    ratio (HERACLES/other survey) giving equal weight to each galaxy.}
\end{deluxetable*}

\begin{figure*}
  \plottwo{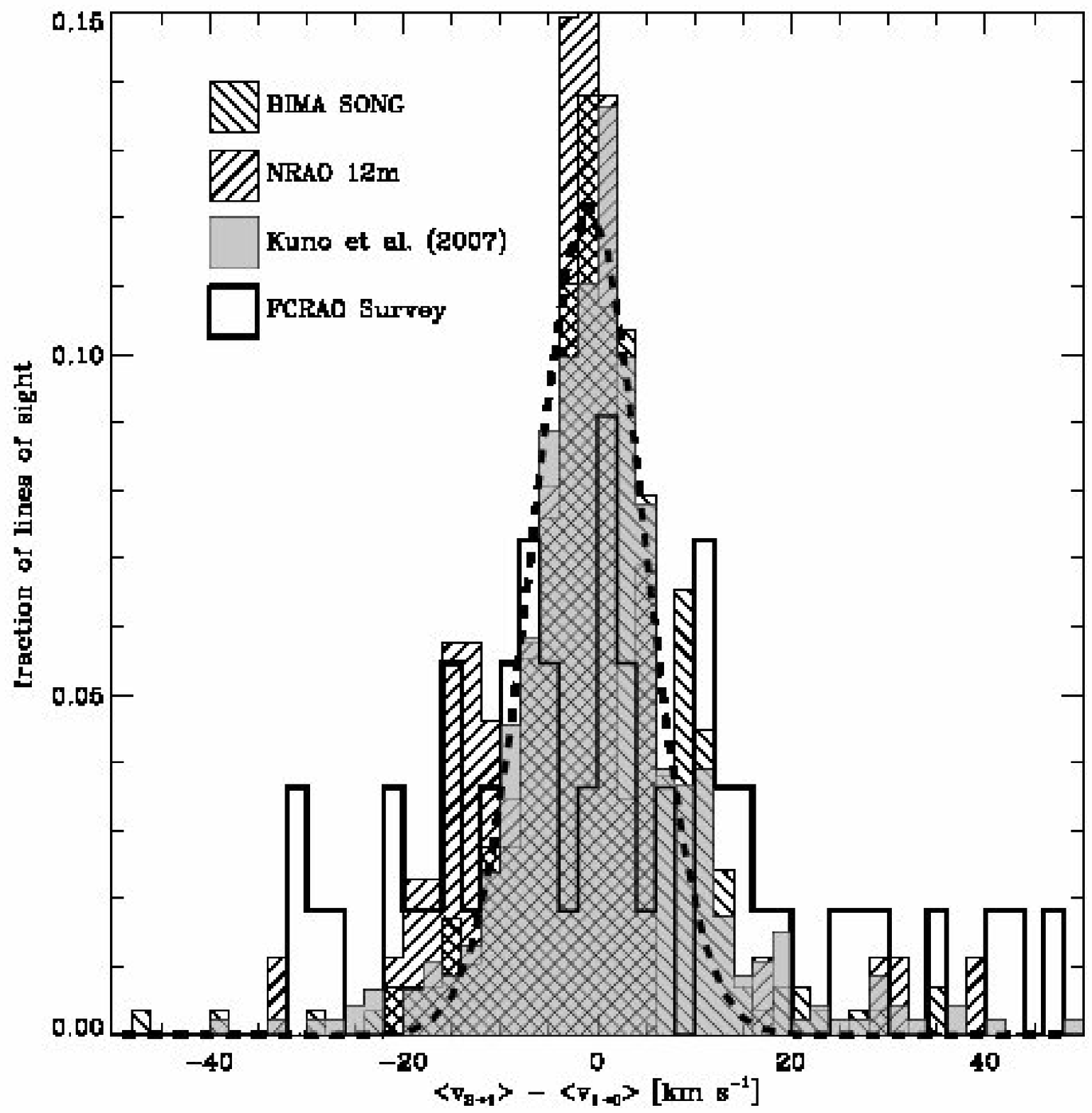}{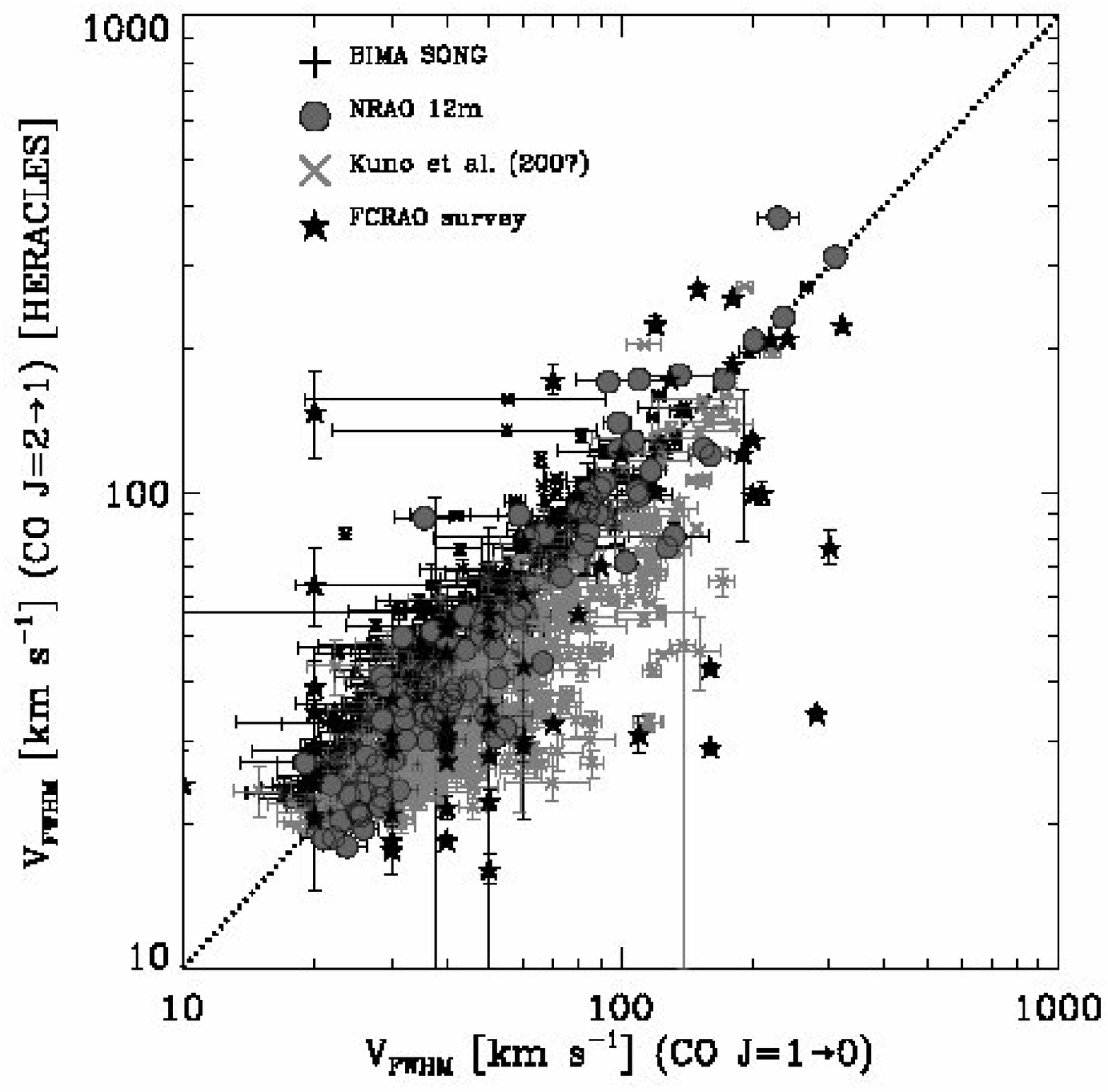}
  \plottwo{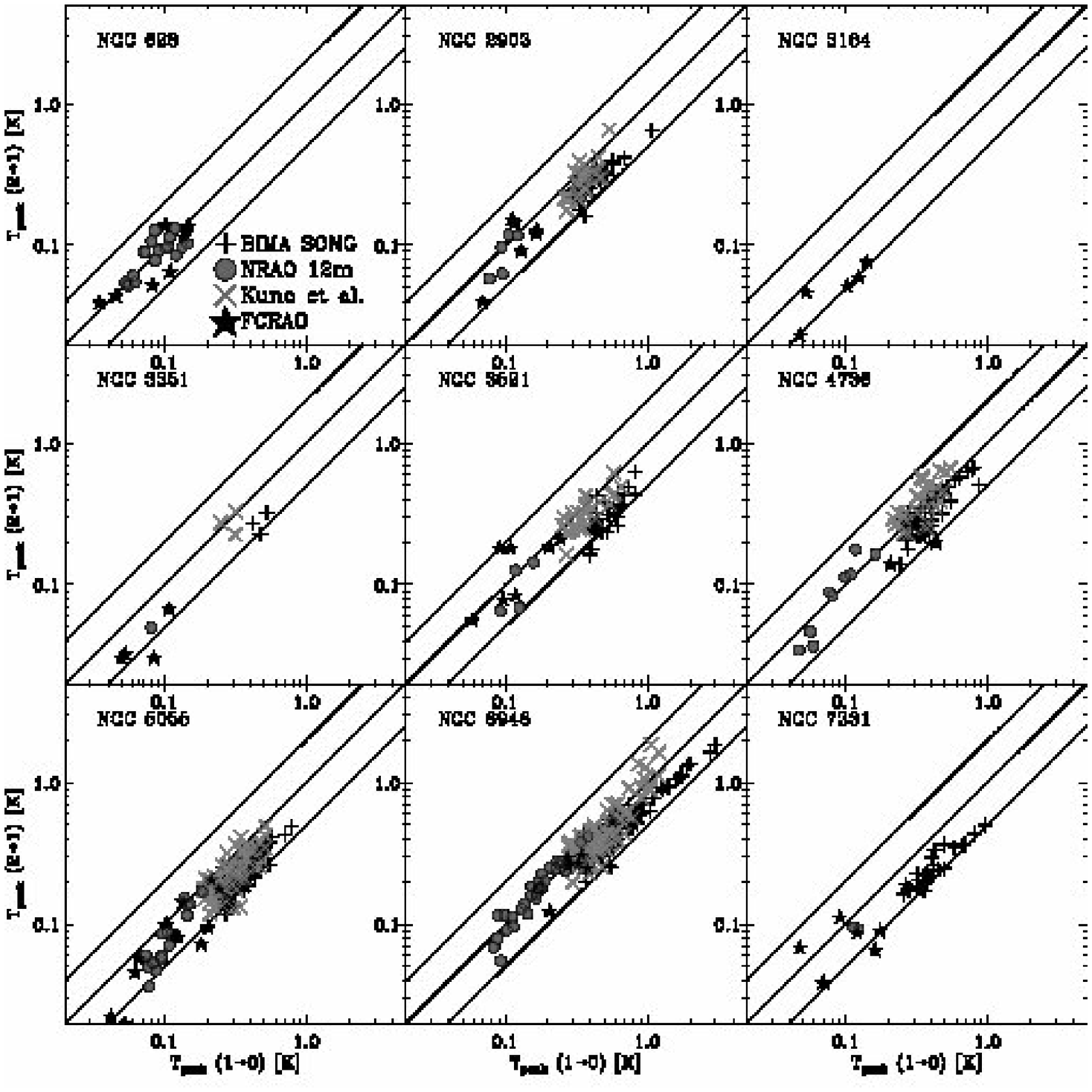}{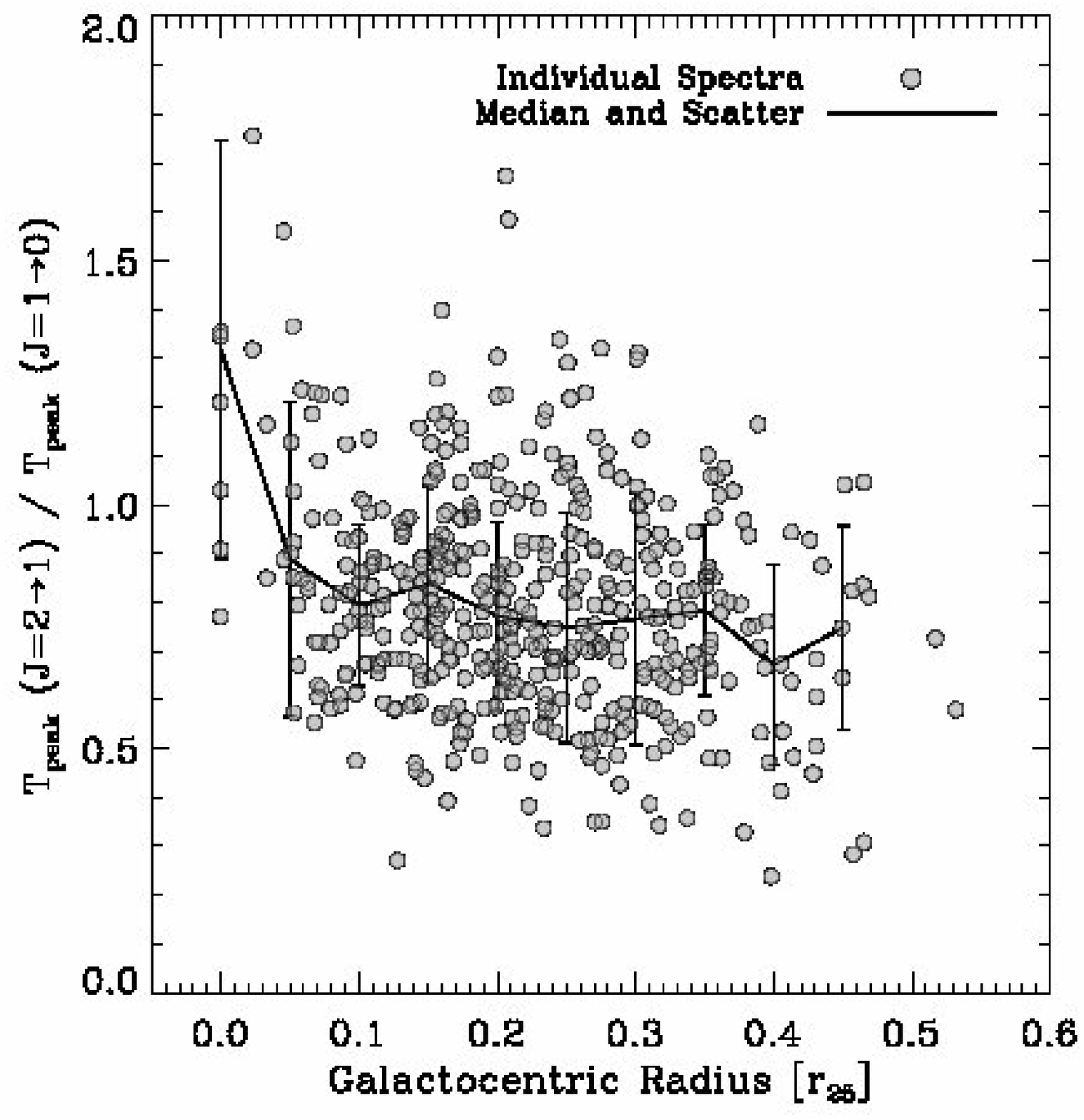}
  \caption{\label{COMPARE21} Comparison of HERACLES to previous CO $J=\jone$
    surveys. ({\em Top Left}) Normalized histograms show the difference
    between intensity weighted mean velocity measured by HERACLES and four
    surveys of CO $\jone$ emission. A thick dashed line shows a Gaussian fit
    to the average of the BIMA SONG, the NRAO 12-m, and \citet{KUNO07}
    histograms (center $-1.2$~km~s$^{-1}$, $1\sigma$ width
    $5.6$~km~s$^{-1}$). ({\em Top Right}) The full line width at half maximum
    ($v_{\rm FWHM}$) measured by HERACLES ($y$-axis) as a function of $v_{\rm
      FWHM}$ measured the same CO $J=\jone$ surveys. The dotted line shows
    equality. ({\em Bottom Left}) Peak temperature (intensity) in the HERACLES
    CO $J=\jtwo$ spectrum as a function of peak temperature in the CO
    $J=\jone$ spectrum at the same position and resolution. From top to
    bottom, solid lines show ratios of $2.0$, $1.0$, and $0.5$. ({\em Bottom
      Right}) Ratio of peak CO $J=\jtwo$ temperature (intensity) measured by
    HERACLES to peak CO $J=\jone$ temperature measured \citet{KUNO07} as a
    function of galactocentric radius normalized to $r_{25}$. Each gray point
    indicates an individual pair of high significance spectra. The black
    points and lines show the median and scatter for the data in bins
    $0.05~r_{25}$ wide with the first bin centered at $0$.}
\end{figure*}

The line width and mean velocity are not expected to vary strongly between the
CO $J=\jtwo$ and $J=\jone$ transition. Therefore comparing our
measurements of these quantities to those from earlier surveys allows a basic
check on our data. Overall, this exercise confirms that we measure the same
basic line shapes found by previous surveys at matched positions and
resolution. We show this in the top two panels of Figure \ref{COMPARE21} and
the first two lines of Table \ref{COMPTAB}, which give mean velocity offset
and line width ratio (with equal weight to each galaxy).

The top left panel of Figure \ref{COMPARE21} shows the distribution of
differences between the intensity-weighted mean CO $J=\jtwo$ (HERACLES)
velocity and the intensity-weighted mean CO $J=\jone$ velocity. Each
comparison survey appears as a separate, normalized histogram. A thick dashed
line shows a Gaussian fit to the average of the BIMA SONG, NRAO 12-m, and
\citet{KUNO07} histograms. This Gaussian is centered at $-1.2$~km~s$^{-1}$ and
has $1\sigma$ width $5.6$~km~s$^{-1}$, identical with the uncertainties to
what we found comparing CO $J=\jtwo$ to \hi .

The top right panel of Figure \ref{COMPARE21} shows the FWHM line width,
$v_{\rm FWHM}$, from HERACLES measured as a function of $v_{\rm FWHM}$ from
other surveys. Error bars show $1\sigma$ uncertainty, estimated by repeatedly
adding the measured noise to each spectrum and re-measuring $v_{\rm FWHM}$
(without the FCRAO spectra, we cannot estimate uncertainties for these
data). As with the mean velocity, there is general good agreement between
HERACLES, BIMA SONG, the NRAO 12-m, and \citet{KUNO07}. There are systematic
differences: BIMA SONG tends to show slightly lower line widths than the other
data sets while the \citet{KUNO07} yield slightly higher line widths. These
have magnitude $\pm 20\%$ and affect the intercomparison of CO
$J=\jone$ data as much as the comparison between HERACLES and the
other surveys.

The outlier this comparison is the FCRAO survey. These data do not agree as
well with our own as the other three data sets. Specifically, there are
several pointings where the mean velocity and line width disagree strongly
with our data. Most of these locations also overlap with BIMA SONG or the
\citet{KUNO07} survey and in these cases, these surveys also disagree with the
FCRAO survey. We note that 1) the FCRAO survey is the only data set from which
we do not ourselves measure the line parameters, so methodology may drive some
of the difference; 2) these are the only data which are not maps, so it is not
easy to tell whether pointing offsets may affect the comparison.

\subsubsection{CO $J=\jtwo / J=\jone$ Line Ratio}
\label{LINERAT}

\begin{deluxetable*}{l c c c c c}
  \tabletypesize{\small} \tablewidth{0pt} \tablecolumns{4}
  \tablecaption{\label{RATTAB} Average $T_{\rm peak}$ Ratio by Galaxy and
    Survey\tablenotemark{a}} \tablehead{ \colhead{Galaxy} & \colhead{FCRAO
      Survey} & \colhead{NRAO 12-m} & \colhead{BIMA SONG} &
    \colhead{\citet{KUNO07}} & \colhead{Other\tablenotemark{b} }} \startdata
  NGC~628 & 0.93 & 1.01 & \nodata & \nodata & $0.44$ \\
  NGC~2841 & 0.72 & \nodata & \nodata & \nodata & \\
  NGC~2903 & 0.91 & 0.90 & 0.68 & 0.82 & $0.6$--$0.9$ \\
  NGC~3184 & 0.59 & \nodata & \nodata & 0.60 & \\
  NGC~3351 & 0.50 & 0.62 & 0.59 & 0.95 & 1.65 \\
  NGC~3521 & 1.14 & 0.81 & 0.59 & 0.86 & \\
  NGC~4736 & 0.70 & 1.02 & 0.83 & 1.12 & \\
  NGC~5055 & 0.71 & 0.77 & 0.64 & 0.79 & \\
  NGC~6946 & 0.85 & 1.03 & 0.77 & 0.92 & $0.8$ \\
  NGC~7331 & 0.82 & 0.79 & 0.60 & \nodata & $0.5$ \\
\enddata
\tablenotetext{a}{Mean ratio of peak CO $J=\jtwo$ temperature to peak
  CO $J=\jone$ temperature.}  
\tablenotetext{b}{Integrated intensity
  ({\em not} peak temperature) ratios drawn from the literature. References:
  {\em NGC~628, NGC~3351, NGC~7331} --- \citet{BRAINE93}; {\em NGC~2903} ---
  \citet{JACKSON91}; {\em NGC~6946} --- \citet{CROSTHWAITE07}}
\end{deluxetable*}

The ratio of CO $J=\jtwo$ measured by HERACLES to CO $J=\jone$
emission measured by previous surveys reflects both the relative calibrations
of the telescopes and the line ratio, $R_{21} =
I_{jtwo}/I_{\jone}$, which depends on the optical depth and
excitation temperature of the gas. Generally speaking, $R_{21} > 1$ indicates
warm, optically thin gas. Optically thick gas produces $R_{21} \approx
0.5$--$1.0$ with the value reflecting the temperature of the gas, though in
low density regions the levels may not be populated according to local
thermodynamic equilibrium \citep[LTE; see, e.g.,][for a more thorough
  discussion]{ECKART90}.

In the bottom left panel of Figure \ref{COMPARE21} we plot the peak
temperatures measured by HERACLES ($y$-axis) as a function of the peak
temperature measured from other surveys at matched position and resolution
($x$-axis). We use the peak temperature because it avoids any concern related
to the line profile or range of integration. We do not plot error bars, but
recall that our condition to include a spectrum in this analysis was that
$T_{\rm peak}$ have SNR $>5$ (meaning the statistical uncertainty on the ratio
is always $1\sigma < 30\%$). Table \ref{RATTAB} compiles the average peak
temperature ratio for each galaxy and survey and notes several measurements
from the literature.

The ratios in Table \ref{RATTAB} span $0.48$--$1.06$, with most values between
$0.6$ and $1.0$. These values agree with previous observations of the Milky
Way and other galaxies. CO emission in the Milky Way comes mostly from
optically thick regions and typical line ratios are $\sim 0.6$ -- $0.8$ in the
disk and $\sim 1$ towards the Galactic center
\citep{SAKAMOTO95,OKA96,OKA98}. In the central regions of other galaxies,
\citet{BRAINE92} found an average line ratio $\sim 0.9$. \citet{CASOLI91}
compile observations on a number of galaxies and report typical values of
$\sim 1$ in galaxy nuclei and $0.5$ -- $0.7$ in galaxy disks.

If the gas is optically thick, a ratio of $0.6$ corresponds to an excitation
temperature of $\sim 5$~K, $0.8$ to $\sim 10$~K, and $0.9$ to $\sim
21$~K. Higher excitation temperatures yield a line ratio $\sim 1$. The average
of all the values in Table \ref{RATTAB} is $0.81$, which is consistent with
optically thick gas with an excitation temperature $\sim 10$~K. One should
treat this temperature with caution: we do not constrain the optical depth
(e.g., using one of the isotopes of CO) and we have no verification that our
sources are in LTE, so sub-thermally excited, low-density envelopes --- which
are observed in Galactic GMCs --- may lower the overall intensity ratio
\citep[e.g.,][]{SAKAMOTO94}.

There is noticeable scatter among the peak temperatures ratios derived from
different data sets for the same galaxy. Some scatter may arise from
differences in resolutions or area considered (the set of spectra with peak
SNR $>5$ varies with comparison survey). If we ignore these explanations,
Table \ref{RATTAB} suggests that there are systematic calibration differences
among the various surveys at the $\sim 10\%$ level and that the calibration of
a given galaxy in a given survey is uncertain by $\sim 15\%$. These
uncertainties seem reasonable for millimeter line observations and agree with
the scatter in comparisons with the FCRAO survey carried out by \citet[][
  their Figure 51]{HELFER03} and \citet[][their Figure 1]{KUNO07}.

Because $R_{21}$ is observed to vary between the center and the disk in both
the Milky Way and other galaxies, the bottom right panel of Figure
\ref{COMPARE21} shows the ratio of $T_{\rm peak}$ in the two transitions as a
function of galactocentric radius. This plot shows only data from HERACLES and
\citet{KUNO07}. These data most closely match our own in observing strategy,
instrumentation, and resolution making them the best option for this
comparison. Black points connected by lines show the median ratio in bins
$0.05~r_{25}$ with the first bin centered at $0$. In the centers of galaxies,
we observe the same trend noted by \citet{CASOLI91} and found in the Milky
Way: the ratio is high ($\sim 1.3$) in the center of the galaxy and then drops
rapidly to $\sim 0.8$, remaining almost constant at this value out to the edge
of our comparison at $\sim 0.45~r_{25}$. This constancy must be interpreted
with care. By selecting only high significance spectra, we bias ourselves to
bright regions. This may have the effect of creating a homogeneous data set
while omitting lower intensity emission that accounts for a significant
fraction of the emission at large radii.

\section{Summary}
\label{SUMMARY}

This paper presents the HERA CO-Line Extragalactic Survey (HERACLES), a new
survey of CO $J=\jtwo$ emission from nearby galaxies obtained using the IRAM
30m. Because on-the-fly mapping mode with the multi-pixel HERA allows us to
efficiently map large areas with good sensitivity and adequate resolution,
HERACLES targets a wider area than previous surveys. Most maps extend to the
edge of the optical disk, defined by the isophotal radius $r_{25}$.

Our sample overlaps THINGS and SINGS, allowing easy comparison to data from
radio to UV wavelengths. One application of this multiwavelength data allows
us to automate our reduction. We use the mean \hi\ velocity from THINGS as a
prior to define baseline fitting regions and are thus able to reduce $10^7$
spectra in a robust, automated manner. We verify the assumption that CO and
\hi\ exhibit the same mean velocity by direct comparison of HERACLES and
THINGS mean velocities, finding little or no systematic difference. We also
confirm that our automated approach matches a by-hand reduction well.

We clearly detect 14 galaxies and place upper limits on CO $jtwo$ emission
from 4 low metallicity, irregular dwarf galaxies.  For a Galactic CO-to-H$_2$
conversion factor, the implied H$_2$ masses are mostly $0.2$--$0.6$ times the
\hi\ mass and $0.03$--$0.1$ times the stellar mass. More detailed analysis of
the relationship between H$_2$, \hi , stars, and star formation in our sample
is presented elsewhere \citep{BIGIEL08,LEROY08}.

We illustrate the distribution of CO in the detected galaxies via channel
maps, azimuthally averaged profiles, and intensity maps. Where we find
emission, the brightness is usually consistent with $\sim 5$--$15\%$ of the
area inside the beam being covered by Galactic GMCs, though in a few regions
the implied surface density averaged over our $\sim 500$~pc beam exceeds that
of a Galactic GMC. The line width and mean velocity that we derive agree
reasonably with previous (CO $J=\jone$) surveys; systematic
differences among surveys do exist, with magnitude $\sim 10$--$20\%$, but
HERACLES never appears to be an outlier.

The ratio of CO $J=\jtwo$ intensity to CO $J=\jone$ intensity
for high significance spectra lies mostly in the range $0.6$ -- $1.0$ with an
average value of $0.8$, comparable to that found in the inner Milky Way. This
could be produced by optically thick gas with an excitation temperature $\sim
10$~K, though this temperature should not be over-interpreted without
constraints on the optical depth or applicability of LTE. This line ratio is
higher ($\sim 1.3$) in the centers of galaxies and then roughly constant as a
function of radius, though we caution that strong selection effects may be at
work.

Our detections include high significance emission from the outer part of the
star-forming disk in several galaxies. We also find bright CO emission
associated with an \hi\ filament outside the main disk of NGC~4736 (radius
$1.33~r_{25}$). The azimuthally averaged intensity usually declines smoothly
even beyond these detections and we are sensitive to only the most massive
individual GMCs. Therefore we would expect to find many more such regions with
improved sensitivity.

Although the channel and integrated intensity maps show a variety of
morphologies, the azimuthally averaged profiles are well-described by
exponential declines. The best-fit scale lengths range from $0.8$ -- $3.2$~kpc
and correlate closely with optical radius, near-IR (stellar mass) scale
length, and UV+IR (star formation) scale length. These results are in good
agreement with previous findings that CO emission closely follows both the
stellar light and distribution of star formation on large scales. The
exponential decline in CO brightness is a combination of decreasing maximum
brightness of CO emission and a decreased filling fraction of bright CO
emission at large radii.

\acknowledgments We thank the anonymous referee for helpful comments. We thank
J\'er\^ome Pety for his help with the CLASS portion of the reduction pipeline,
and the IRAM~30-m staff for their help with the observations, In particular
operators Juan Luis Santar\'en, Fr\'ed\'eric Damour, Enrique Lobato, and
Mariano Espinosa and support astronomers Gabriel Paubert, Rebeca Aladro, and
Denise Riquelme. FB acknowledges support from the Deutsche
Forschungsgemeinschaft (DFG) Priority Program 1177. AU has been supported
through a Post Doctoral Research Assistantship from the UK Science \&
Technology Facilities Council. The work of WJGdB is based upon research
supported by the South African Research Chairs Initiative of the Department of
Science and Technology and National Research Foundation. We acknowledge use
of: the NASA/IPAC Extragalactic Database (NED) which is operated by the Jet
Propulsion Laboratory, California Institute of Technology, under contract with
the National Aeronautics and Space Administration; the HyperLeda catalog,
located on the World Wide Web at
http://www-obs.univ-lyon1.fr/hypercat/intro.html; and NASA's Astrophysics Data
System (ADS).

\bibliography{akl}
\end{document}